\documentclass[aps,prd,groupedaddress]{revtex4}
\def\be{\begin{equation}}
\def\ee{\end{equation}}
\def\bea{\begin{eqnarray}}
\def\eea{\end{eqnarray}}
\input{epsfig.sty}
\usepackage{graphicx}
\begin{document}
\title{Real and Virtual Nucleon Compton Scattering in the Perturbative Limit}
\author{Richard Thomson, Alex Pang\protect\footnote{ 
Present address: C-BASS LLC, 335 Madison 
Avenue 19th Floor, New York, NY 10017}
and Chueng-Ryong Ji\\
Department of Physics, North Carolina State University,\\
Raleigh, NC 27695-8202}
\begin{abstract}
\noindent
We present the results of calculations analyzing nucleon Compton scattering to lowest 
order using perturbative QCD (pQCD) methods. Two scenarios are considered: (1) the 
incoming photon is real; and (2) the incoming photon is virtual. The case of a real 
photon has been previously analyzed at least 5 times using pQCD, but no two results are 
in agreement. Here it is shown that our result agrees with that of Brooks and Dixon 
published in 2000. The case of a virtual photon has been previously analyzed only once 
using pQCD. However, doubt has been cast on the validity of that result. The results 
presented here for virtual photon are believed to be more reliable. Some consideration is 
given of how to compare these results with experiment. Following the lead of Brooks and 
Dixon, for the proton, this involves normalizing the cross section using the Dirac proton 
form factor, which we also calculate. Finally, there is a 
comparison of our results with recent experiments. 
\end{abstract}

\date{\today}

\maketitle

\section{Introduction}
\label{sect.I}
\noindent
Nucleon Compton scattering provides an important experimental probe of the structure of both the 
proton and the neutron. One area where it is possible to perform a theoretical analysis of such experiments 
is in the high energy limit where the asymptotic perturbative behavior is applicable. This perturbative 
domain is the focus of our paper. We address both real Compton scattering 
where the incoming photon is real and virtual Compton scattering where the incoming photon is virtual. 
For a more general overview of all aspects of nucleon Compton scattering, the reader is referred to the review 
written by Vanderhaeghen \cite{V}. In the high energy limit and at large momentum transfer, 
it is possible to use the formalism developed by Brodsky and Lepage \cite{Lep80}, in which the scattering amplitude is 
factorized as a convolution of a perturbative 
hard scattering kernel with a non-perturbative distribution amplitude for the three valence quarks. Several models 
are available for the distribution amplitude, based on QCD sum rules. It is anticipated that the results presented 
here can shed some light on the validity of these models. 

The case of real Compton scattering has been computed a number of times \cite{FM,FZ,KN,BD,VG}, each time with different 
results. We found that our own results do agree quite well with the most recent calculation made by Brooks and Dixon \cite{BD}. 
The virtual Compton scattering has been calculated only once previously \cite{FZ}. The virtual calculation is 
considerably more complex than the real calculation, due to the following reasons. First, there are over twice as many 
Feynman diagrams that need to be considered. Second, the pole structure occuring in the virtual case is more complex than 
in the real case. Third, for the virtual case there is also the need to consider longitudinal/temporal 
polarization of the incoming photon. Generation of the relevant Feynman diagrams has been done using a software package, 
FeynGen \cite{COMPUTE}. The corresponding hard scattering amplitudes have also been automatically generated using a 
software package, FeynComp \cite{SpTech}. Both FeynGen and FeynComp are symbolic software packages developed in Maple 
at NC State University. 

In recent years serious criticisms have been raised about the validity of 
the Brodsky-Lepage (BL) approach, at experimentally accessible energies. Also the models used for the distribution amplitude (DA) 
have been questioned. Much of the criticism has focussed on the calculation of the Dirac proton form factor. Using the BL approach and 
a highly asymmetric DA such as that of Chernyak, Oglobin and Zhitnitsky (COZ) \cite{COZ}, the form factor can be calculated with 
a result quite close to experiment. However, this calculation is questionable:\\
(a) are such asymmetric DAs realistic?\\
(b) since the dominant contribution is coming from the end-point regions where pQCD breaks down, the calculation appears to be 
logically inconsistent.

Regarding the first question, the authors recognize that, for the pion, it is now widely accepted that the DA should resemble the 
asymptotic DA. However, there is no such consensus for the DA of the proton. If one models the proton as a quark-diquark 
configuration, it appears quite likely that the proton DA will not be symmetrical. Since there is considerable doubt about the 
correct form of the DA, we present our results in two formats: the first, presents specific results for cross-sections and phases, based on specific DAs defined in the literature, such as COZ; and the second tabulates results that allow cross-sections and phases to be calculated for any DA which can be written as a combination of low order Appell polynomials. Due to the large size of the Appell polynomial tables, they are not included in the version of this paper submitted to 
Physical Review D, but they are included in this online version of the paper.  

Regarding the second question, there have been efforts to modify the form factor calculation to ensure its logical consistency. 
Specifically, these efforts have involved inclusion of transverse quark momenta and the use of a Sudakov form factor (see 
\cite{LS} for the pion, \cite{LI,BKBS,KLSJ} for the proton). There are differences in the results of these papers, ranging from 
those concluding that the modified perturbative calculation is close to the experimental data, to those concluding that it is 
only a small part of the experimental data, the rest coming from soft contributions (Feynman mechanism). There does not appear to 
be a consensus, although the authors would agree with the opinion of Sterman and Stoler \cite{SS} that soft non-perturbative 
processes probably play an important role for much of the existing data.

This logical consistency concern regarding the proton form factor is also relevant to 
proton Compton scattering, since there is a direct relationship between the Compton 
amplitude and the form factor \cite{BL2}. Our calculation addresses this concern 
based on the following points.\\
(a) When applied to numerical calculations, Li and Sterman \cite{LS} indicate that inclusion of Sudakov corrections produces an 
effect very similar to the use of a frozen coupling constant. As explained in Section II, our method of calculation is 
equivalent to the use of a frozen coupling constant.\\ 
(b) We have looked at the gluon momenta for specific diagrams that make the largest contributions to the amplitude. We have 
calculated $\left<|q^2|\right>$ for such diagrams, where $q$ is the gluon 4-momentum and $\left<...\right>$ means averaging over the DA. In the case of 
COZ, we find that $\left<|q^2|\right> > \sim 0.02s$, where $s$ is the Mandelstam invariant. We may reasonably expect experiments in the near 
future with $s=20$ GeV$^2$ which would give $\left<|q^2|\right> > \sim 0.40$ GeV$^2$. Taking a value of $0.1$ GeV for $\Lambda_{QCD}$, we have that 
$\left<|q^2|\right> >> \Lambda_{QCD}^2$, so, in this case, we can expect that the non-perturbative end-point regions will not make a significant 
contribution.\\
(c) In order for pQCD to be valid, the condition $|q^2|>\mu^2$ is often used, where $\mu^2$ represents a threshold based on 
$\Lambda_{QCD}$. However, it has been shown \cite{JPS} that this condition can be made less stringent. For $q^2=kx_1y_1$, 
for example, where $k$ is some constant, $|q^2|>\mu^2$ implies that all $(x_1,y_1)$ are to be excluded that lie below the hyperbola $kx_1y_1=\mu^2$. 
Light front analysis shows that there is a substantial area below the curve where $x\sim y$ for which the gluon light front energy 
is above the threshold. Furthermore, in the case of the pion form factor, if this additional area is allowed, the calculated 
value tends to the value calculated with no threshold at much lower momentum transfer. For example, with a very conservative 
threshold of $1$ GeV$^2$, at momentum transfer of $10$ GeV$^2$, the calculated value is already about half of the value calculated 
with no threshold. 

The other sections of this paper are organized as follows. Section II gives details of the method used in the calculations. 
In Section III we present our results for both real and virtual Compton scattering. There is also a discussion of the 
normalization of the results and, related to that, our calculation of the Dirac proton form factor. Section IV  
compares our results with experimental results for both real Compton scattering and virtual Compton scattering. 
Section V gives our conclusions. There are four appendices: Appendix A details the calculation 
for a specific diagram; Appendix B gives details of the 'folding method' used to evaluate principal part integrals; 
Appendix C gives details of the method used to calculate the Dirac proton form factor; and Appendix D tabulates results for different Appell polynomials, allowing cross section and phase to be calculated for any distribution amplitude.  

\section{Calculations}
\label{sect.II}
\noindent
In perturbative QCD, the helicity amplitude $\cal{M}$ for nucleon Compton scattering is given by
\begin{equation}
\label{convol}
{\cal M}^{\lambda\lambda^\prime}_{hh^\prime}(s,t)=\sum_{d,i}\int[dx][dy]
\phi_i(x_1,x_2,x_3)T^{(d)}_i(x,h,\lambda;y,h^\prime,\lambda^\prime)
\phi^\star_i(y_1,y_2,y_3).
\end{equation}
Here $\lambda$ ($\lambda^\prime$) and $h$ ($h^\prime$) are the helicity of the initial (final) photon and nucleon states, respectively;
$(x_1,x_2,x_3)$ and $(y_1,y_2,y_3)$ are the momentum fractions carried by the three quarks in the initial and final states, respectively;
$\phi_i$ is the distribution amplitude for the $i$th spin/flavor
state in the nucleon; and $T^{(d)}_i$ is the hard scattering amplitude for a specific Feynman diagram, $d$. 
The sum is over all possible diagrams and over the spin/flavor states. 
In leading-twist approximation, only valence collinear quarks contribute. The integration measure $[dx]$ implies integration over 
$x_1,x_2$ and $x_3$ subject to the constraint $x_1+x_2+x_3=1$. 

It is convenient to rewrite the hard scattering amplitude as
\begin{equation}
T^{(d)}_i(x,h,\lambda;y,h^\prime,\lambda^\prime)=
C^{(d)} g^4 Z^{(d)}_i \tilde{T}^{(d)}(x,h,\lambda;y,h^\prime,\lambda^\prime),
\end{equation}
where $C^{(d)}=4/9$ is the common color factor, g is the QCD coupling
constant, $Z^{(d)}_i$ is the product of the electric charges of the
struck quarks, and $\tilde{T}^{(d)}(x,h,\lambda;y,h^\prime,\lambda^\prime)$
being the rest of the expression. 
The advantage is that 
$\tilde{T}^{(d)}(x,h,\lambda;y,h^\prime,\lambda^\prime)$
is now a color- and flavor-independent quantity.
In the rest of the paper, when there is no confusion, we frequently
refer to $\tilde{T}$ as the "hard scattering amplitude".

\subsection{Hard Scattering Amplitude}
\begin{figure}
\centerline{\epsfig{figure=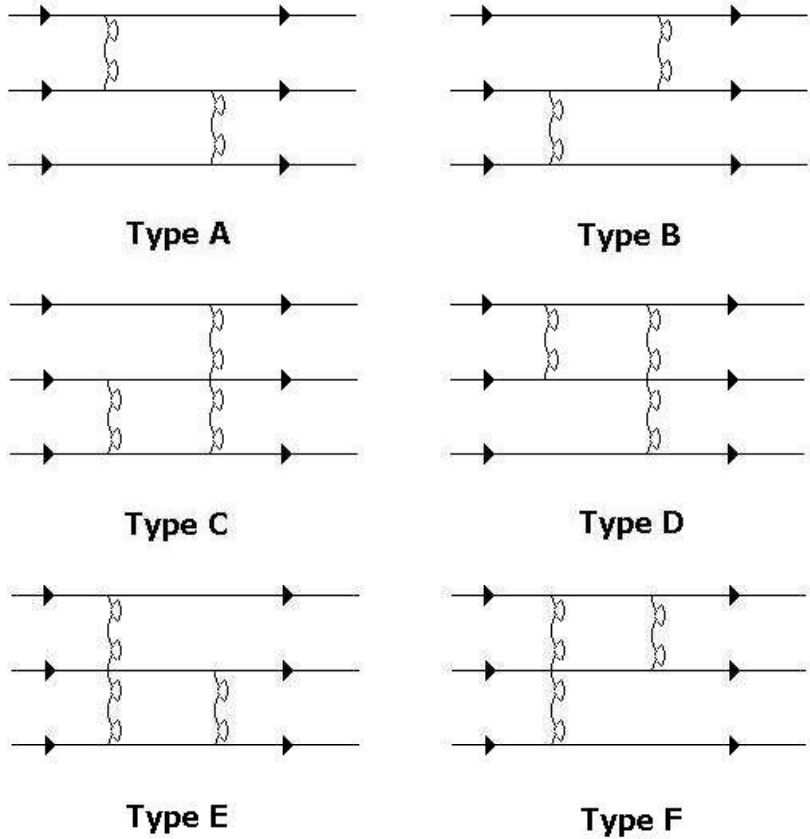,height=5in}}
\caption{Classification of diagrams with two gluons attached 
\label{vnc_gluon}}
\end{figure}

\noindent
To leading order, the hard scattering amplitude is given by
the scattering amplitude of the collinear valence quarks with a  
minimum number of gluon exchanges. 
For comparison, we use the same notation
and classification as in Ref.\cite{KN}. All the possible Feynman
diagrams can be classified according to the 
arrangement of the gluon lines in the diagram, as shown in Fig. \ref{vnc_gluon}.
There are 56 ways of attaching the two photons in groups A to F. 
Therefore altogether there are 6x56 = 336 Feynman
diagrams which can contribute to the hard-scattering amplitude. 

To simplify the calculations, one can exploit symmetry
among the diagrams. Diagrams in groups B, D and F
may be obtained from A, C, and E, respectively, by the interchange:
\begin{equation}
{\cal E}: x_1\leftrightarrow x_3, y_1\leftrightarrow y_3, e_1\leftrightarrow e_3.
\end{equation}
By making use of this symmetry, one has only to calculate the diagrams in groups A, C and E. The contributions for groups B, D and F are then accounted for by doubling the results for groups A, C and E. For real Compton scattering, time-reversal symmetry is also available. However, since this 
symmetry does not apply in the case of virtual Compton scattering, it has not been used 
in the calculations (except as a means for checking the results for real Compton scattering).

In order to calculate the hard scattering amplitudes using the FeynComp package, symbolic definitions must be provided for the incoming/outgoing momenta and for the polarization vectors of the incoming/outgoing photon. Working in the center-of-mass frame, the incoming (outgoing)
nucleon and photon momenta, $p$ and $q$ ($p^\prime$ and $q^\prime$)
are given by
\begin{eqnarray}
p &=& P(1,0,0,1),\\
q &=& P(\sqrt{1-Q^2/P^2},0,0,-1), \\
p^\prime &=& E(1,\sin\theta,0,\cos\theta), \\
q^\prime &=& E(1,-\sin\theta,0,-\cos\theta),
\end{eqnarray}
in Cartesian components. The angle $\theta$ is the scattering angle. 
$E$ is the center-of-mass energy and is related to the
Mandelstam invariant in the $s$-channel by
$s=(p+q)^2=(p^\prime+q^\prime)^2=4E^2$, where we have neglected
masses, since we are working in the leading-twist approximation. $P$ represents the magnitude of the 3-momentum of the 
incoming nucleon and photon. $Q^2$ is defined as $-q^2$. $P$ can be related to $s$ and $Q$ by applying energy conservation 
to give
\begin{equation}
P+P\sqrt{1-Q^2/P^2}=\sqrt{s}.
\end{equation}

The transverse photon polarization vectors,
denoted by $\epsilon_i$ and $\epsilon_f$ for incoming and
outgoing photon respectively, are given by
\begin{eqnarray}
\epsilon_i(\uparrow)&=&{1\over \sqrt{2}}(0,1,-i,0), \\
\epsilon_i(\downarrow)&=&{-1\over \sqrt{2}}(0,1,i,0), \\
\epsilon_f(\uparrow)&=&{1\over \sqrt{2}}(0,\cos\theta,-i,-\sin\theta),\\
\epsilon_f(\downarrow)&=&{-1\over\sqrt{2}}(0,\cos\theta,i,-\sin\theta).
\end{eqnarray}
We define a quantity $R=1+Q^2/s$, which we refer to as the virtuality parameter. Therefore, $R=1$ refers
to the real case ($Q^2=0$) and $R=2$ refers to the deeply virtual case ($Q^2$ large). 

For the virtual photon case, we must additionally define a longitudinal/temporal polarization vector for the 
incoming photon. This vector must be orthogonal to the two transverse polarization vectors and its spatial components 
must be parallel to the momentum of the incoming photon. In addition, we impose the Lorenz condition, 
$k^\mu\epsilon_\mu=0$, to obtain
\begin{eqnarray}
\epsilon_i(L)&=&{1\over \sqrt{4R-4}}(R,0,0,R-2). 
\end{eqnarray} 
Note that this polarization vector is normalized to +1, whereas the transverse 
polarizations are normalized to -1.

With the kinematic definitions above, FeynComp has been used to generate the hard scattering amplitudes for all 336 diagrams. 
It turns out that only 192 of these diagrams have a non-zero amplitude. Exactly one half of these diagrams (96 diagrams) belong to the 
groups A, C and E. A check was made of the hard scattering expressions by setting $R=1$. For this case, it was verified that  
the hard scattering expressions reduce to the 
expressions given in Table III and Table VI of Ref. \cite{KN}, with the noted corrections in Ref. \cite{BD}.

\subsection{Distribution amplitude}
\noindent
Our calculations consider the case of an incoming proton with helicity $h=+1$. In the leading twist approximation, the outgoing proton will 
have the same helicity as the incoming proton. Results for $h=-1$ are easily obtained by using the result
\begin{equation}
{\cal M}^{\lambda\lambda^\prime}_{--}(s,t)={\cal M}^{\bar{\lambda}\bar{\lambda}^\prime}_{++}(s,t),
\end{equation}
where overbar denotes opposite helicity. With the collinear approximation, a proton state with
helicity $h=+1$ is
\begin{equation}
\label{flip}
|p_\uparrow>={f_N\over 8\sqrt{6}}\int[dx]\sum_i\phi_i(x_1,x_2,x_3)
|i;x_1,x_2,x_3>,
\end{equation}
where $f_N$ is the nucleon decay constant and $x_i$
(with $x_1+x_2+x_3=1$) are the
momentum fractions carried by the three quarks.
The spin-flavor states are
\begin{eqnarray}
|1;x_1,x_2,x_3>&=&|u_\uparrow(x_1)u_\downarrow(x_2)d_\uparrow(x_3)>,\\
|2;x_1,x_2,x_3>&=&|u_\uparrow(x_1)d_\downarrow(x_2)u_\uparrow(x_3)>,\\
|3;x_1,x_2,x_3>&=&|d_\uparrow(x_1)u_\downarrow(x_2)u_\uparrow(x_3)>.
\end{eqnarray}
The distribution amplitudes are $\phi_i (i=1,2,3)$ and only one of them,
say $\phi_1$, is an independent degree of freedom. The others are given by
\begin{eqnarray}
\phi_2(x_1,x_2,x_3)&=&-\left[\phi_1(x_1,x_2,x_3)+\phi_1(x_3,x_2,x_1)\right],\\
\phi_3(x_1,x_2,x_3)&=&\phi_1(x_3,x_2,x_1).
\end{eqnarray}
The state with helicity $h=-1$ is obtained by flipping the helicity
of all the quarks in Eq. (\ref{flip}); neutron states are obtained by
switching up and down quarks and multiplying the state by
the factor -1.

There are several common models for the distribution amplitude
$\phi_1$ from QCD sum rules. The ones which have been used in our calculations are those of 
Chernyak, Oglobin and Zhitnitsky (COZ) \cite{COZ} and King and Sachrajda (KS) \cite{KS}:
\begin{eqnarray}
\phi_{COZ}&=&120x_1x_2x_3(5.880-25.956x_1-20.076x_3+36.792x^2_1+19.152x^2_3+25.956x_1x_3),\\
\phi_{KS}&=&120x_1x_2x_3(8.40-26.88x_1-35.28x_3+35.28x^2_1+37.80x^2_3+30.24x_1x_3).
\end{eqnarray}
In addition we have calculated using the asymptotic distribution amplitude, defined simply as 
\begin{eqnarray}
\phi_{ASY}&=&120x_1x_2x_3.
\end{eqnarray}

The distribution amplitude is dependent on 
the renormalization scale $Q$, i.e. $\phi(x_i,Q)$ is a function of
$Q$ as well. A discussion of the evolution of the distribution 
amplitude and the running of $\alpha_s$ is given in Section II-D.
 
\subsection{Convolution of hard scattering amplitude with
distribution amplitudes}
As noted previously, the symmetry operation ${\cal E}$ can be used to simplify the
calculation by restricting the sum to diagrams in groups A, C and E and then multiplying the result 
by 2 to account for the groups B, D and F. Thus
\begin{equation}
{\label M}
{\cal M}=2{4\over 9}g^4 \sum_{d\in A\cup C\cup E}\int[dx][dy]
\tilde{T}^{(d)}(x;y)
\times\sum_{i}\left[Z^{(d)}_i\phi_i(x)\phi^\star_i(y)\right].
\end{equation}
Since the distribution amplitudes are polynomials of the form
\begin{equation}
\phi(x_1,x_2,x_3)=120x_1x_2x_3(b_0+b_1x_1+b_3x_3+b_{11}x^2_1
+b_{13}x_1x_3+b_{33}x^2_3),
\end{equation}
the factor $\sum_{i}Z^{(d)}_i\phi_i(x)\phi^\star_i(y)$ is given
by
\begin{equation}
\sum_i Z^{(d)}_i\phi_i(x)\phi^\star_i(y)=
(120)^2\sum_{m_1,m_3,n_1,n3}{\cal C}^{(d)}(m_1,m_3|n_1,n_3)
x^{m_1+1}_1x_2x^{m_3+1}_3y^{n_1+1}_1y_2y^{n_3+1}_3,
\end{equation}
where the coefficients ${\cal C}^{(d)}(m_1,m_3|n_1,n_3)$ are given by, for example,
\begin{eqnarray}
{\cal C}^{(d)}(0,0|0,0)&=&Z_1b^2_0+4Z_2b^2_0+Z_3b^2_0,\\
{\cal C}^{(d)}(0,1|0,0)&=&Z_1b_0b_3+2Z_2b_0(b_1+b_3)+Z_3b_0b_1.
\end{eqnarray}
%\begin{table}
%\caption{Matrix elements of the coefficient matrix
%${\cal C}^{(d)}(m_1,m_3|n_1,n_3)$}
%\epsfysize=7.in
%\epsffile{scv.ps}
%\label{scv}
%\end{table}
Therefore, Eq.(\ref{M}) becomes
\begin{eqnarray}
{\cal M}=2{4\over 9}(4\pi\alpha_{em})(4\pi\alpha_s)^2 
\left[{120f_N\over 8\sqrt{6}}\right]^2
\sum_{d\in A\cup C\cup E}
\sum_{m_1,m_3,n_1,n3}{\cal C}^{(d)}(m_1,m_3|n_1,n_3)
I^{(d)}(m_1,m_3;n_1,n_3),
\label {Mexpanded}
\end{eqnarray}
where
\begin{equation}
I^{(d)}(m_1,m_3;n_1,n_3)=\int [dx][dy]x^{m_1+1}_1x_2x^{m_3+1}_3
y^{n_1+1}_1y_2y^{n_3+1}_3 \tilde{T}^{(d)}(x,y).
\end{equation}
To evaluate the integral $I^{(d)}(m_1,m_3;n_1,n_3)$, the first
step is to eliminate one of the momentum fractions from
the integration measure $[dx]=\delta(1-x_1-x_2-x_3)dx_1dx_2dx_3$ and
$[dy]=\delta(1-y_1-y_2-y_3)dy_1dy_2dy_3$ by using the formula:
\begin{eqnarray}
\label{F}
\int [dx]F(x_1,x_2,x_3)&=&\int^1_0dx_1\int^{\bar{x}_1}_0 dx_3
\tilde{F}(x_1,x_3) \\
&=&\int^1_0dx_1\int^1_0dx_3 \bar{x}_1\tilde{F}(x_1,\bar{x}_1x_3),\nonumber
\end{eqnarray}
where $\tilde{F}(x_1,x_3)=F(x_1,1-x_1-x_3,x_3)$ and
$\bar{x_i}=1-x_i$. Next, the integrals are separated into real and imaginary 
parts by splitting the poles using the well-known result
\begin{equation}
{1\over{f(\alpha,\beta)+i\epsilon}}=P{1\over{f(\alpha,\beta)}}-i\pi \delta (f(\alpha,\beta)),
\end{equation}
where $\alpha \in \{ x_1,x_2,x_3,\bar{x_1},\bar{x_2},\bar{x_3}\}$, 
$\beta \in \{ y_1,y_2,y_3,\bar{y_1},\bar{y_2},\bar{y_3}\}$, 
$f(\alpha,\beta)$ is the pole expression, 
linear in $\alpha$ and $\beta$, appearing in the denominator of $\tilde{T}$, and $P$ indicates principal value. 
After manual integration of the delta functions, $I^{(d)}$ reduces to a sum of 
integrals of the form
\begin{equation}
\label{IT}
J^{(d)} =\int^0_1 dx_1dx_3\int^0_1 dy_1dy_3 x_1^{k_1}
\bar{x}_1^{\bar{k}_1} x_3^{k_3}\bar{x}_3^{\bar{k}_3} y_1^{l_1}
\bar{y}_1^{\bar{l}_1} y_3^{l_1}\bar{y}_3^{\bar{l}_3}
\prod^n_i P{1\over f_i(\alpha,\beta)},
\end{equation}
where $f_i(\alpha,\beta)$ are the poles,
$n$ is the number of poles and $k_i,\bar{k}_i,l_i,\bar{l}_i$
are integers.
Since many diagrams share the same pole structure, it is
more convenient to group diagrams with the same pole structure
and evaluate the integral $J$ for the whole group.
Table \ref{pole} lists the classification of the diagrams
according to the number of poles and their structures. In the table $<y,x>$ is an abbreviation for 
$y(1-Rs^2x)-Rc^2x$ where $s=\sin{\theta\over 2}$ and $c=\cos{\theta\over 2}$.

\begin{table}
\caption{Classification of the Feynman diagrams according
to their pole structure}
\label{pole}
\begin{tabular}{llll}
\hline
no. of poles & type & Pole structures \\
\hline
1 & 1a & $(1-\bar{x}_1R)$  \\
  & 1b & $<y_1,\bar{x}_3>$  \\
  & 1c & $<\bar{y}_1,1>$  \\
2 & 2a & $(1-\bar{x}_1R)<\bar{y}_1,\bar{x}_1>$  \\
  & 2b & $(1-\bar{x}_3R) <y_1,x_1>$  \\
  & 2c & $<y_1,x_1><y_1,\bar{x}_3>$  \\
  & 2d & $<y_1,x_1><\bar{y}_3,x_1>$  \\
  & 2e & $<\bar{y}_3,1><y_1,\bar{x}_3>$  \\
  & 2f & $<\bar{y}_1,1><\bar{y}_1,\bar{x}_1>$  \\
3 & 3a & $(1-\bar{x}_1R)<\bar{y}_1,\bar{x}_1><\bar{y}_1,x_3>$ \\
  & 3b & $(1-\bar{x}_3R)<y_1,x_1><\bar{y}_3,x_1>$ \\
  & 3c & $(1-\bar{x}_1R)<\bar{y}_1,\bar{x}_1><y_3,\bar{x}_1>$ \\
  & 3d & $<\bar{y}_1,1><\bar{y}_1,\bar{x}_1><\bar{y}_1,x_3>$ \\
  & 3e & $<\bar{y}_3,1><y_1,x_1><y_1,\bar{x}_3>$ \\
  & 3f & $<\bar{y}_1,1><\bar{y}_1,\bar{x}_1><y_3,\bar{x}_1>$ \\
  & 3g & $<\bar{y}_1,\bar{x}_1><\bar{y}_1,x_3><y_3,x_3>$ \\
  & 3h & $<y_1,x_1><y_1,\bar{x}_3><\bar{y}_3,\bar{x}_3>$ \\
4 & 4a & $(1-\bar{x}_1R)<\bar{y}_1,\bar{x}_1><\bar{y}_1,x_3><y_3,x_3>$\\
& 4b & $(1-\bar{x}_3R)<y_1,x_1><y_1,\bar{x}_3><\bar{y}_3,\bar{x}_3>$\\
& 4c & $<\bar{y}_1,1><\bar{y}_1,\bar{x}_1><\bar{y}_1,x_3><y_3,x_3>$\\
& 4d & $<\bar{y}_3,1><y_1,x_1><y_1,\bar{x}_3><\bar{y}_3,\bar{x}_3>$ \\
\hline
\end{tabular}
\end{table}

Some of the diagrams do not have exactly the same pole
structure. But after relabeling or changing integration variables, 
they share the same pole structure.
The basic strategy now is to integrate $J$ analytically
over as many momentum fractions as possible
for each type of pole structure in Table \ref{pole}.
An explicit example of the diagram $A51$ in the group $3F$ is
presented in the Appendix A.
Also, the treatment of the principal value integrations
is shown in the Appendix B.

In the real case, as shown in Ref.\cite{KN}, all the integrals
can be reduced to one or two-dimensional integral with at
most one pole. Therefore, one has only to consider how to 
handle the single pole principal integral. 
However, in the virtual case,
one cannot always reduce the integral to a single pole, so one 
must handle multiple-pole principle value integrals.
Moreover, the pole structure becomes more complicated in
the virtual case. More propagators can go on shell in the virtual
case and new pole structures appear.
For example, all the poles in the real case take the form
$\alpha(1-\beta s^2)-\beta c^2 +i\epsilon$, where
$\alpha$ and $\beta$ are momentum fraction variables and 
$s=\sin{\theta\over 2}$ and $c=\cos{\theta\over 2}$.
But in the virtual case, these poles become
$\alpha(1-\beta Rs^2)-\beta Rc^2+i\epsilon$. There are also
additional poles which do not appear in the real case, taking the form 
$(1-\alpha R)+i\epsilon$.

\subsection{Running of $\alpha_s$ and evolution of the distribution amplitude}
\noindent

$\alpha_s$ and the distribution amplitude evolution are $Q^2$-dependent variables appearing inside the integrations. 
It is possible to take them outside the integrations by setting them to average values, although it is not clear what 
these average values should be. However, note that in general we focus on dimensionless ratios, e.g. normalizing 
cross sections using the form factor or making a ratio of cross sections to find $K_{LL}$, the polarization transfer. 
If we make the assumption 
that the averages will be similar for both numerator and denominator, it appears that these ratios will not be 
greatly affected by the running of $\alpha_s$ or the DA evolution. Here we mention some 
specific calculations of the variations we expect due to running of $\alpha_s$ and the DA evolution which further 
justifies the approach of using average values.

Concerning the running of $\alpha_s$, our approach is based on using the 'frozen' coupling constant when the 
gluon momentum is small. This approach has some justification, since numerically it is expected to give results 
similar to those obtained using Sudakov corrections \cite{LS}. The approach is based on the use of an effective 
gluon mass $m_g$, giving
\begin{equation}
\alpha_s(Q^2)=\frac{4\pi}{\beta \text{log}[(Q^2+4m_g^2)/\Lambda_{QCD}^2]}
\end{equation}
The frozen value of $\alpha_s$ (i.e. setting $Q^2=0$) is sensitive to the choice of $m_g$ and $\Lambda_{QCD}$. Cornwall concludes 
\cite{Cornwall} that $m_g$ lies in the range $0.5\pm0.2$ GeV. Taking $\Lambda_{QCD}=0.1$ GeV and $\beta=9$ 
(assuming 3 flavors), leads to the conclusion
\begin{equation}
   0.26<\alpha_s(\text{frozen})<0.39.
\end{equation}
This can be taken as the upper bound of $\alpha_s$ in our integrations. The lower bound, in the kinematic region that we consider, 
can be approximated by taking $Q^2=s$, which for $s=12$ GeV$^2$ results in 
\begin{equation}
   0.19<\alpha_s(\text{lower bound})<0.20.
\end{equation}
Thus our average $\alpha_s$ is in the range
\begin{equation}
   0.19<\bar{\alpha}_s<0.39.
\end{equation}

Concerning the DA evolution, with the variation in $\alpha_s$ suggested, i.e. $0.19<\alpha_s<0.39$, one can calculate 
the corresponding variation for each Appell term in the DA expansion. For the $A_0$ term, there is no evolution; 
in general the $A_i$ term includes an evolution factor of
\begin{equation}
   [\alpha_s(\bar{Q}^2)/\alpha_s(\mu_0^2)]^{(b_i/\beta)},
\end{equation}
where $\bar{Q}^2=min(x_iQ)^2$, $\mu_0^2$ corresponds to a typical value of $Q^2$ in the interaction, $\beta=9$ 
(assuming 3 flavors), and $b_i = 20/9,24/9,32/9$ for $A_1,A_2,A_3$ respectively. The $A_3$ term will have a greater 
variation than $A_1$ or $A_2$, so here we concentrate on $A_3$. We can conclude that the evolution factor is greater than
\begin{equation}
    (0.19/0.39)^{(32/81)}=0.75,
\end{equation}
and less than
\begin{equation}
    (0.39/0.19)^{(32/81)}=1.33.
\end{equation}
The actual variation will be considerably less than this range, since $\mu_0^2$ should lie somewhere in the middle 
of the range, rather than at the ends. 

\section{Numerical Results}
\label{sect.III}
\noindent

In this section we describe the results of our calculations for both real and virtual photon. We give the results 
for three specific distribution amplitudes: Chernyak, Oglobin, and Zhitnitsky (COZ), King and Sachrajda (KS) and 
the asymptotic distribution amplitude. Since there is considerable doubt about the validity of the COZ and KS
distribution amplitudes, we have also tabulated results for the first four Appell polynomials ($A_0$, $A_1$, 
$A_2$, $A_3$) in Appendix D. This Appendix also includes instructions how to use these results to 
find cross sections and phases for any distribution amplitude which can be written as a linear combination of 
the Appell polynomials. Tables for $A_4$ and $A_5$ will be added at a later date. 

\subsection{Real Photon Results}
\begin{figure}
\centering
\includegraphics[angle=270,width=3.3in]{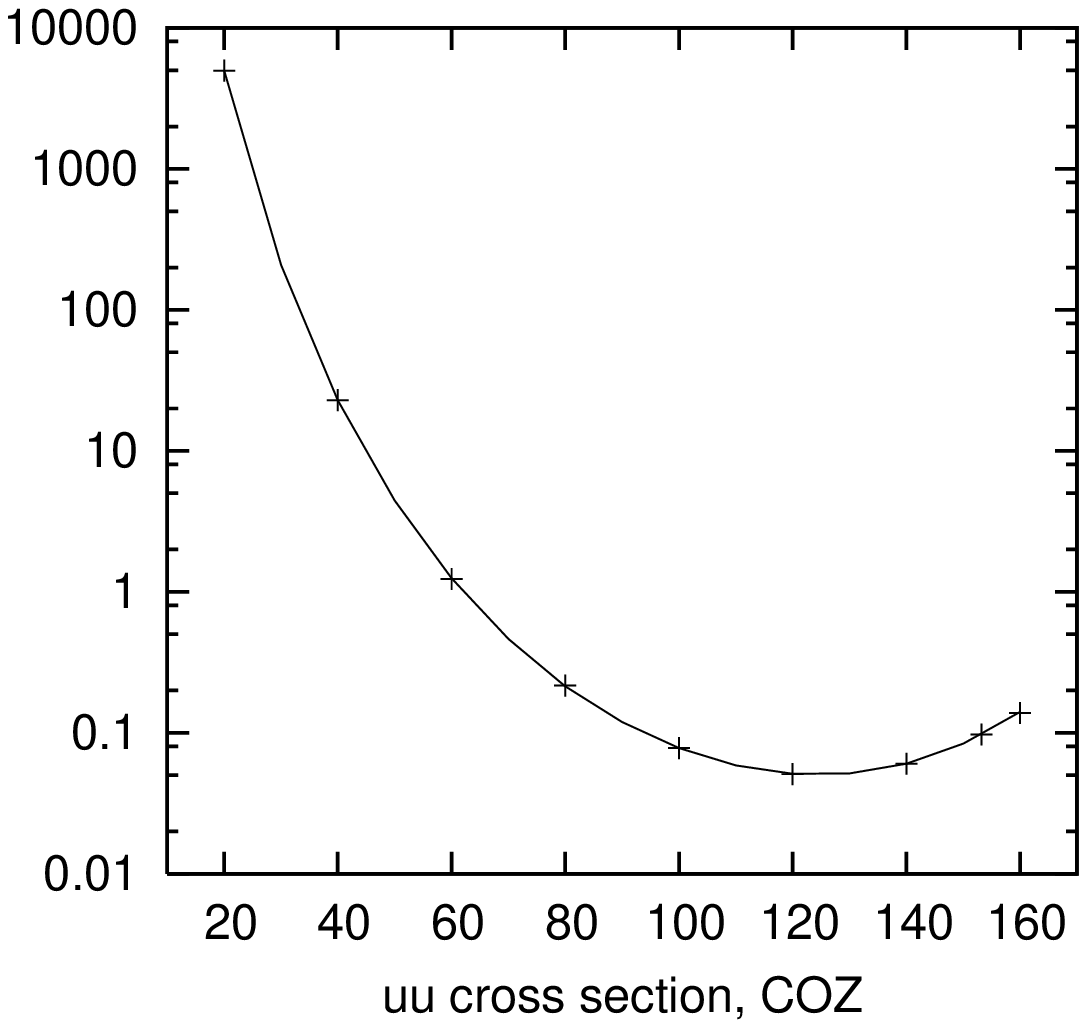}
\includegraphics[angle=270,width=3.3in]{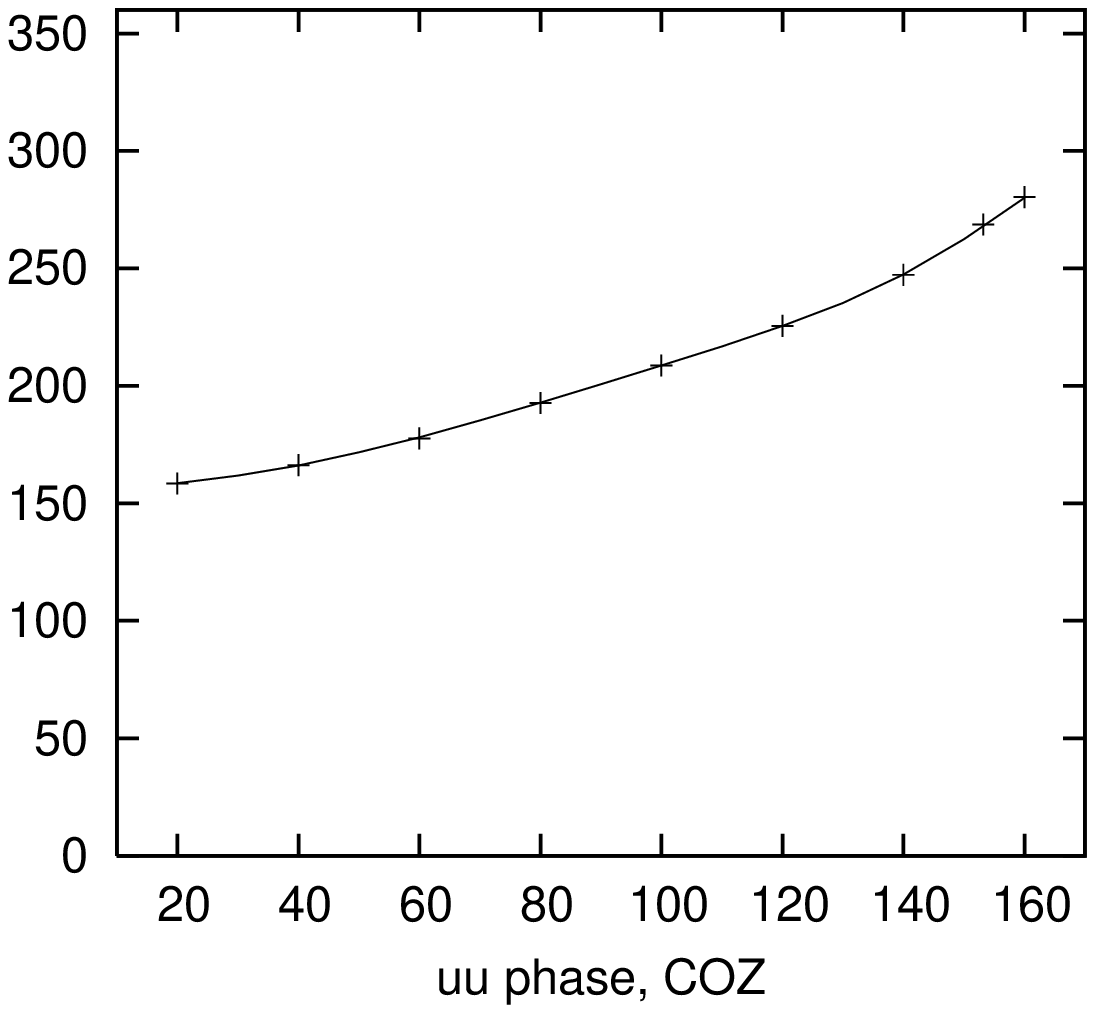}
\includegraphics[angle=270,width=3.3in]{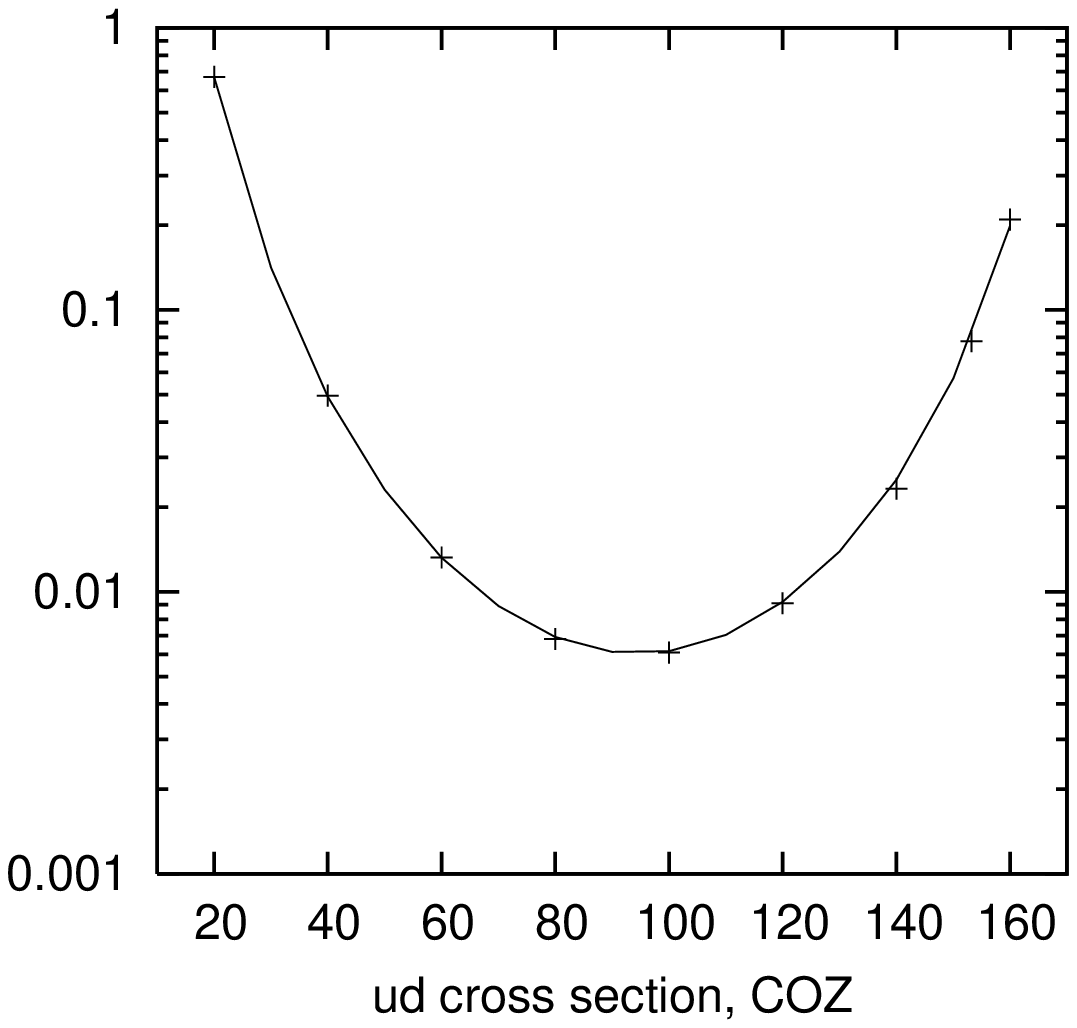}
\includegraphics[angle=270,width=3.3in]{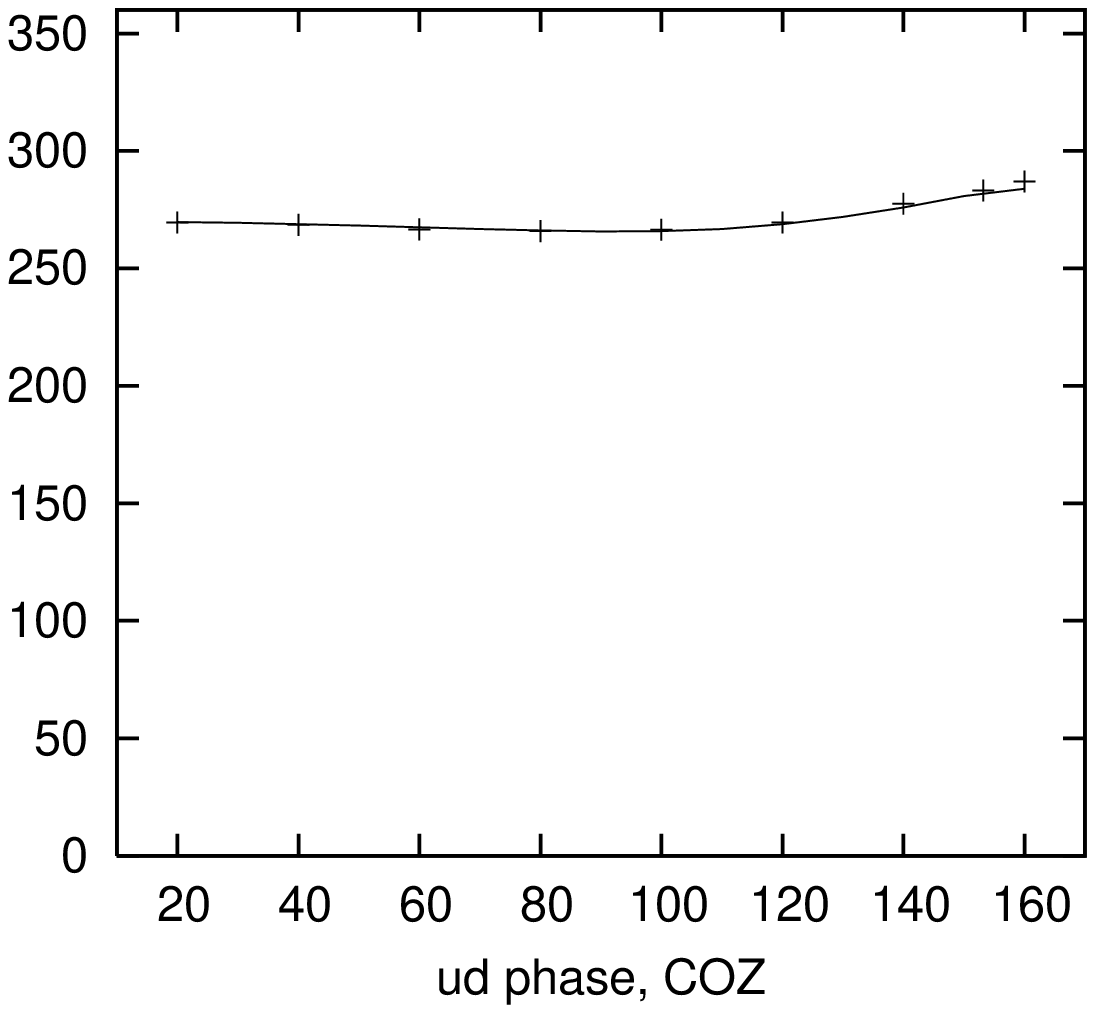}
\includegraphics[angle=270,width=3.3in]{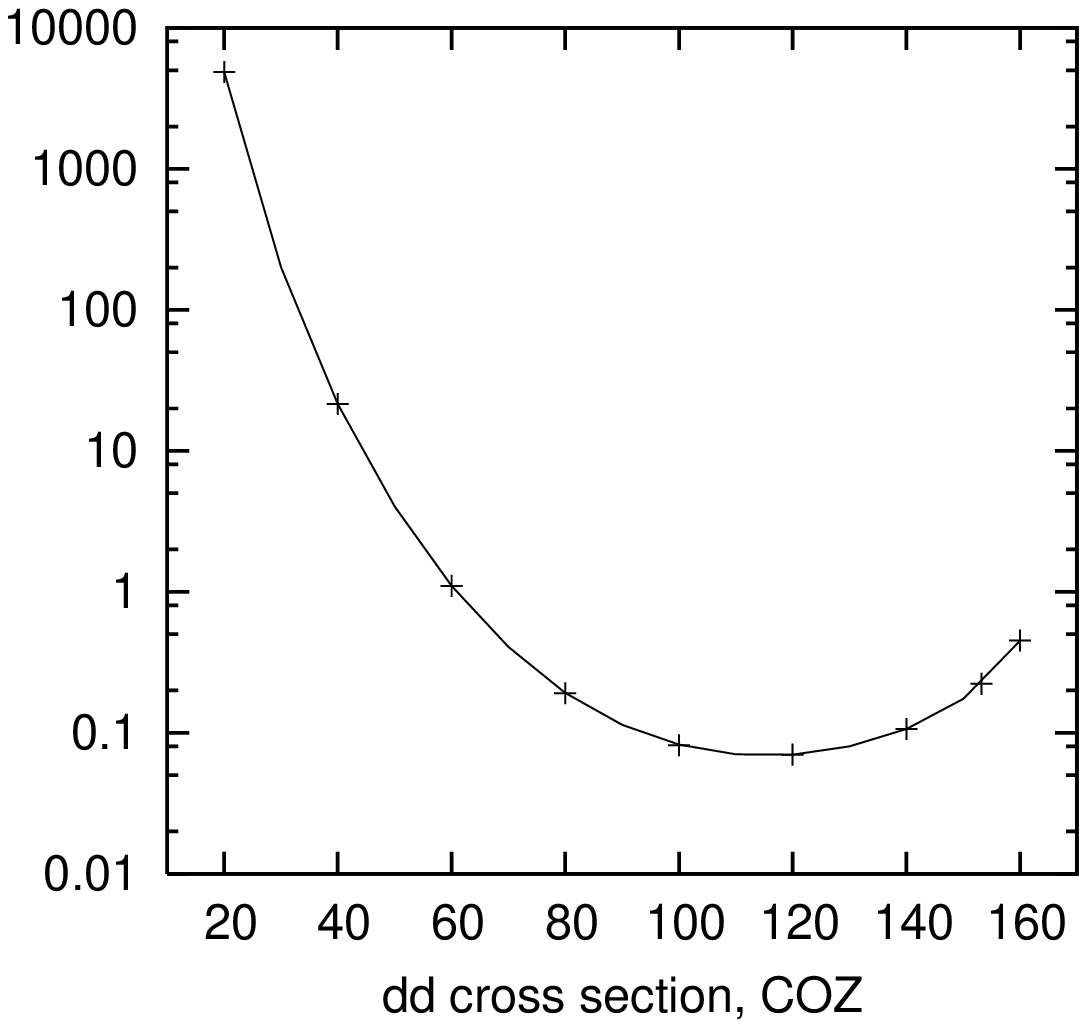}
\includegraphics[angle=270,width=3.3in]{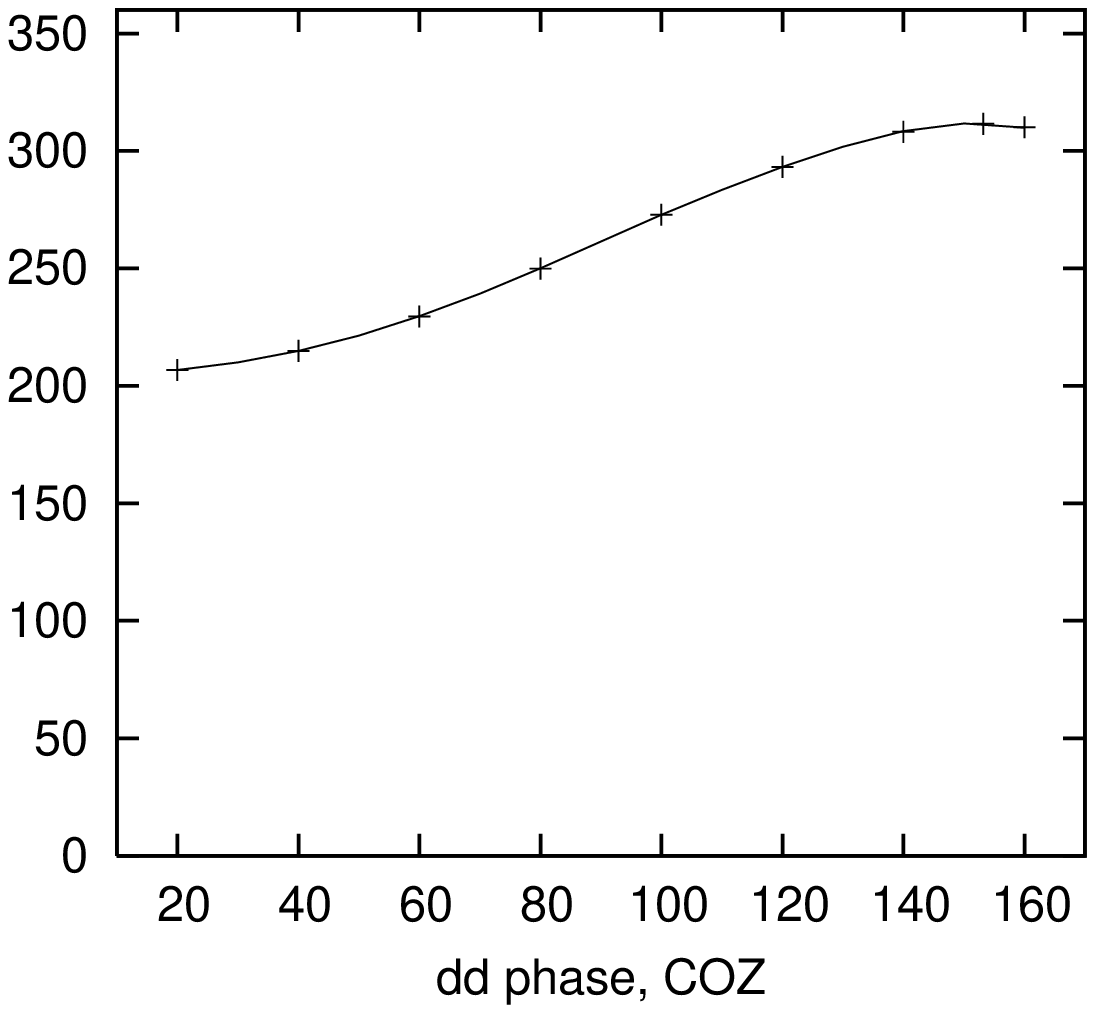}
\caption{Cross section and phase for the COZ distribution (real Compton scattering). Our results are shown 
as  a solid line. The results of Brooks and Dixon (+) are also shown. The vertical axis is 
$s^6d\sigma/dt$ ($10^4$ nb GeV$^{10}$) for cross section plots  and angle in degrees for phase plots. The horizontal 
axis is center of mass scattering angle for all plots. Note that results for 'down' to 'up' (du) polarization 
are not shown since they are identical to 'up' to 'down' (ud).  
\label{cozreal}}
\end{figure}

\noindent
The spin-polarized cross section is given by
\begin{equation}
{d\sigma^{\lambda\lambda^\prime}_{hh^\prime}\over dt}=
{1\over 16\pi s^2}|{\cal M}^{\lambda\lambda^\prime}_{hh^\prime}(s,t)|^2.
\end{equation}
Plots of cross section and phase against the center-of-mass scattering angle $\theta$ are shown in 
Fig. \ref{cozreal} for different polarizations, for
the Chernyak, Oglobin, and Zhitnitsky (COZ) distribution amplitude (solid line). For cross section we plot
$s^6d\sigma/dt$ rather than $d\sigma/dt$.
We computed the amplitudes in steps of $10^o$ for
$20^o\leq\theta\leq 160^o$. Also shown in the figure are the results of 
Brooks and Dixon \cite{BD}. It can be seen that our results are in excellent 
agreement with the work of Brooks and Dixon. This agreement is significant since none of 
the previous results are in agreement. Also, it has to be commented that there was no  collaboration 
between ourselves and Brooks and Dixon. Furthermore, our method of 
integration (the 'folding method' explained in Appendix B) is completely different 
from Brooks and Dixon, who use a contour deformation method. Based on these observations, the authors 
therefore have a very high degree of confidence in the results (and those of Brooks and Dixon). 
We have also calculated the cross section and phase for the King and Sachrajda (KS) distribution 
amplitude and for the asymptotic distribution amplitude. Again these results are in excellent agreement 
with the results of Brooks and Dixon.

In order to directly compare with the results of Brooks and Dixon,
we use the same approximations as theirs.
First, as explained in Section II, we use an average value of $\alpha_s$, which we set to 
$\bar{\alpha}_s=0.3$, in accordance with Brooks and Dixon. There does not appear to be a strong 
justification of this value, although we note that it does lie within our expected range, 
0.19 to 0.39. Second, in section II, we showed that the evolution of the distribution amplitude 
is expected to contribute a factor in the range 0.75 to 1.33. Again, in accordance with Brooks and 
Dixon, we set this factor to an averaged value of 1.0. Finally, we must fix the value of the 
nucleon decay constant. We have used the value $f_N=(5.2\pm0.3)\times 10^{-3}$ GeV$^2$
suggested by QCD sum rules, which again is in accordance with Brooks and Dixon.

We emphasise that, setting these values as described here, is done purely to allow a direct 
comparison of our results with earlier calculations. In Section IV, Comparison with Experiment, 
we do not make use of these values. Instead we normalize our results using the proton form factor, 
as explained later (see Eq. (\ref{Ratio})).

\subsection{Virtual Photon Results}
\begin{figure}
\centering
\includegraphics[angle=270,width=3.3in]{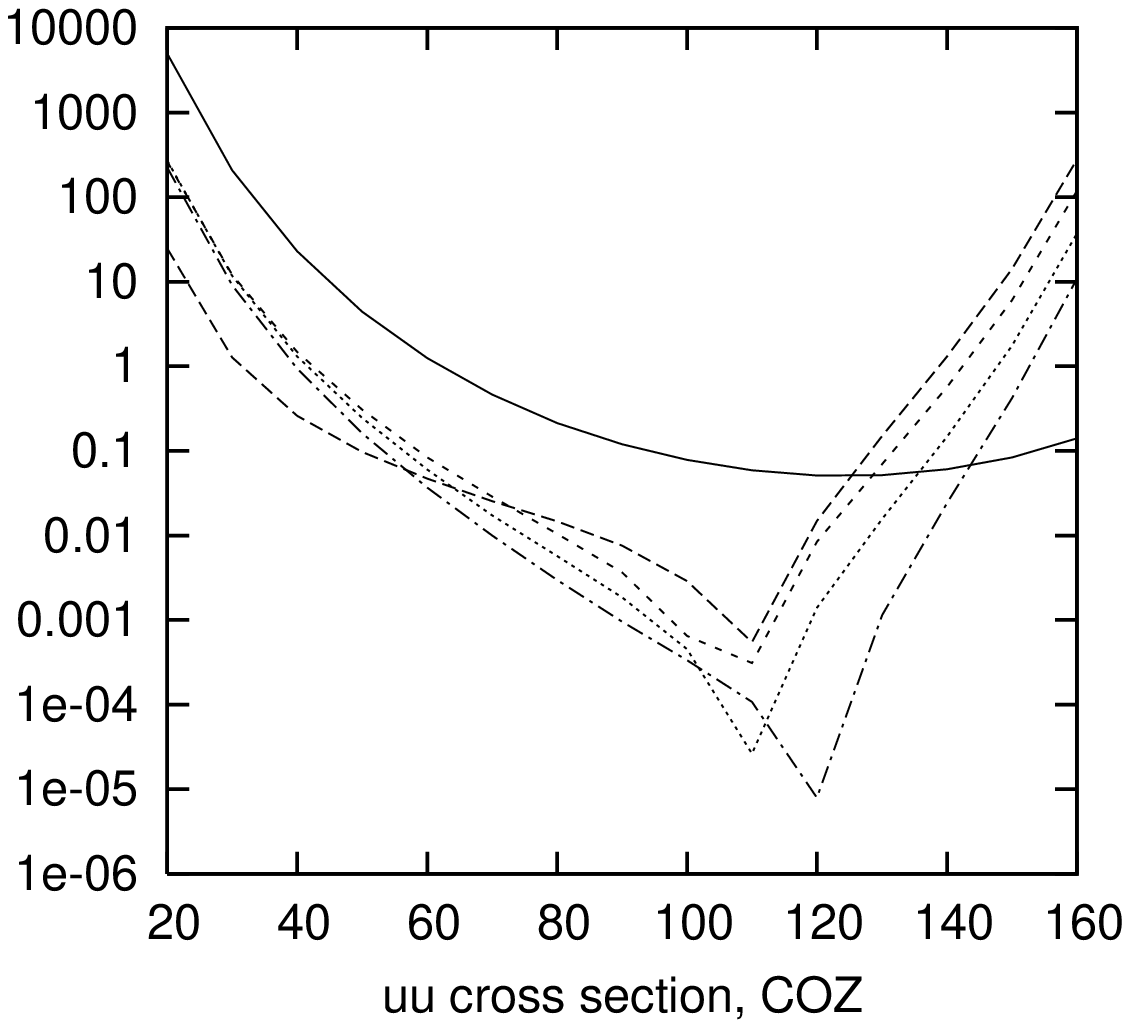}
\includegraphics[angle=270,width=3.3in]{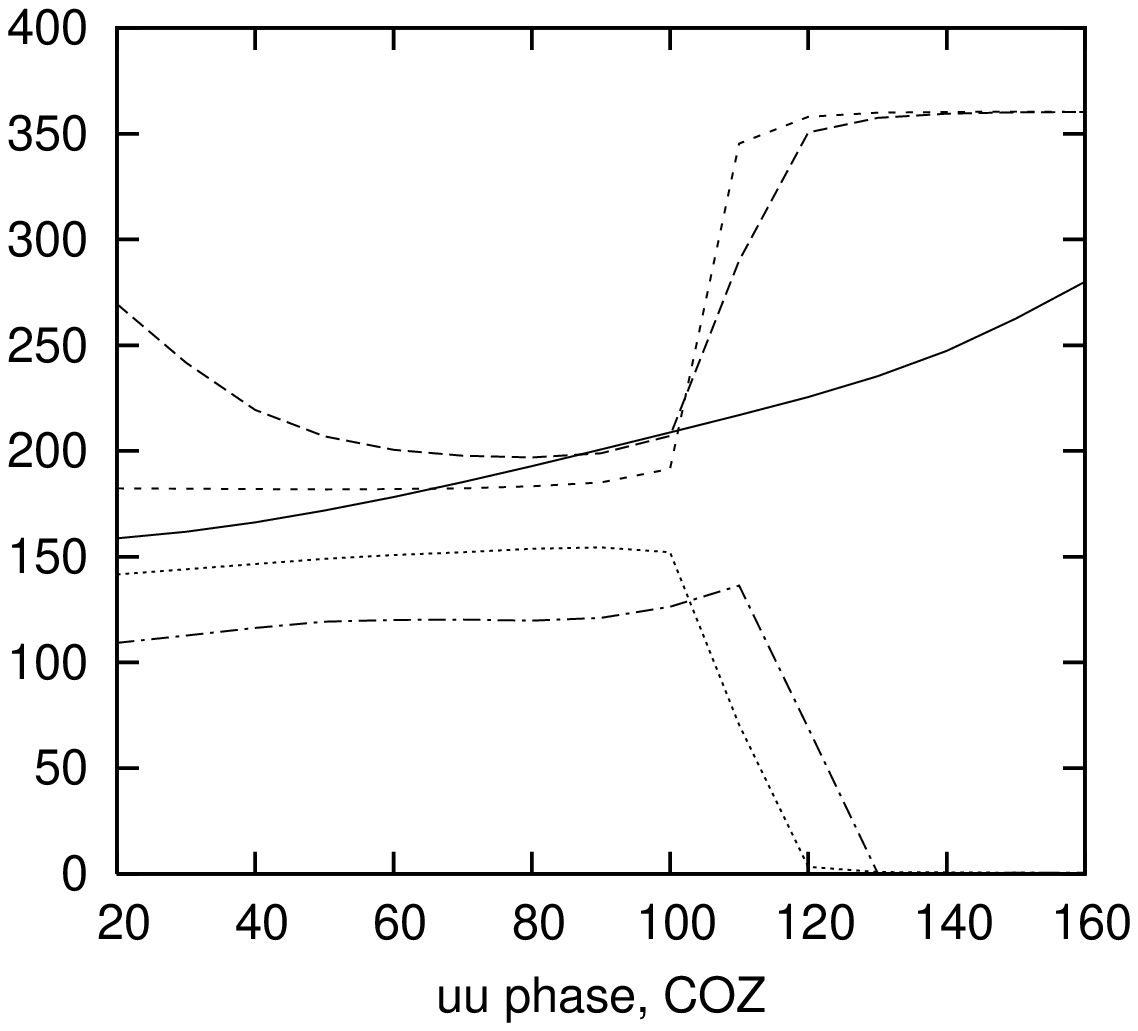}
\includegraphics[angle=270,width=3.3in]{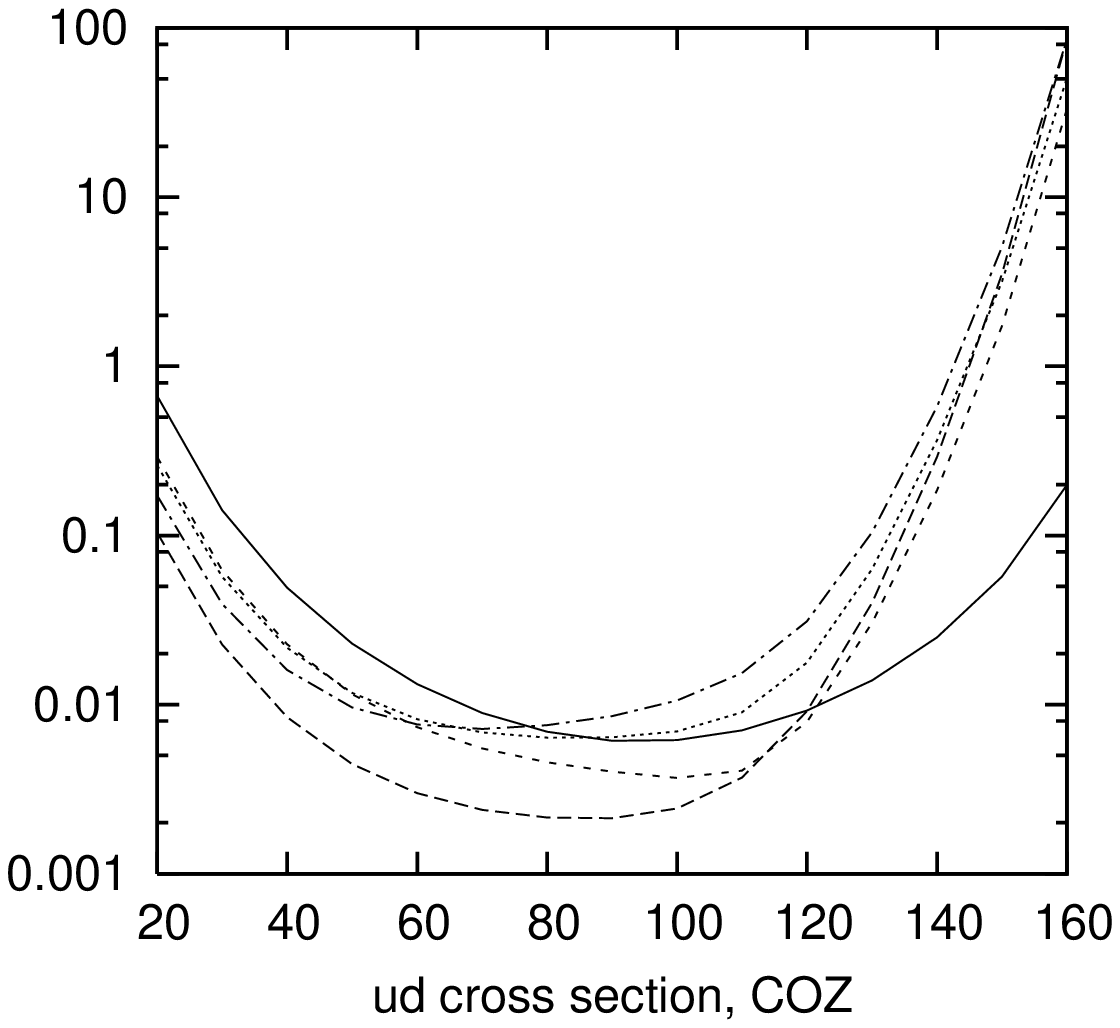}
\includegraphics[angle=270,width=3.3in]{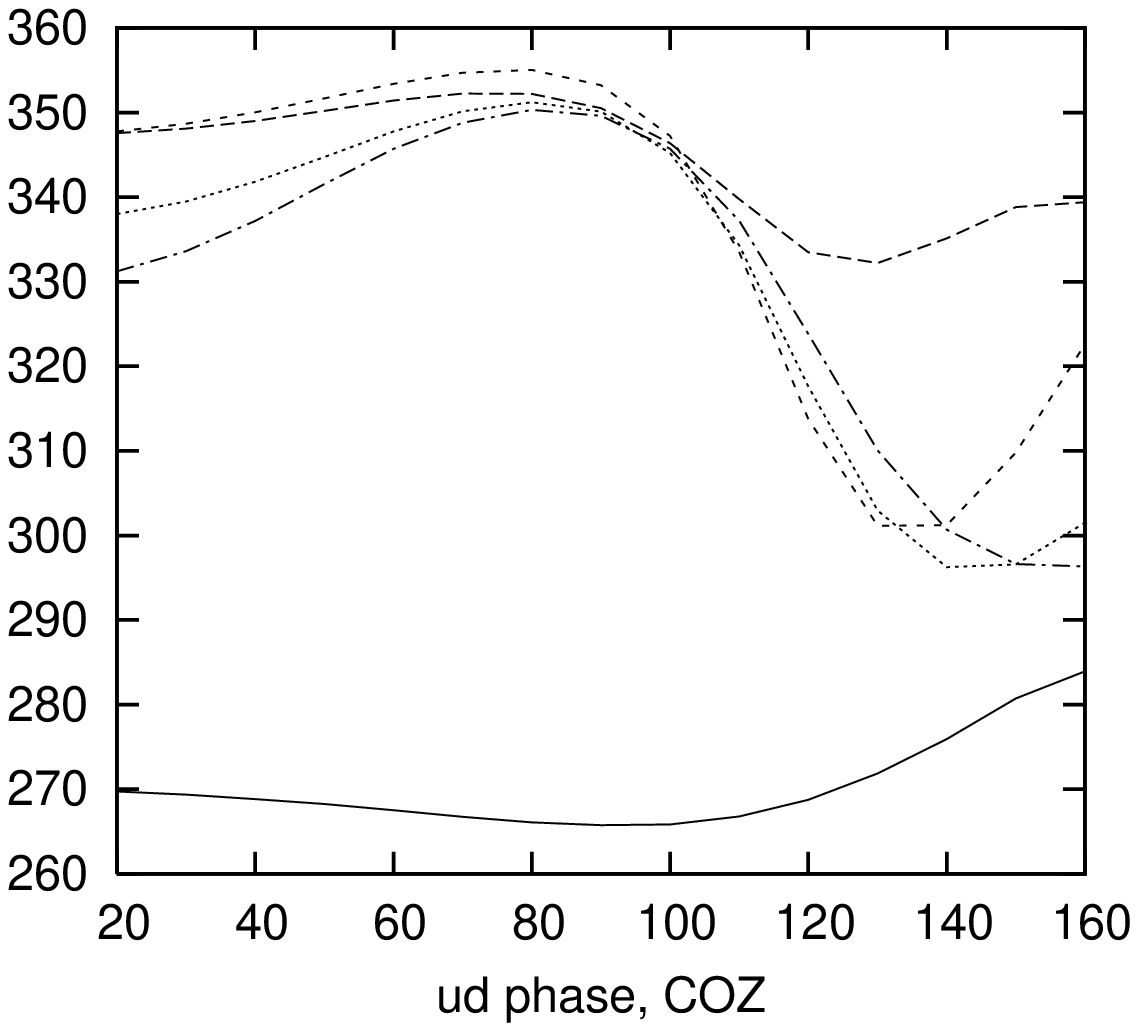}
\includegraphics[angle=270,width=3.3in]{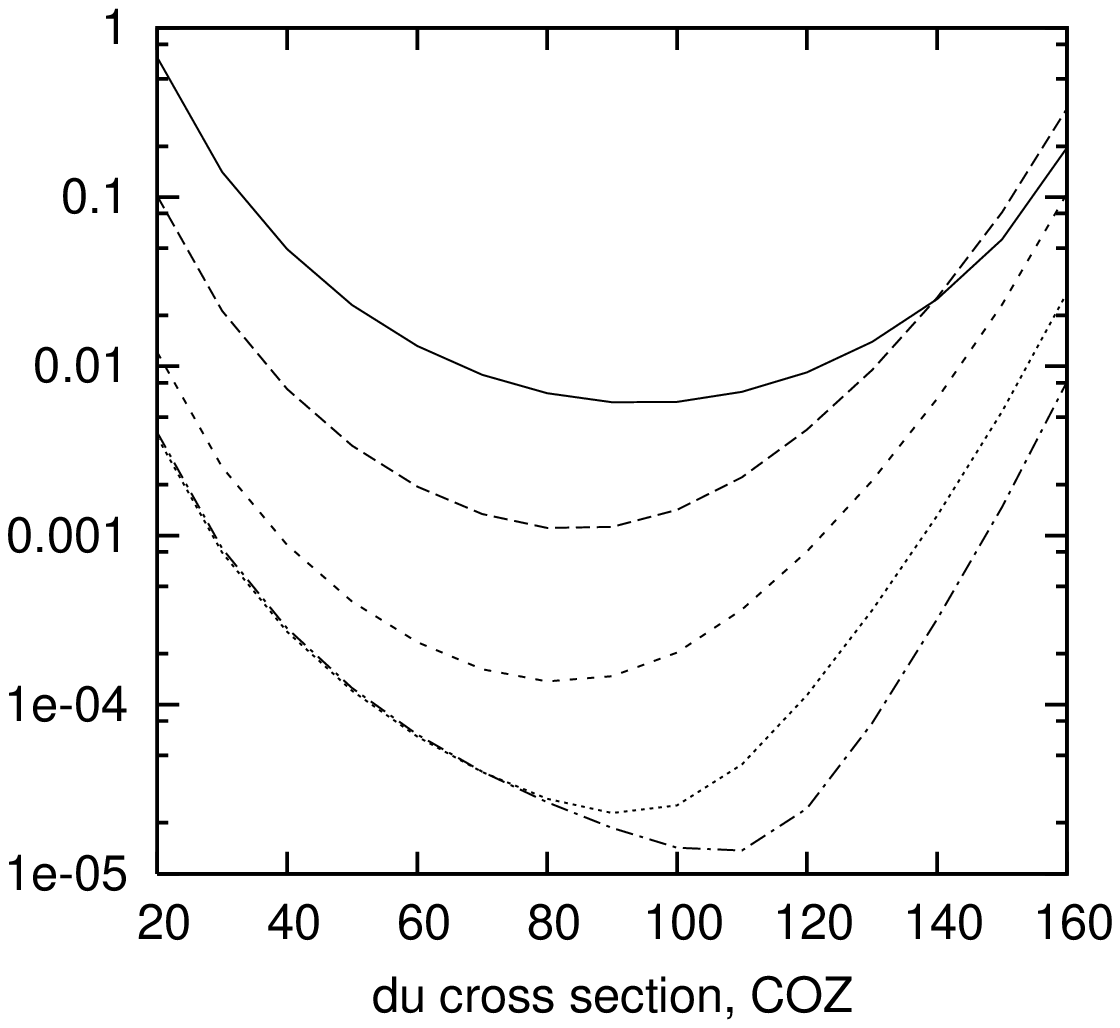}
\includegraphics[angle=270,width=3.3in]{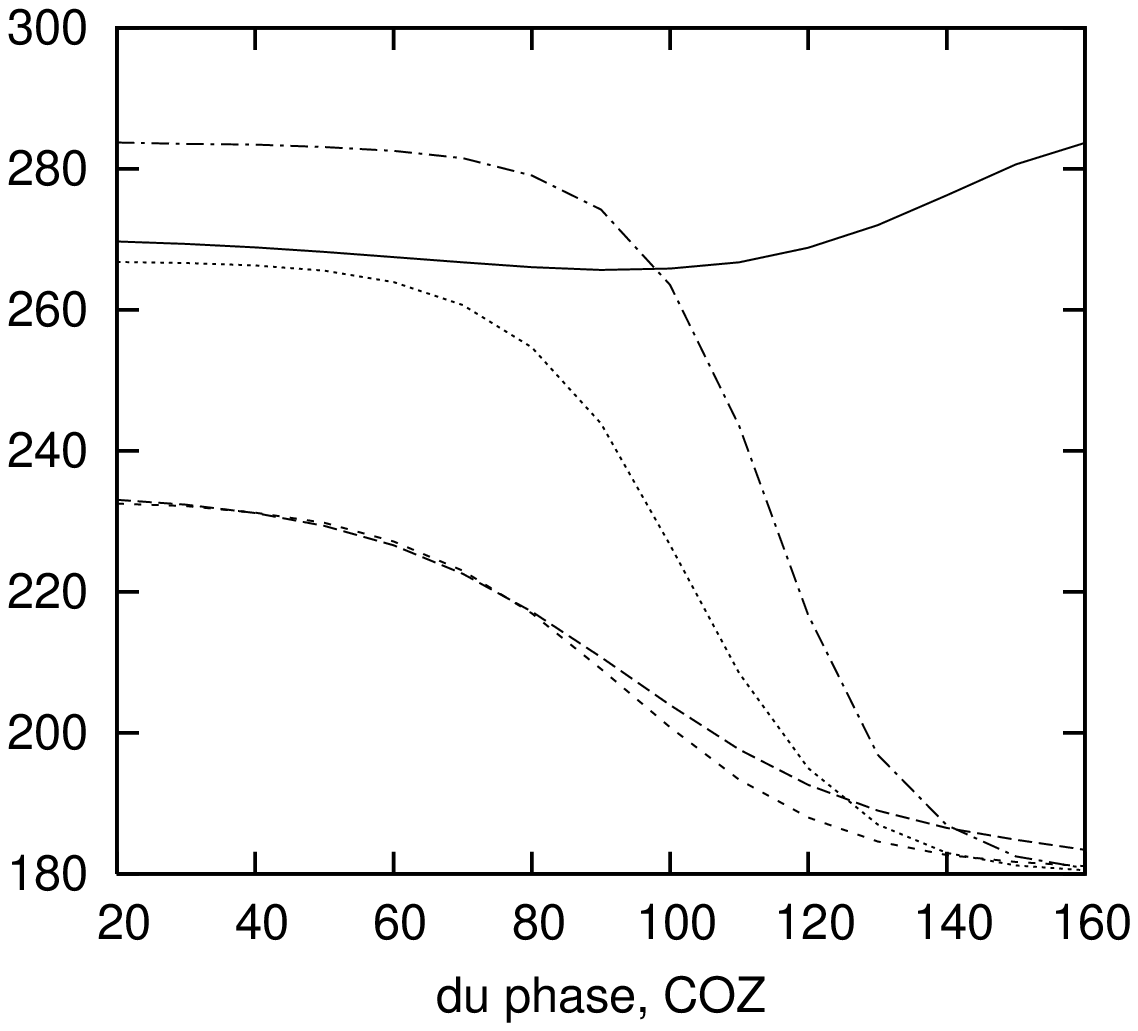}
\caption{Cross section and phase for the COZ distribution amplitude (1 of 2). The vertical axis is 
$s^6d\sigma/dt$ ($10^4$ nb GeV$^{10}$) for cross section plots  and angle in degrees for phase plots. The horizontal 
axis is center of mass scattering angle for all plots. Different values of R are shown as follows: R=1.00 (solid),
R=1.25 (longer dashes), R=1.50 (shorter dashes), R=1.75 (dots), R=2.00 (dashes/dots). 
\label{coz1}}
\end{figure}

\begin{figure}
\centering
\includegraphics[angle=270,width=3.3in]{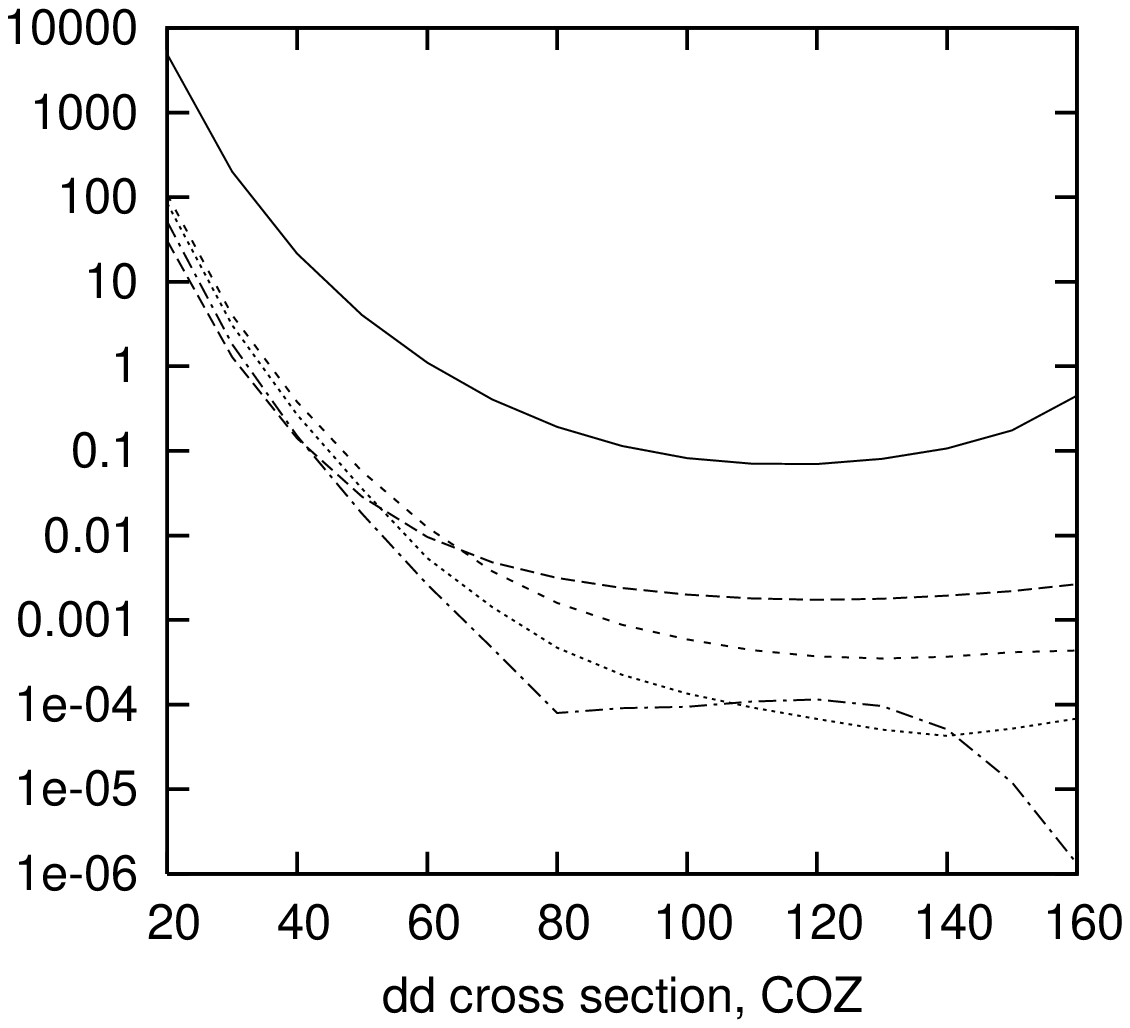}
\includegraphics[angle=270,width=3.3in]{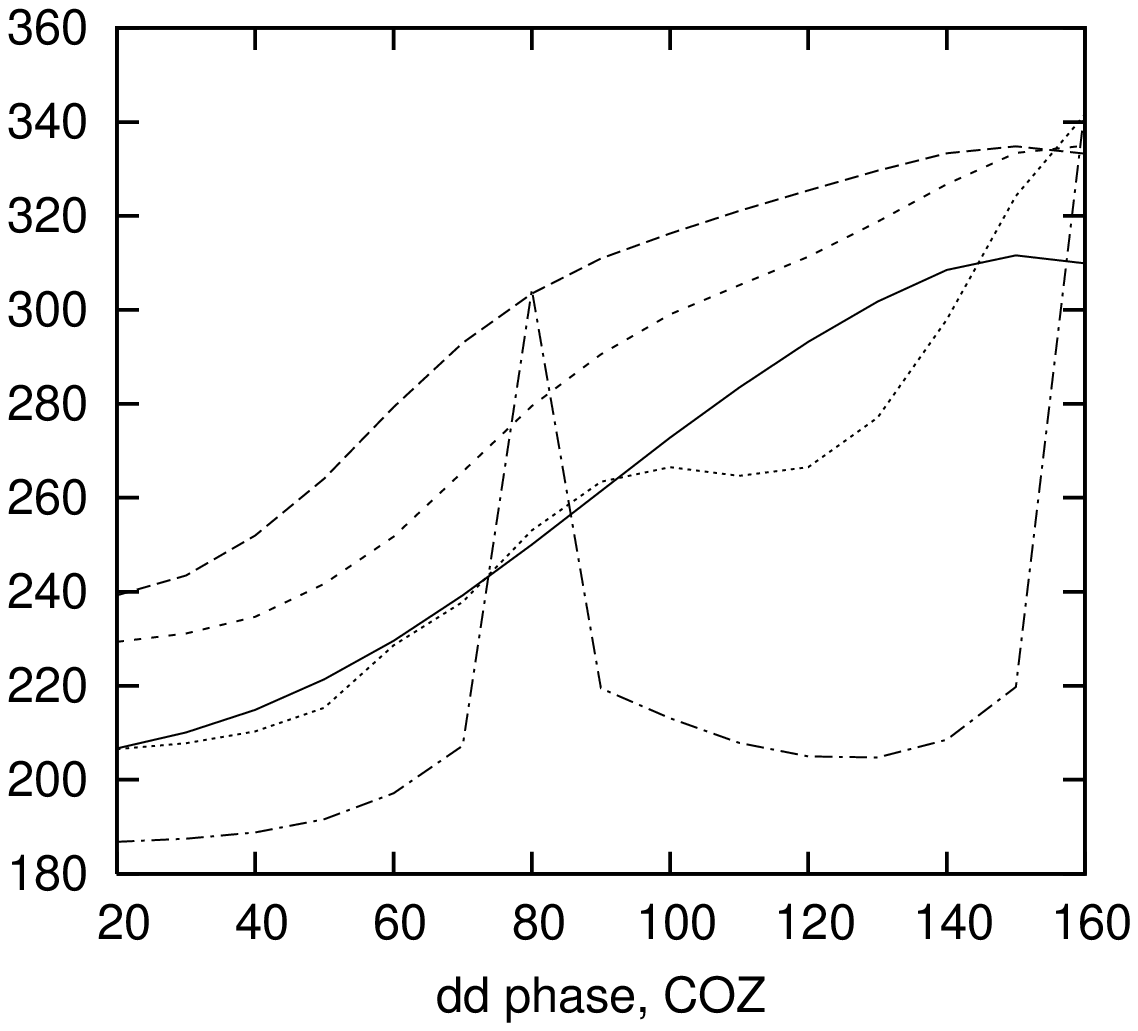}
\includegraphics[angle=270,width=3.3in]{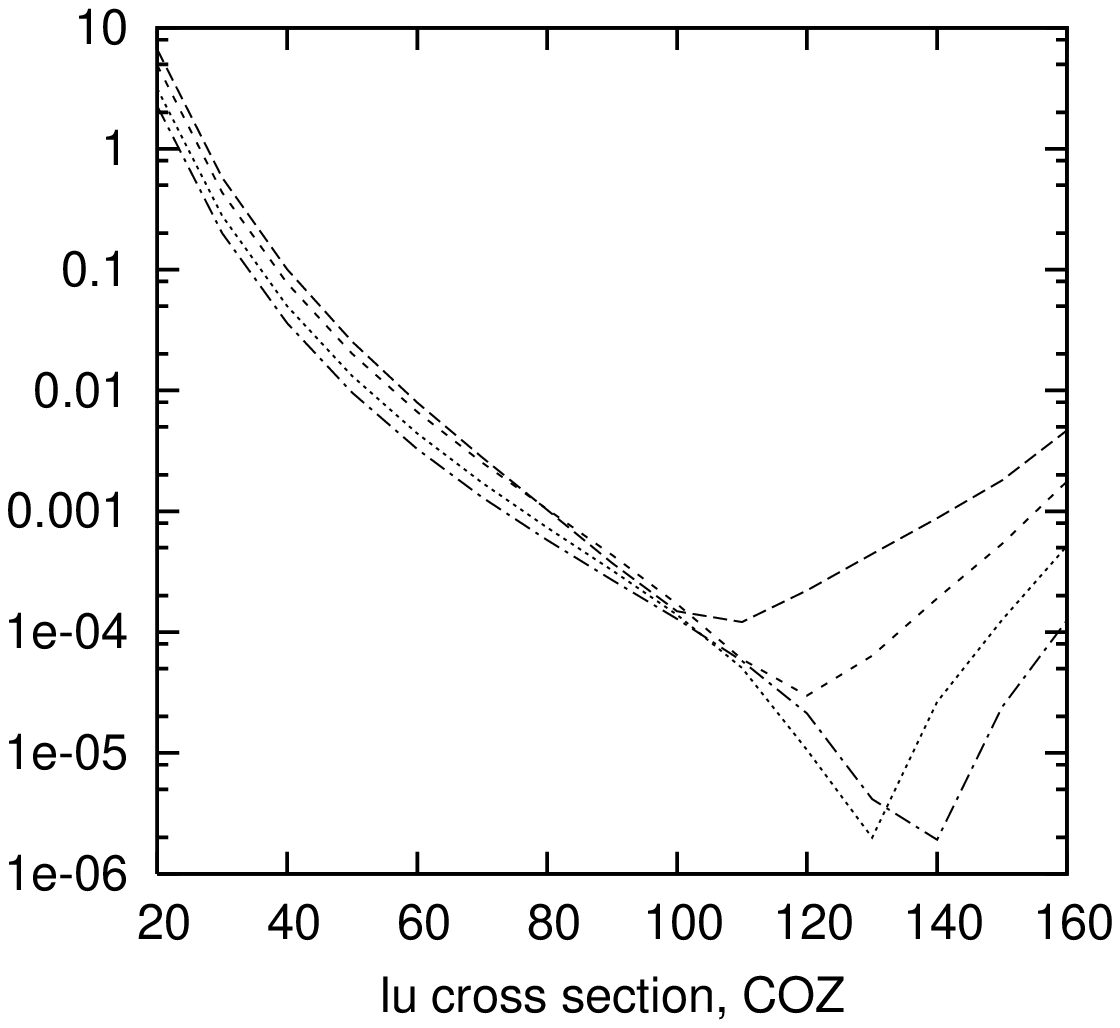}
\includegraphics[angle=270,width=3.3in]{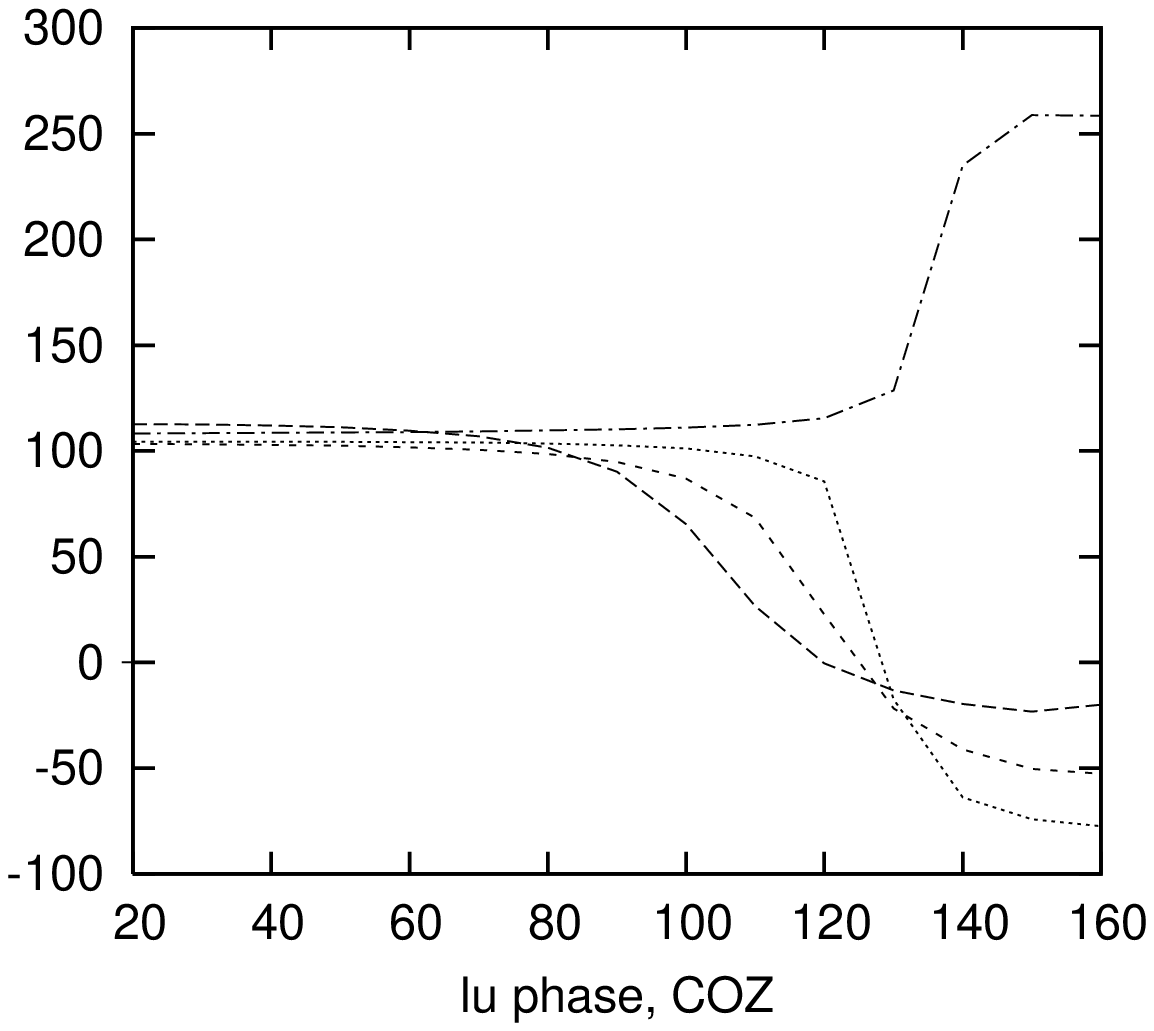}
\includegraphics[angle=270,width=3.3in]{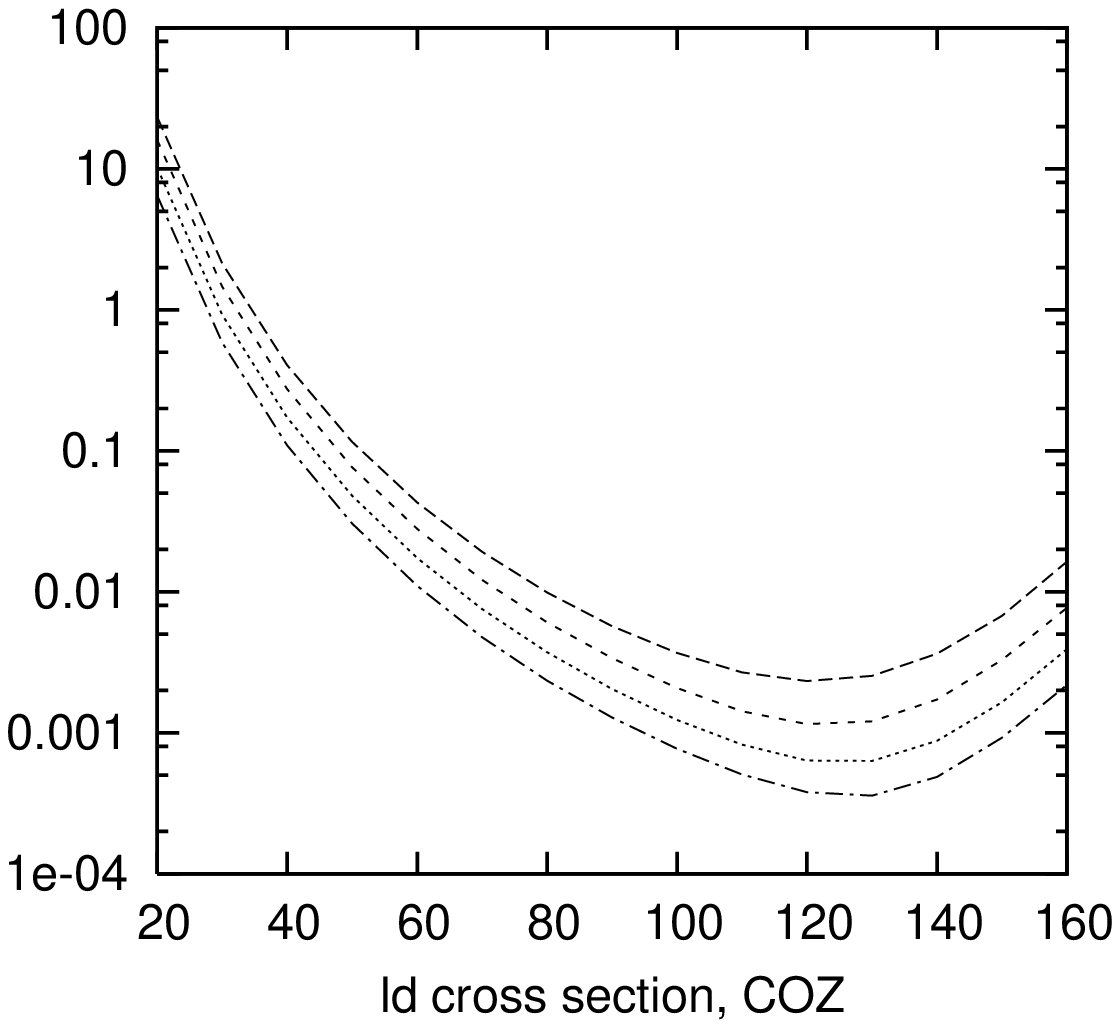}
\includegraphics[angle=270,width=3.3in]{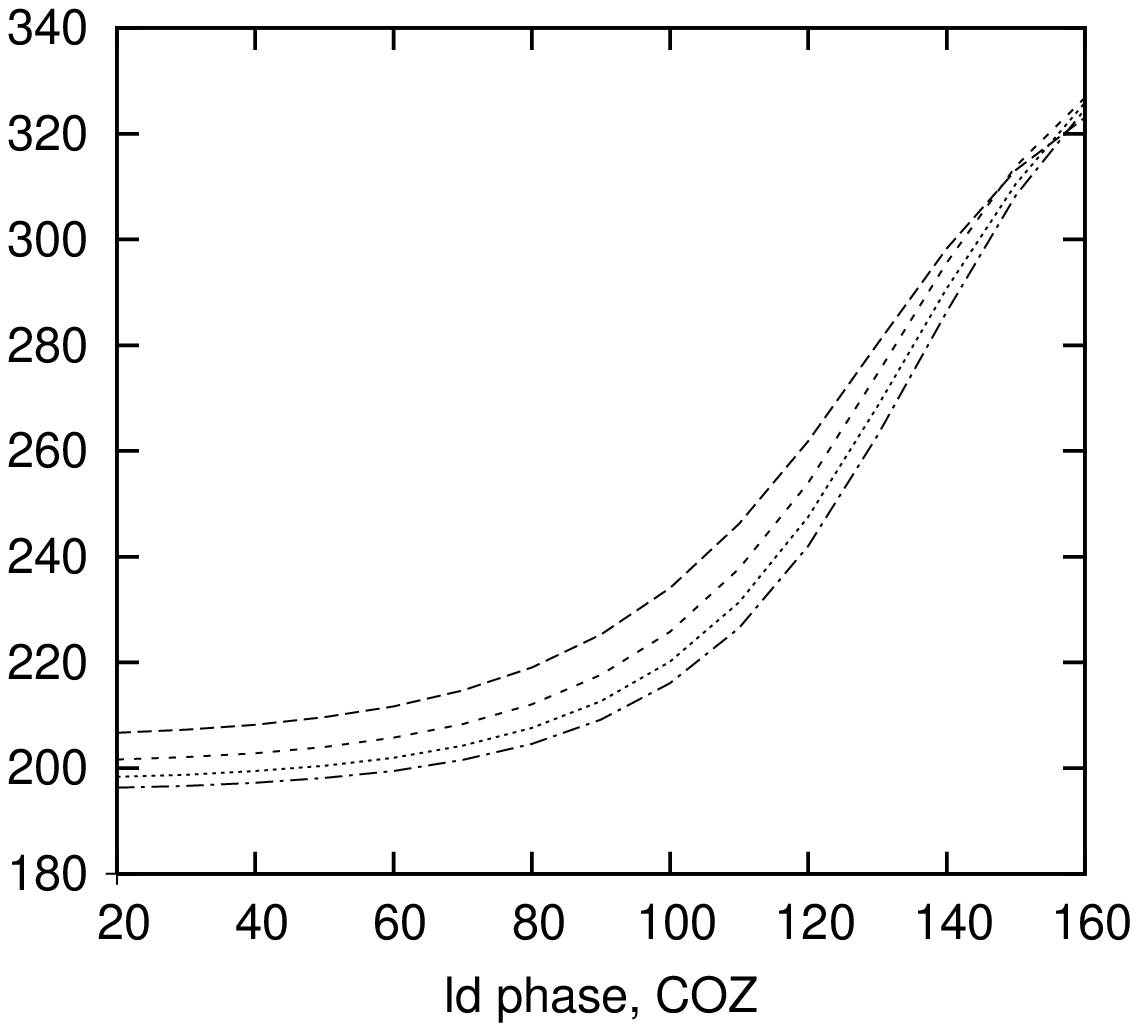}
\caption{Cross section and phase for the COZ distribution amplitude (2 of 2). The vertical axis is 
$s^6d\sigma/dt$ ($10^4$ nb GeV$^{10}$) for cross section plots  and angle in degrees for phase plots. The horizontal 
axis is center of mass scattering angle for all plots. Different values of R are shown as follows: R=1.00 (solid),
R=1.25 (longer dashes), R=1.50 (shorter dashes), R=1.75 (dots), R=2.00 (dashes/dots). 
\label{coz2}}
\end{figure}

\begin{figure}
\centering
\includegraphics[angle=270,width=3.3in]{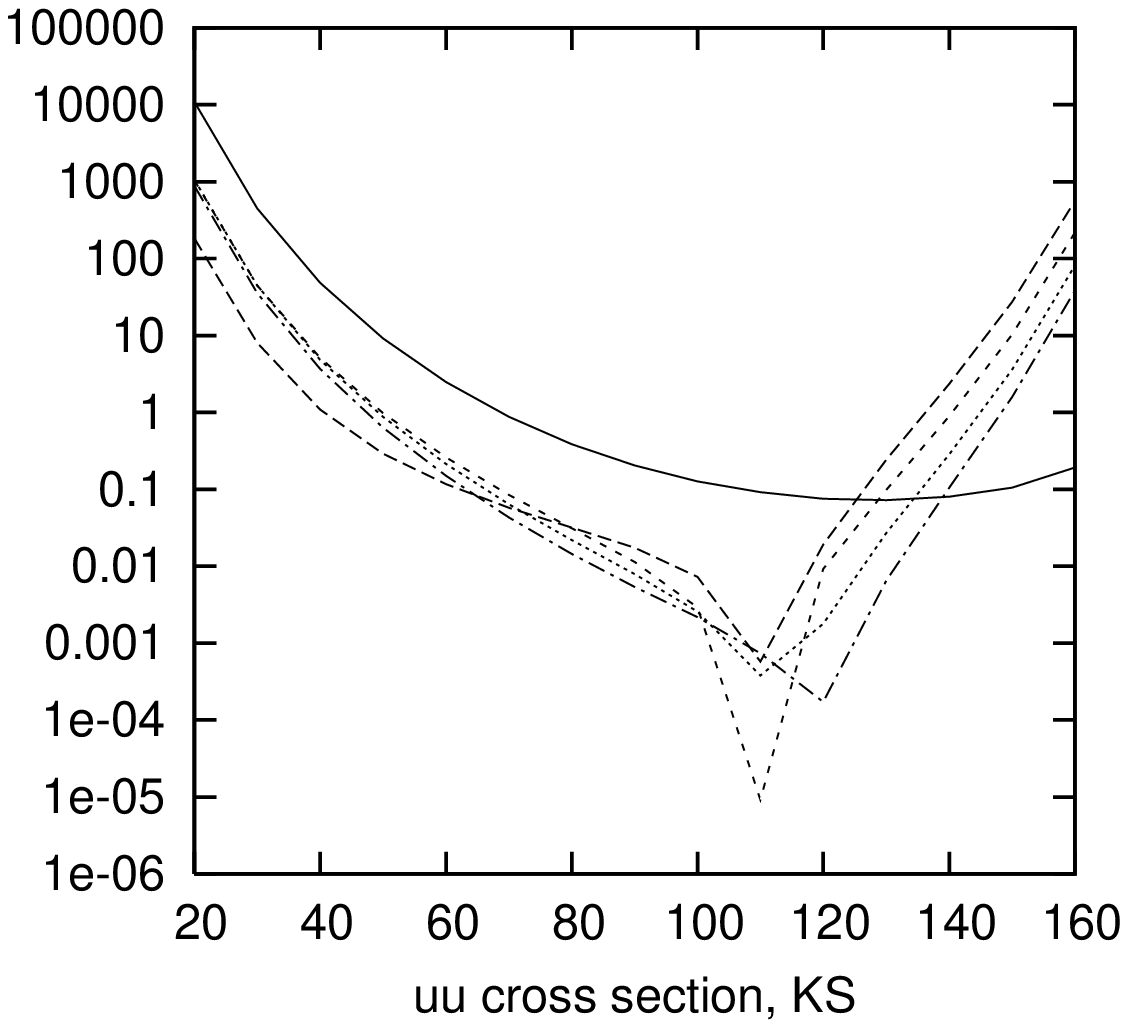}
\includegraphics[angle=270,width=3.3in]{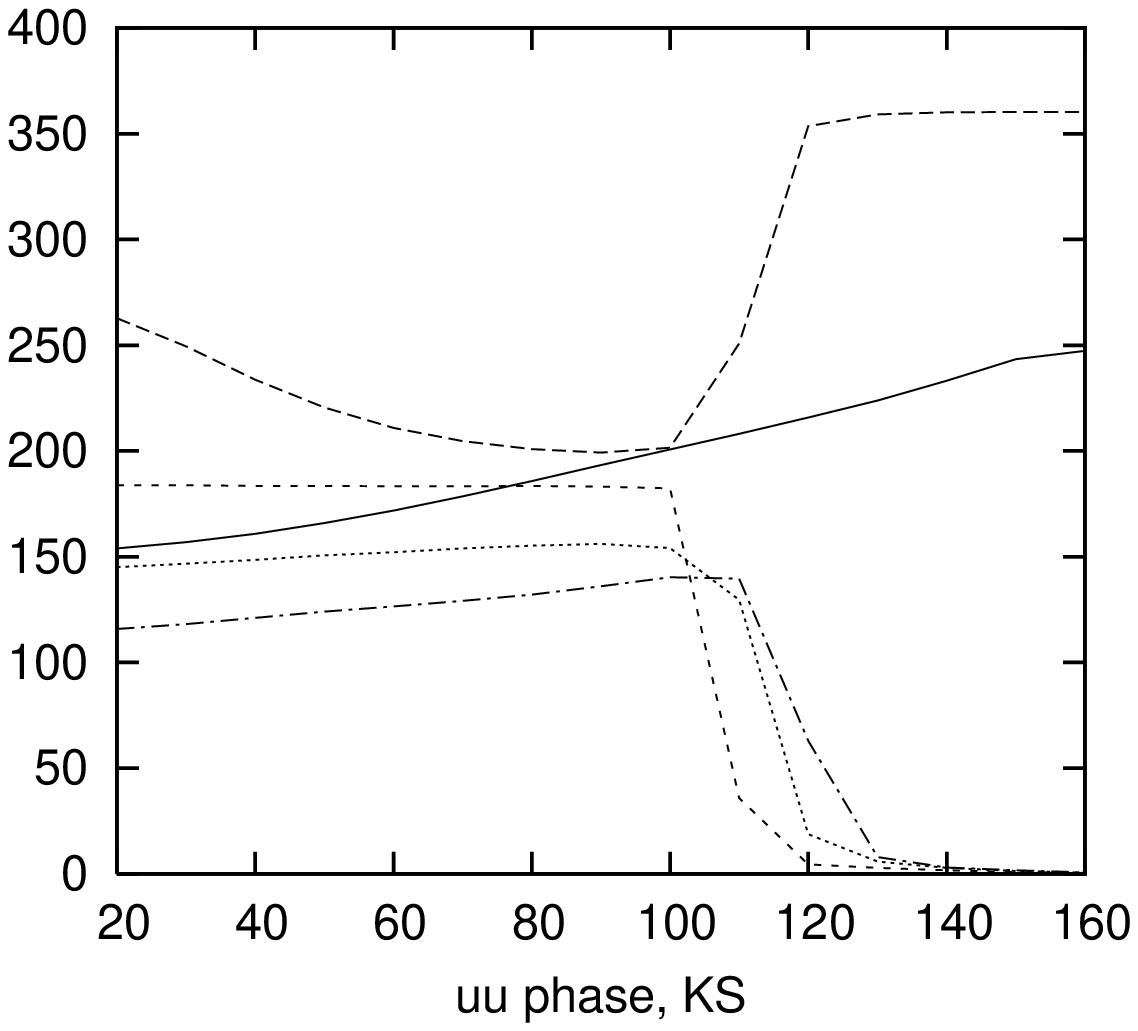}
\includegraphics[angle=270,width=3.3in]{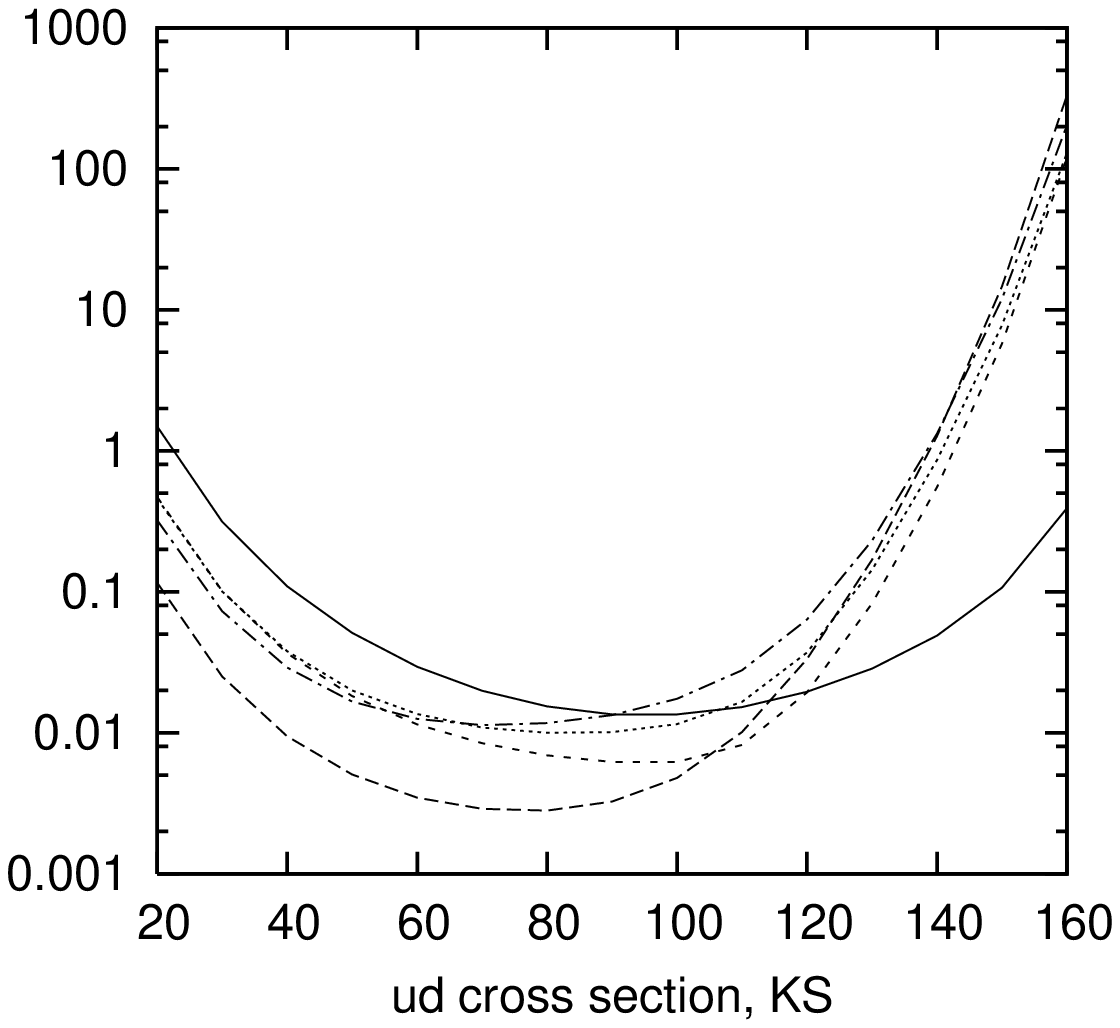}
\includegraphics[angle=270,width=3.3in]{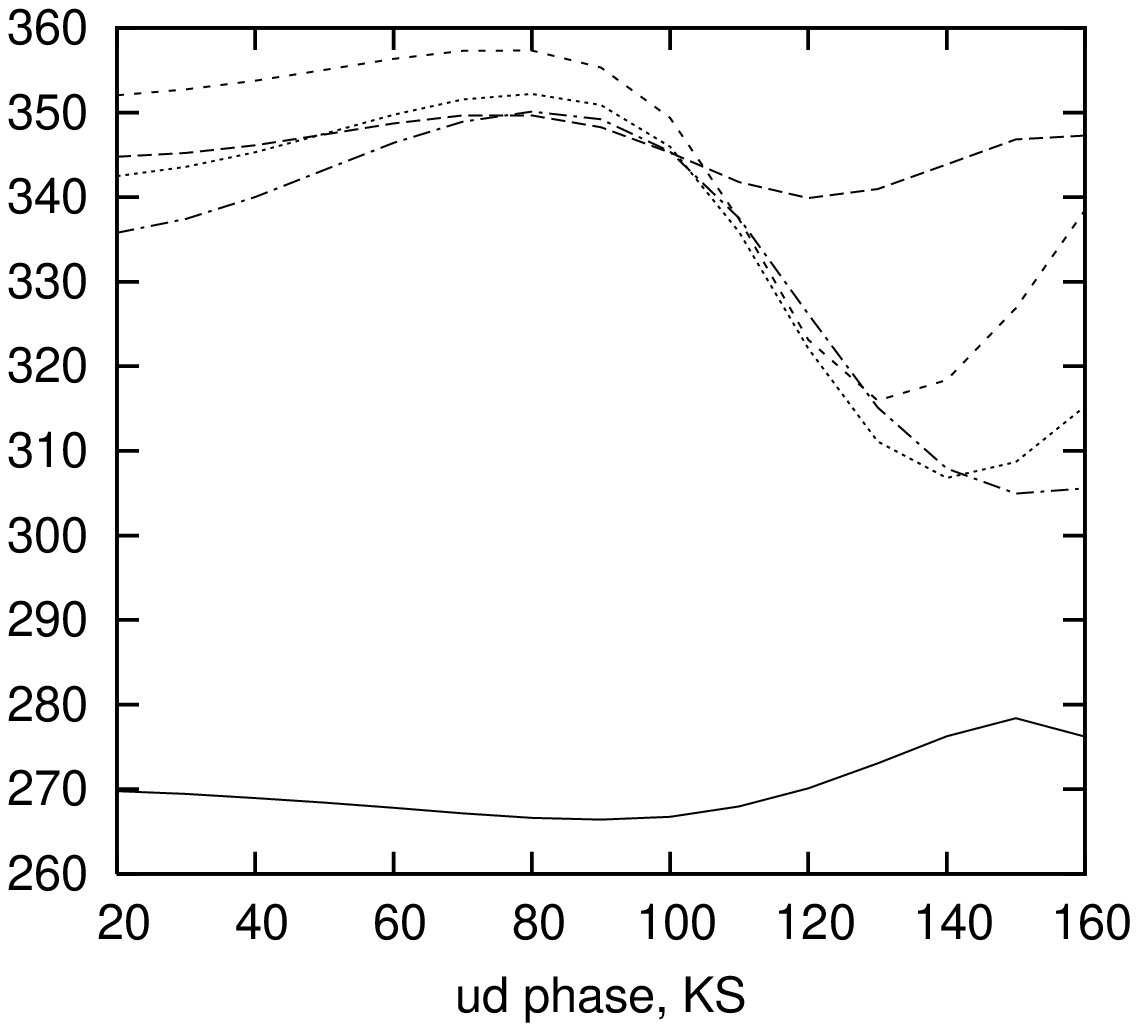}
\includegraphics[angle=270,width=3.3in]{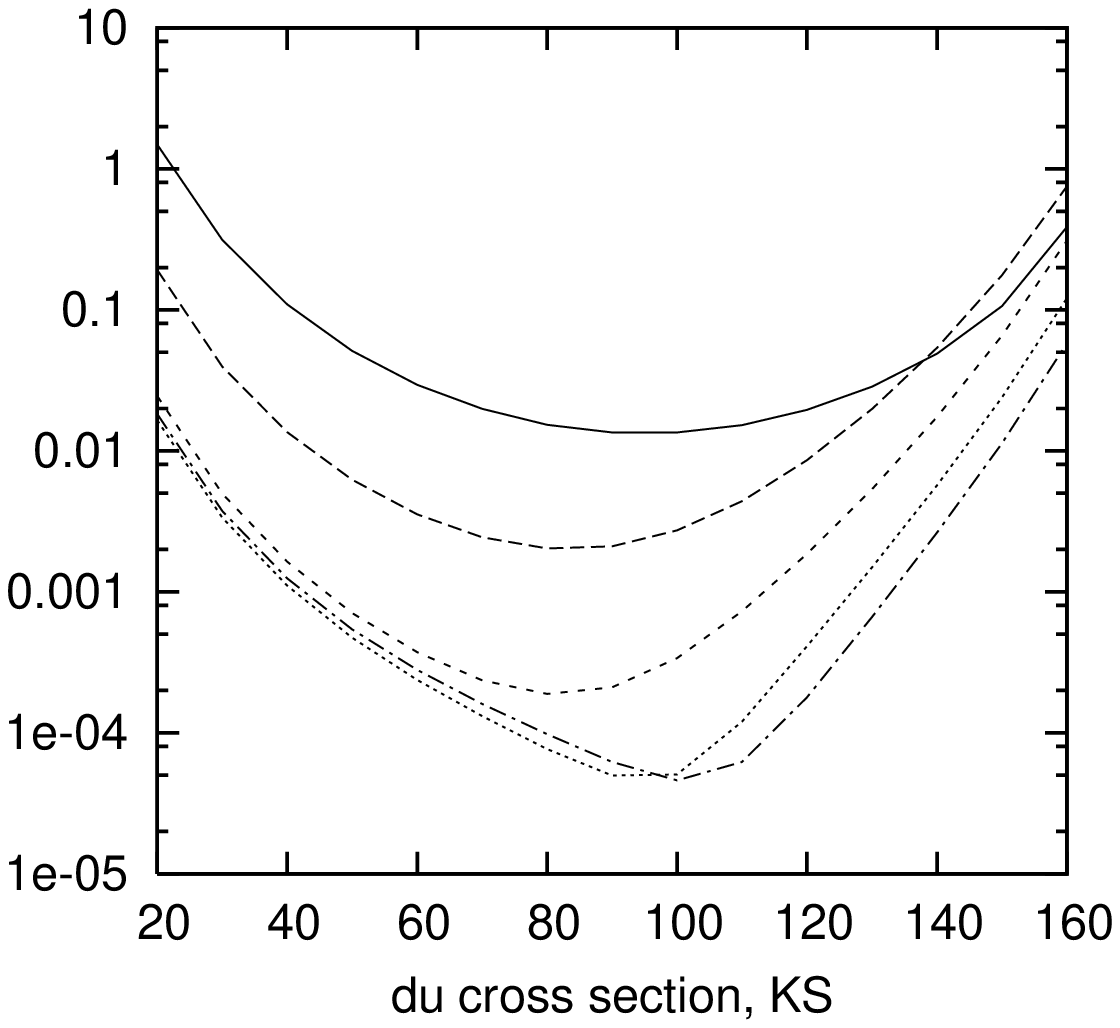}
\includegraphics[angle=270,width=3.3in]{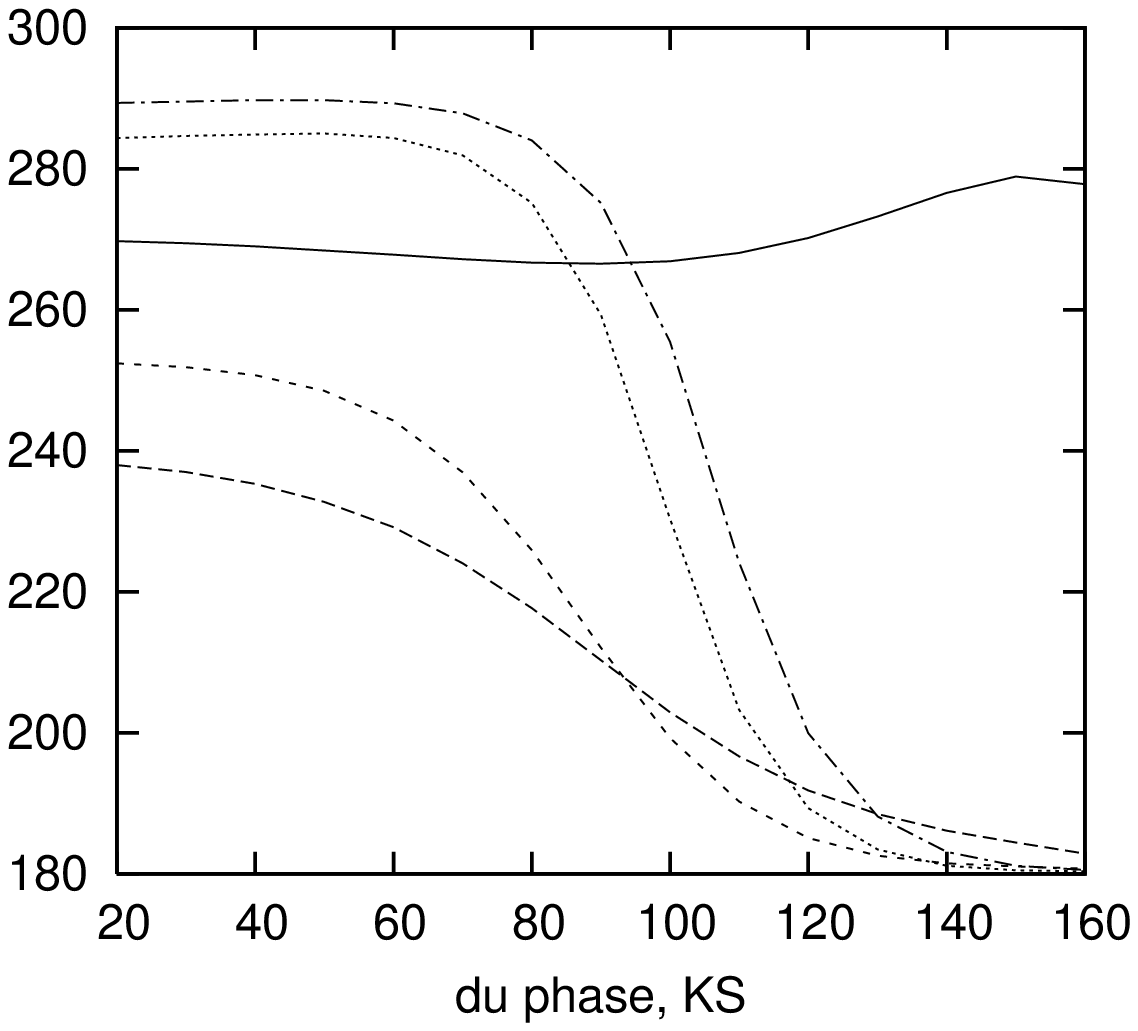}
\caption{Cross section and phase for the KS distribution amplitude (1 of 2). The vertical axis is 
$s^6d\sigma/dt$ ($10^4$ nb GeV$^{10}$) for cross section plots  and angle in degrees for phase plots. The horizontal 
axis is center of mass scattering angle for all plots. Different values of R are shown as follows: R=1.00 (solid),
R=1.25 (longer dashes), R=1.50 (shorter dashes), R=1.75 (dots), R=2.00 (dashes/dots). 
\label{ks1}}
\end{figure}

\begin{figure}
\centering
\includegraphics[angle=270,width=3.3in]{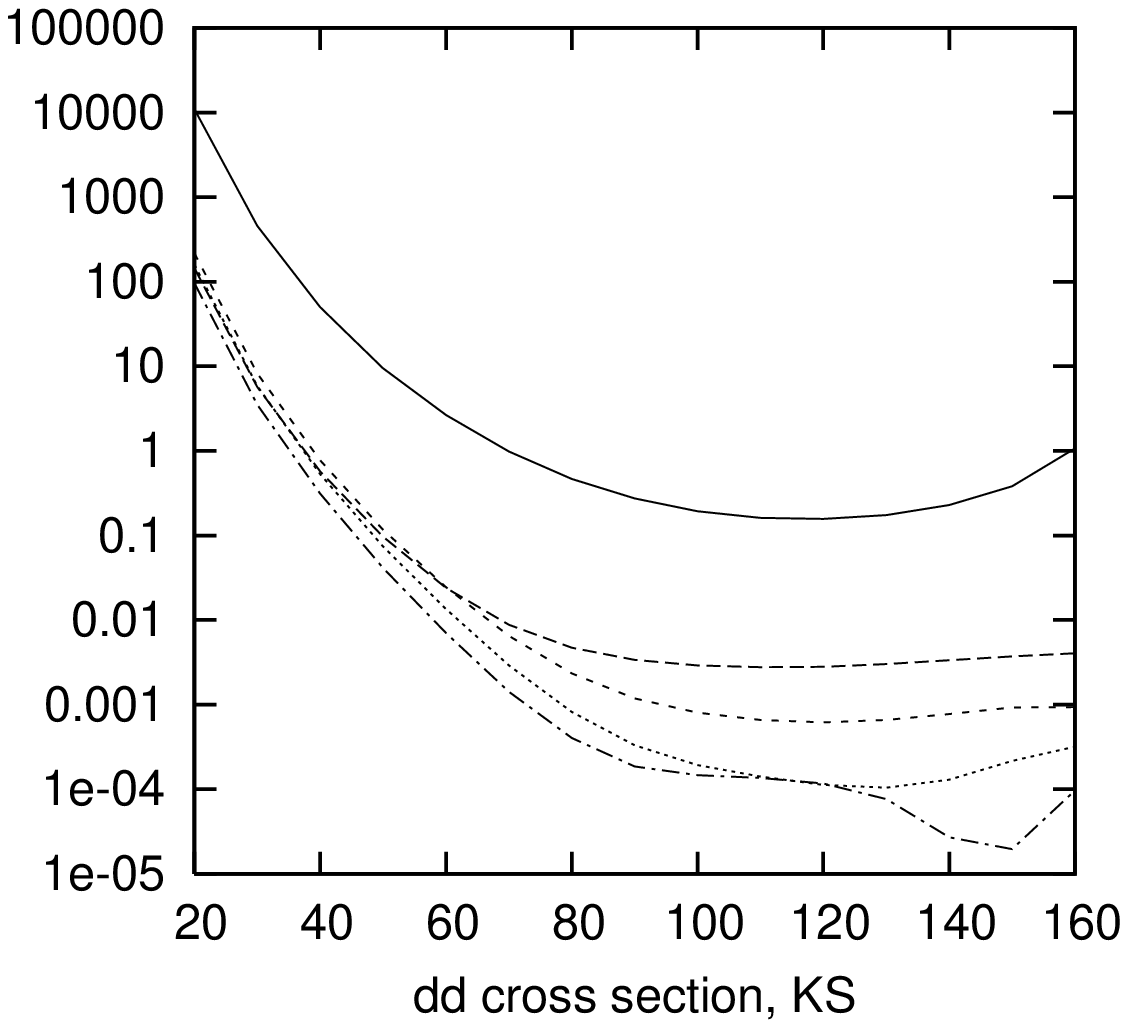}
\includegraphics[angle=270,width=3.3in]{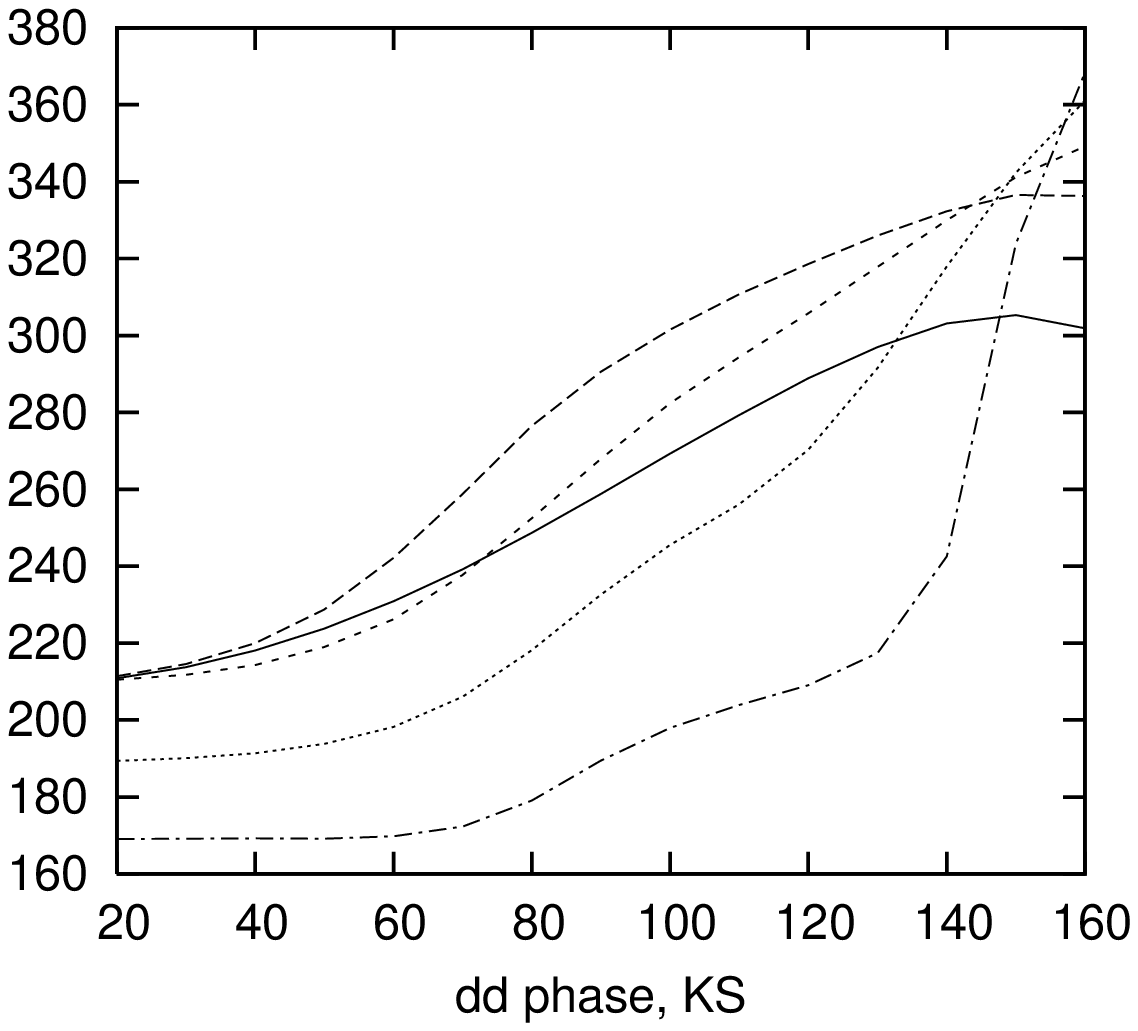}
\includegraphics[angle=270,width=3.3in]{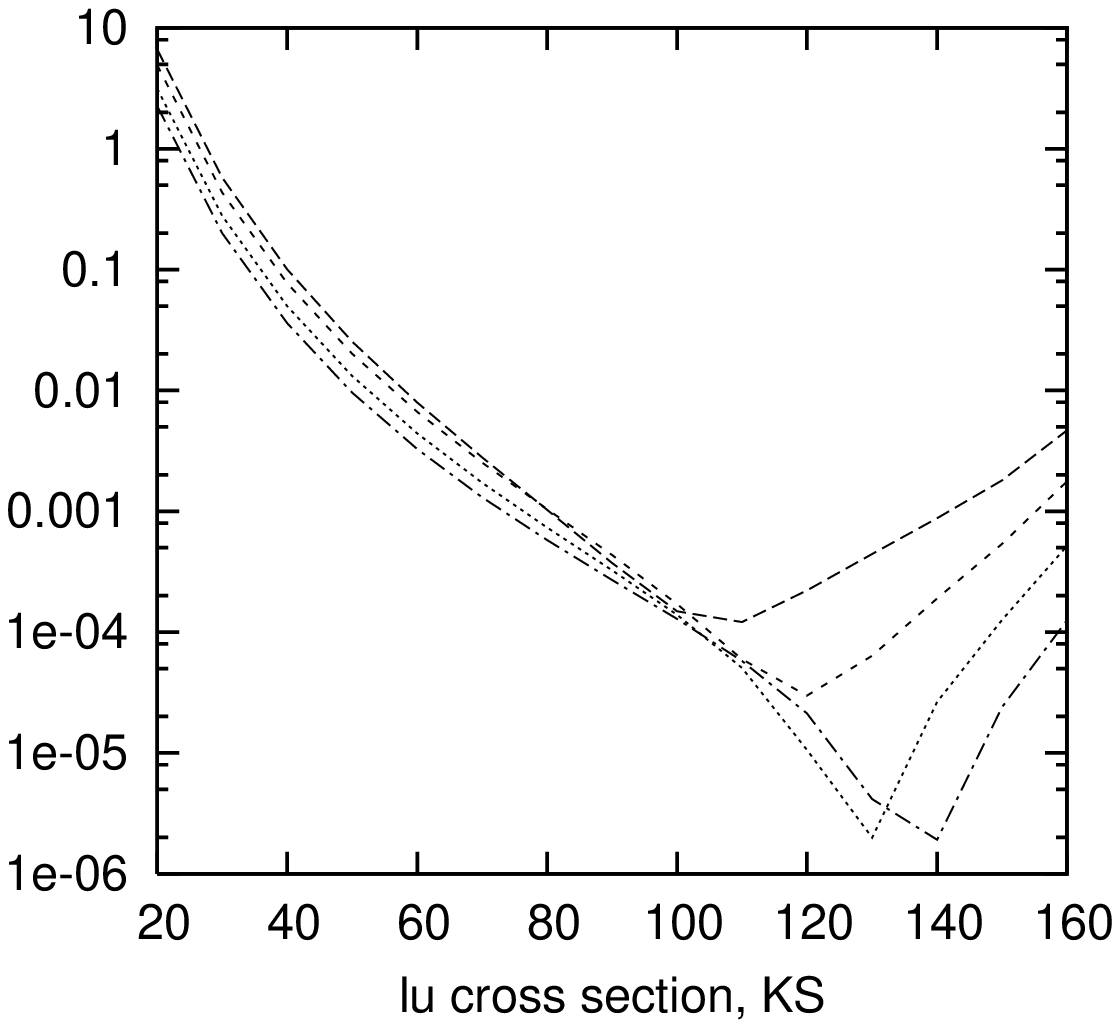}
\includegraphics[angle=270,width=3.3in]{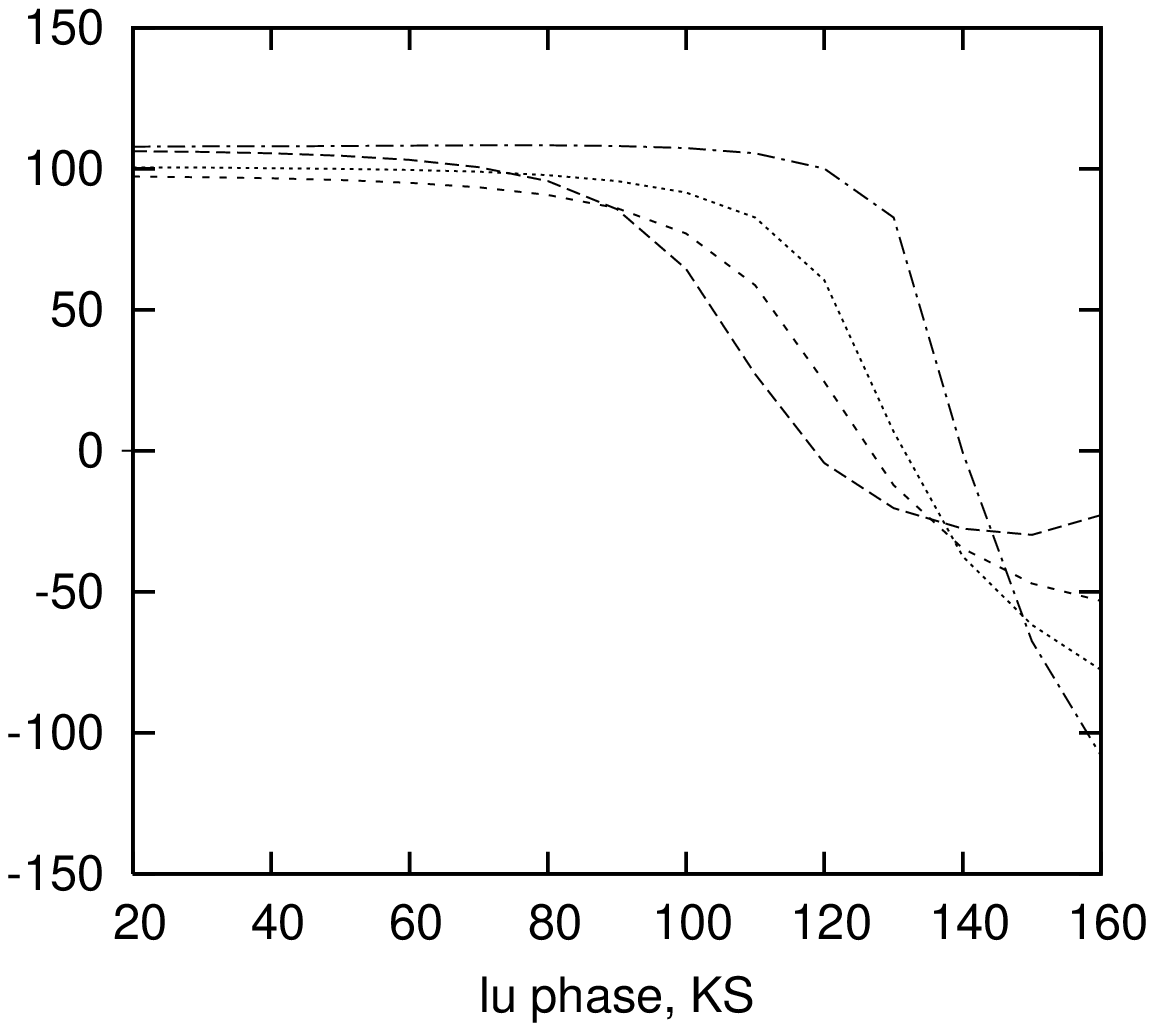}
\includegraphics[angle=270,width=3.3in]{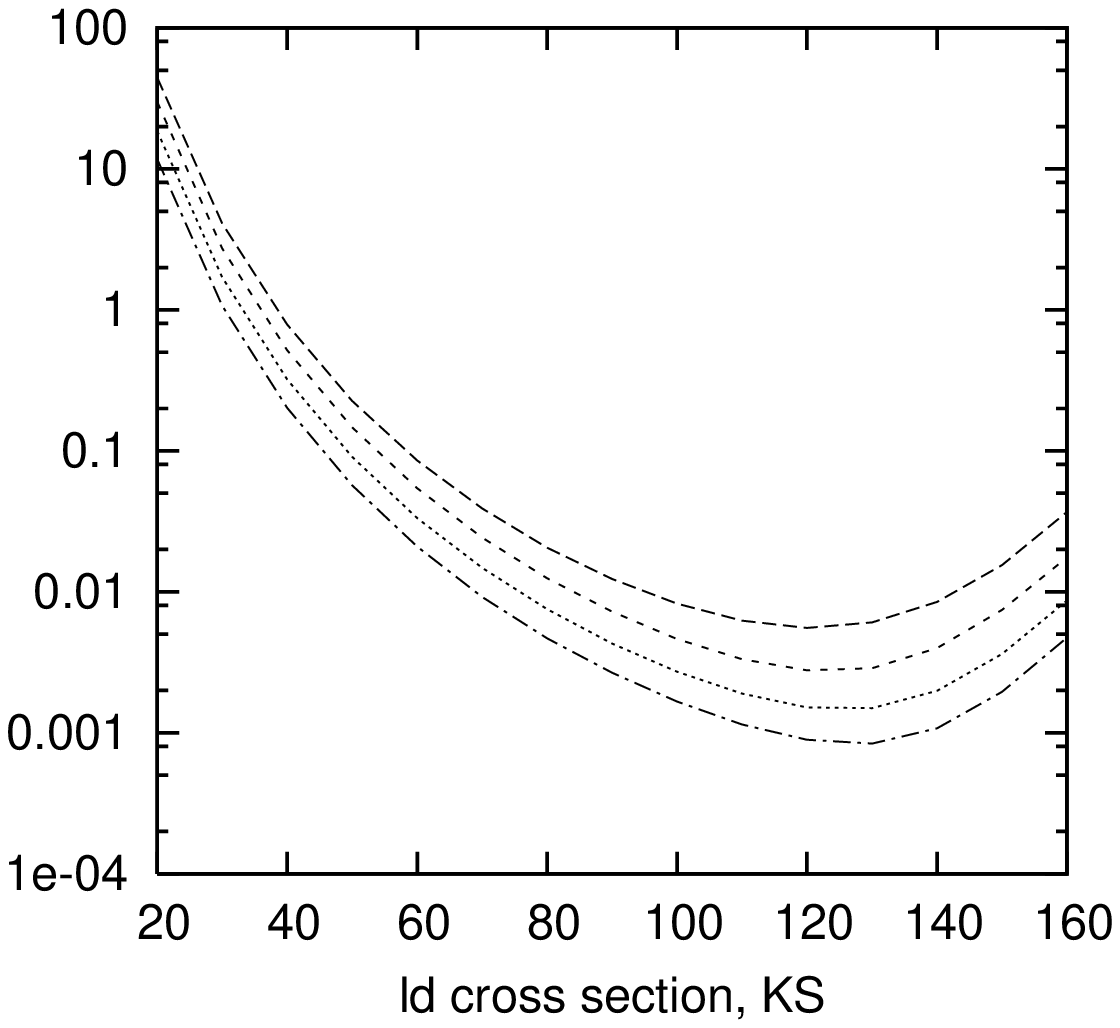}
\includegraphics[angle=270,width=3.3in]{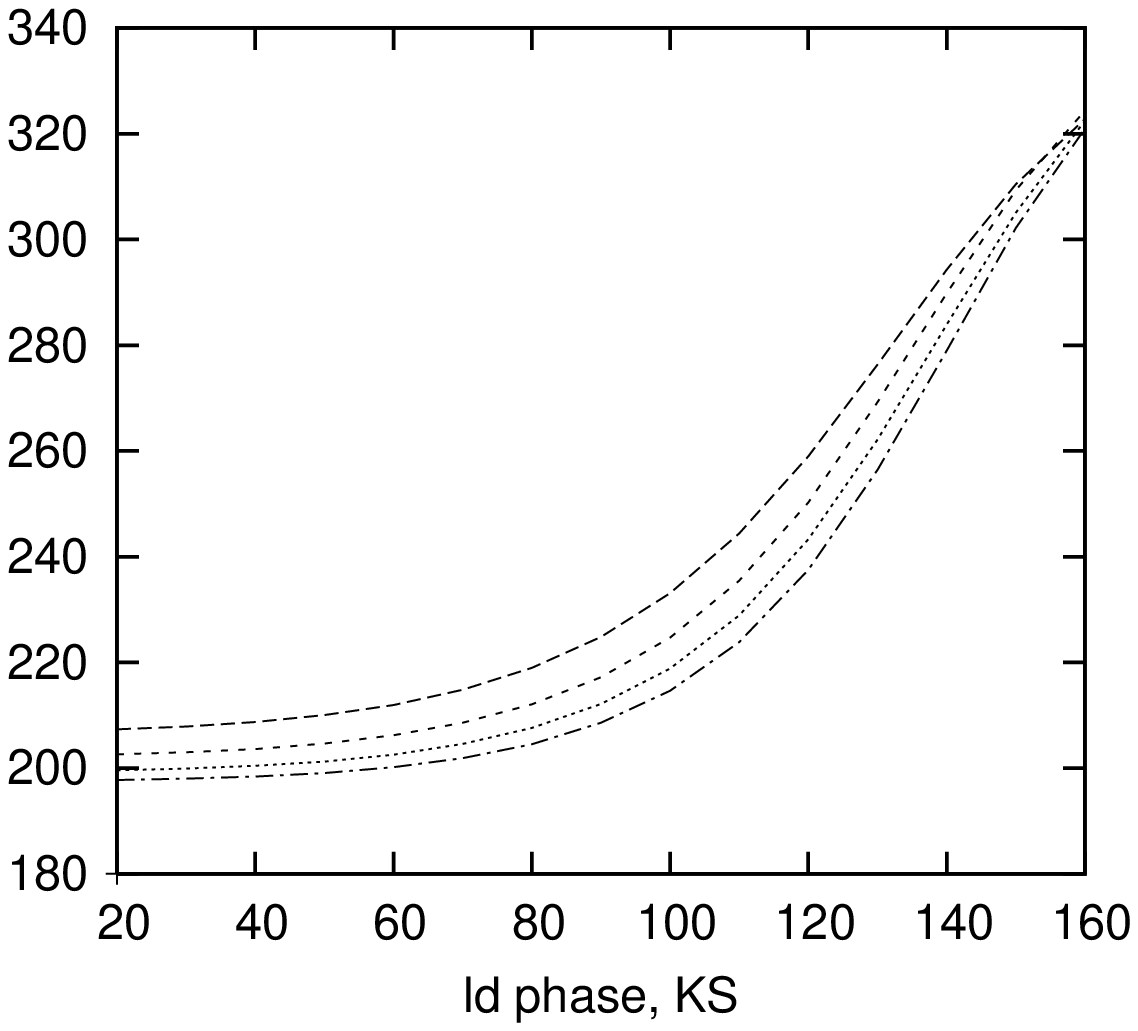}
\caption{Cross section and phase for the KS distribution amplitude (2 of 2). The vertical axis is 
$s^6d\sigma/dt$ ($10^4$ nb GeV$^{10}$) for cross section plots  and angle in degrees for phase plots. The horizontal 
axis is center of mass scattering angle for all plots. Different values of R are shown as follows: R=1.00 (solid),
R=1.25 (longer dashes), R=1.50 (shorter dashes), R=1.75 (dots), R=2.00 (dashes/dots). 
\label{ks2}}
\end{figure}

\begin{figure}
\centering
\includegraphics[angle=270,width=3.3in]{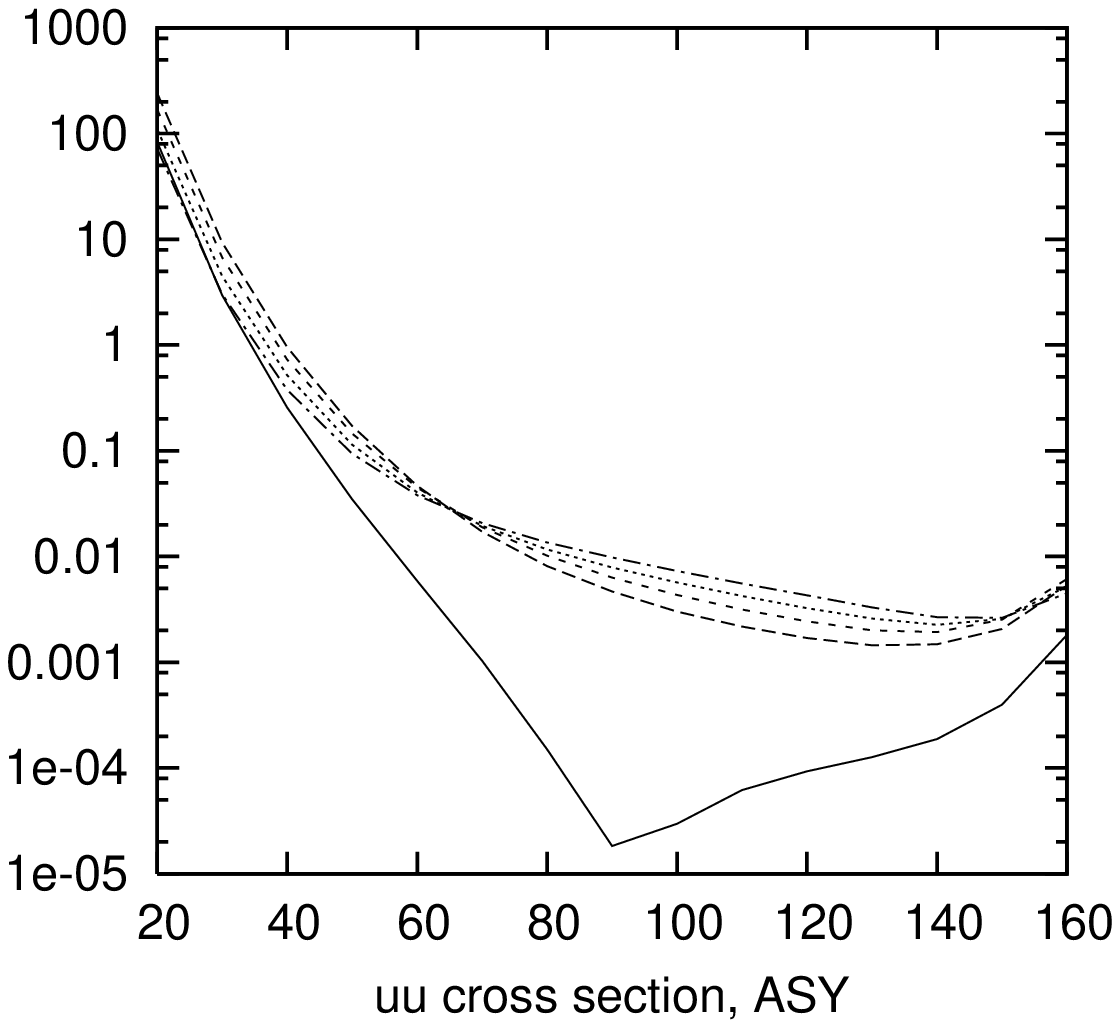}
\includegraphics[angle=270,width=3.3in]{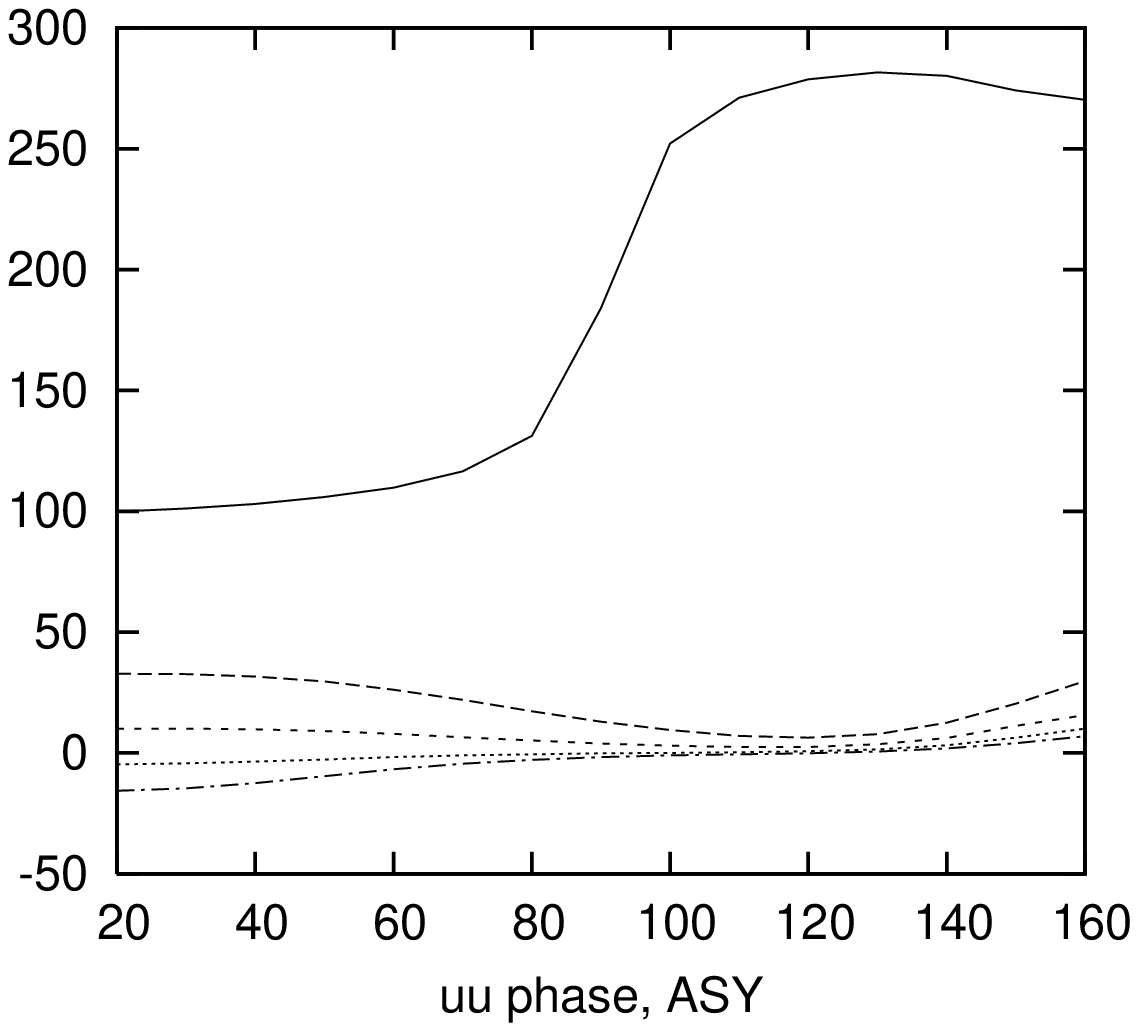}
\includegraphics[angle=270,width=3.3in]{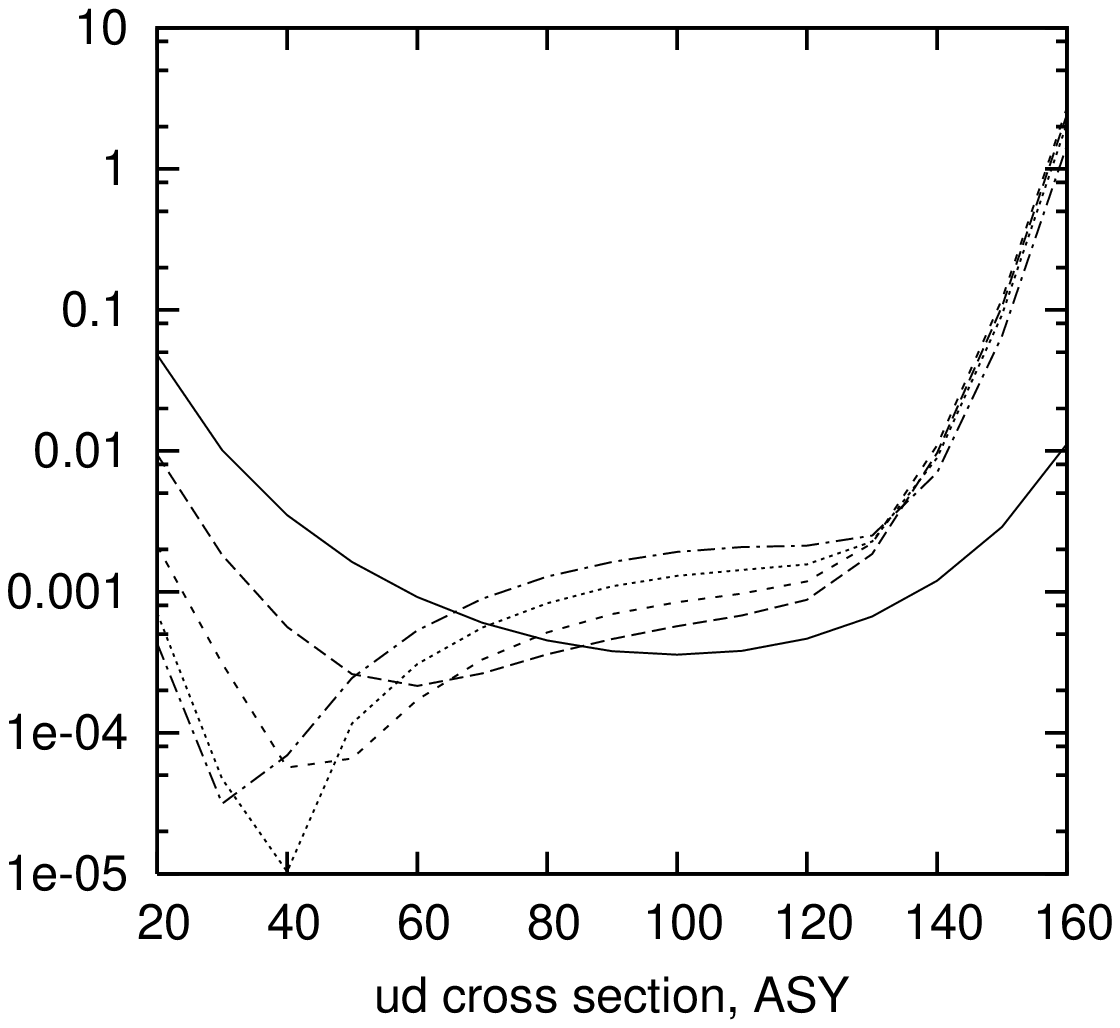}
\includegraphics[angle=270,width=3.3in]{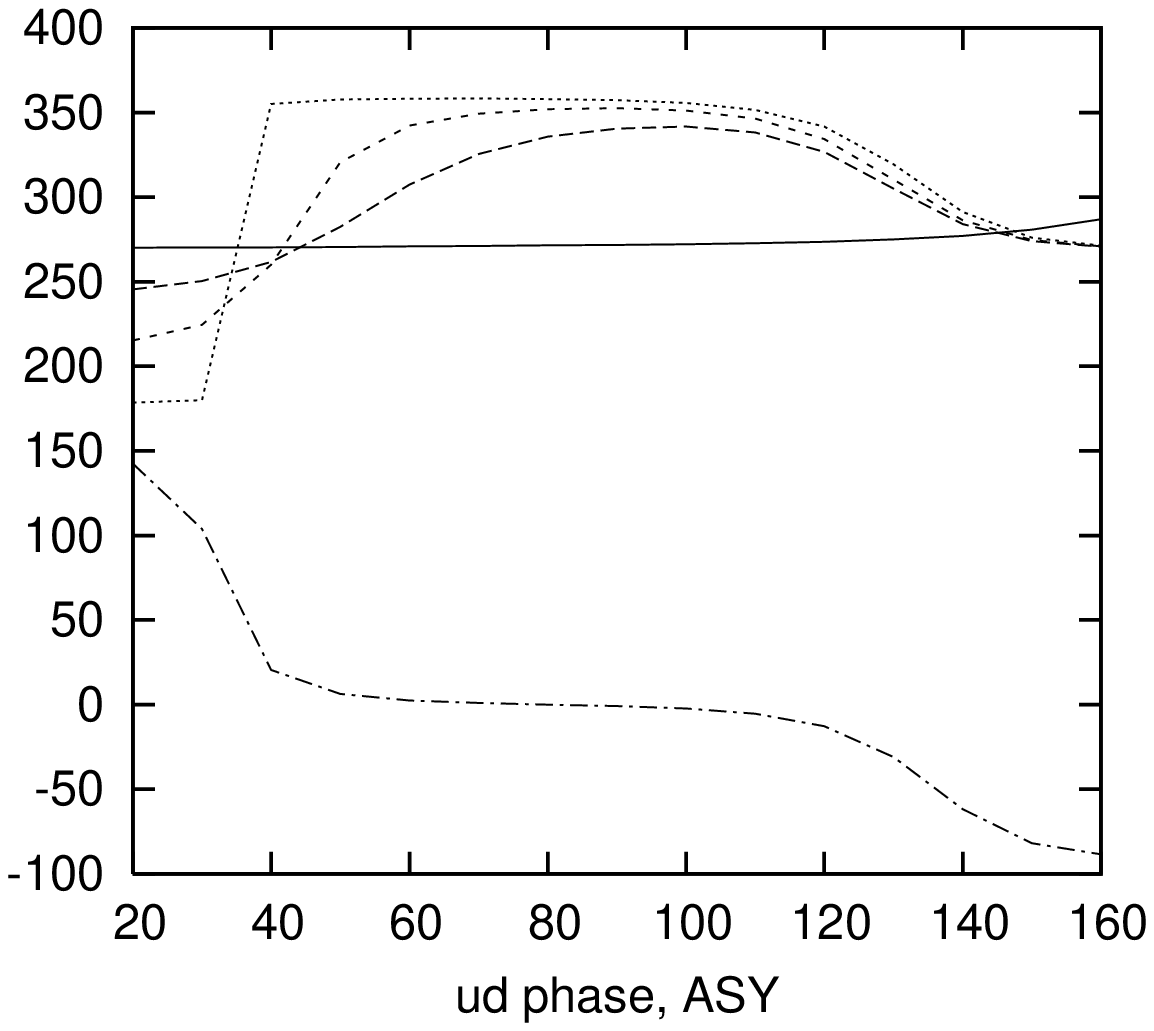}
\includegraphics[angle=270,width=3.3in]{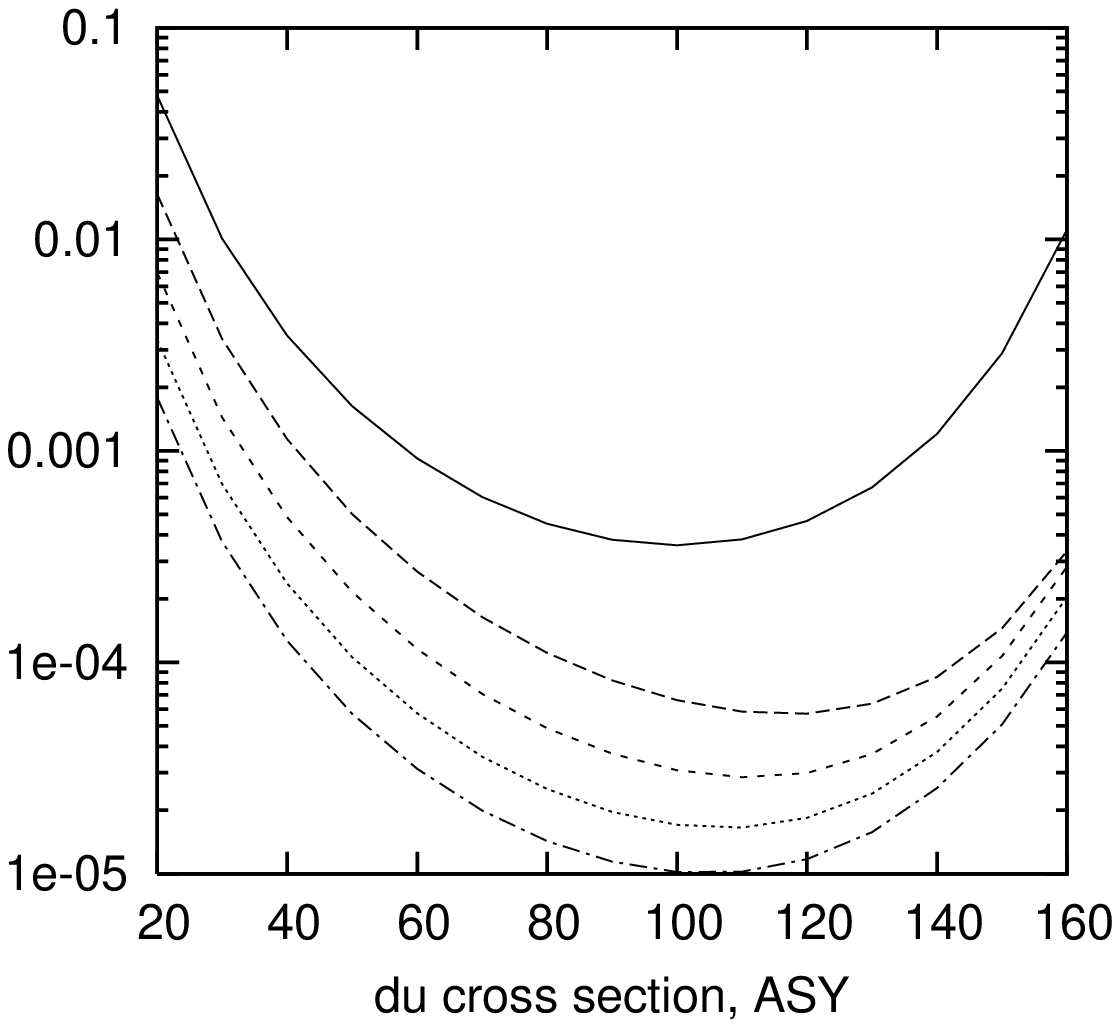}
\includegraphics[angle=270,width=3.3in]{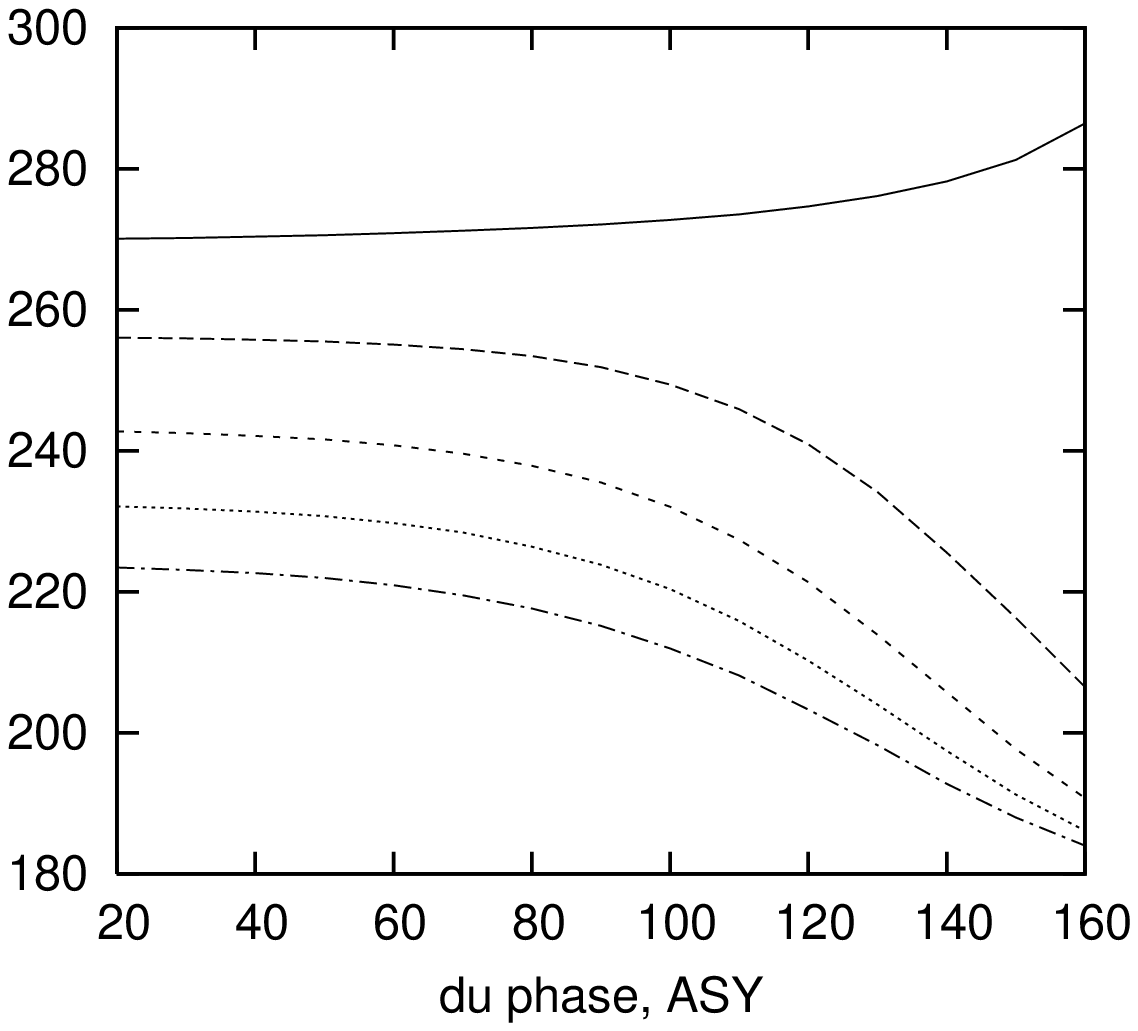}
\caption{Cross section and phase for the ASY distribution amplitude (1 of 2). The vertical axis is 
$s^6d\sigma/dt$ ($10^4$ nb GeV$^{10}$) for cross section plots  and angle in degrees for phase plots. The horizontal 
axis is center of mass scattering angle for all plots. Different values of R are shown as follows: R=1.00 (solid),
R=1.25 (longer dashes), R=1.50 (shorter dashes), R=1.75 (dots), R=2.00 (dashes/dots). 
\label{asy1}}
\end{figure}

\begin{figure}
\centering
\includegraphics[angle=270,width=3.3in]{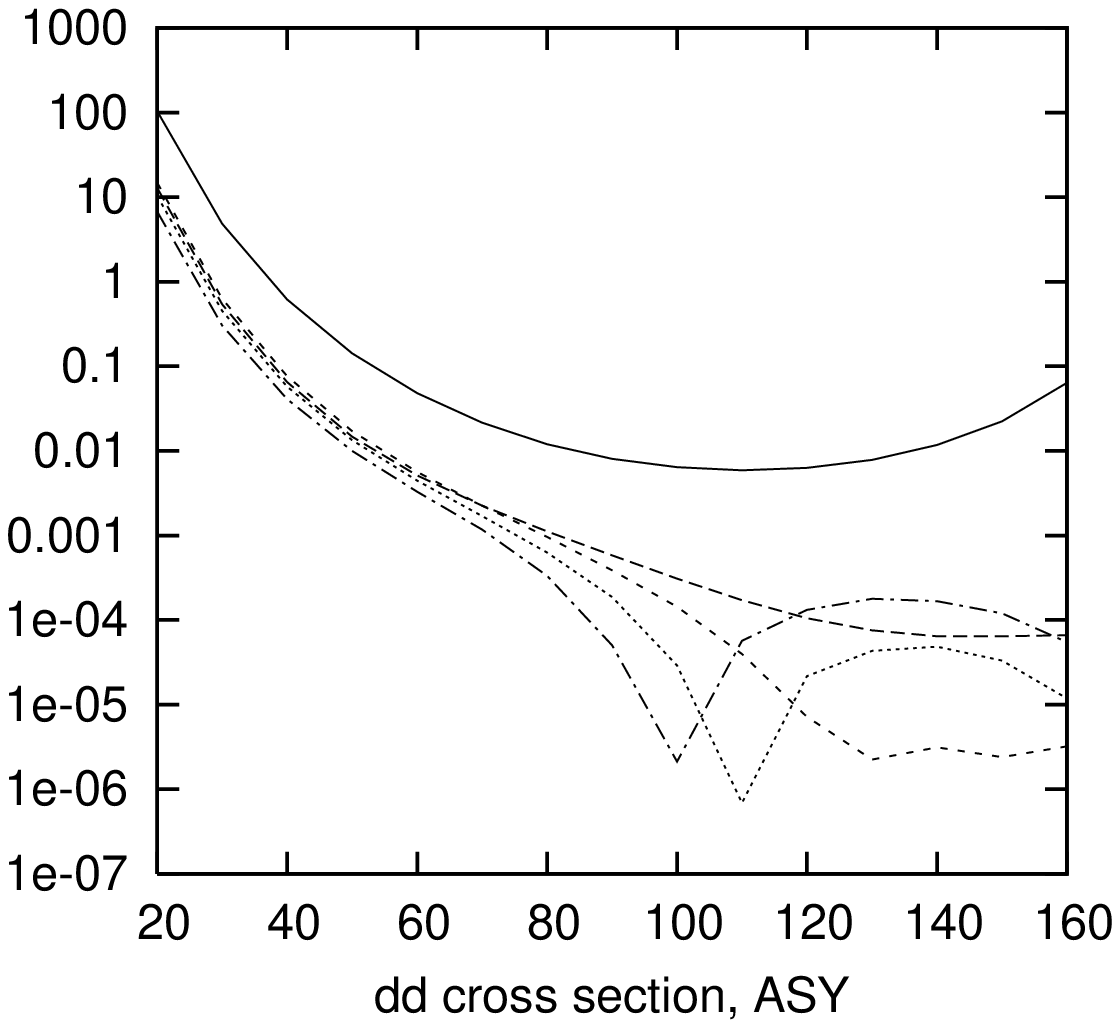}
\includegraphics[angle=270,width=3.3in]{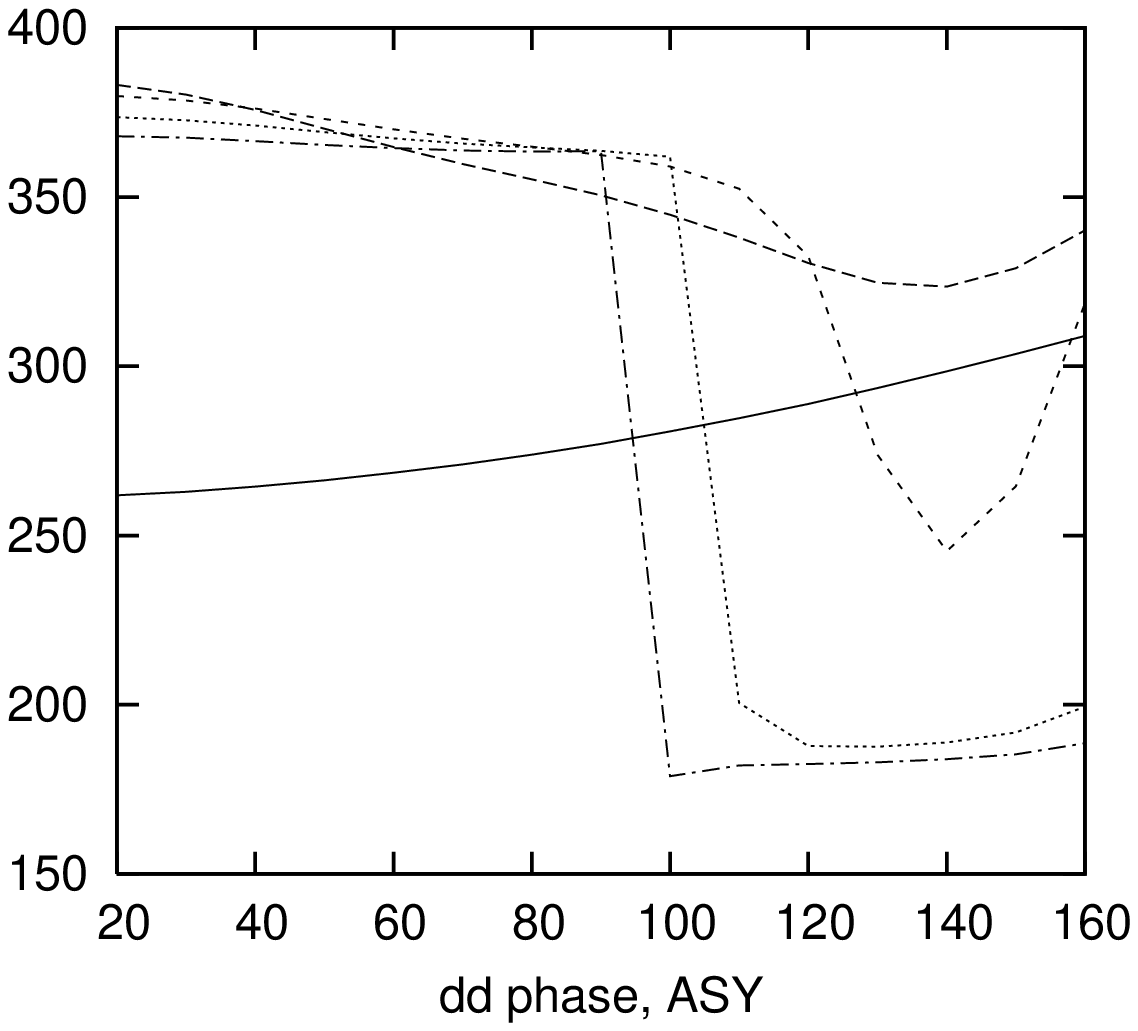}
\includegraphics[angle=270,width=3.3in]{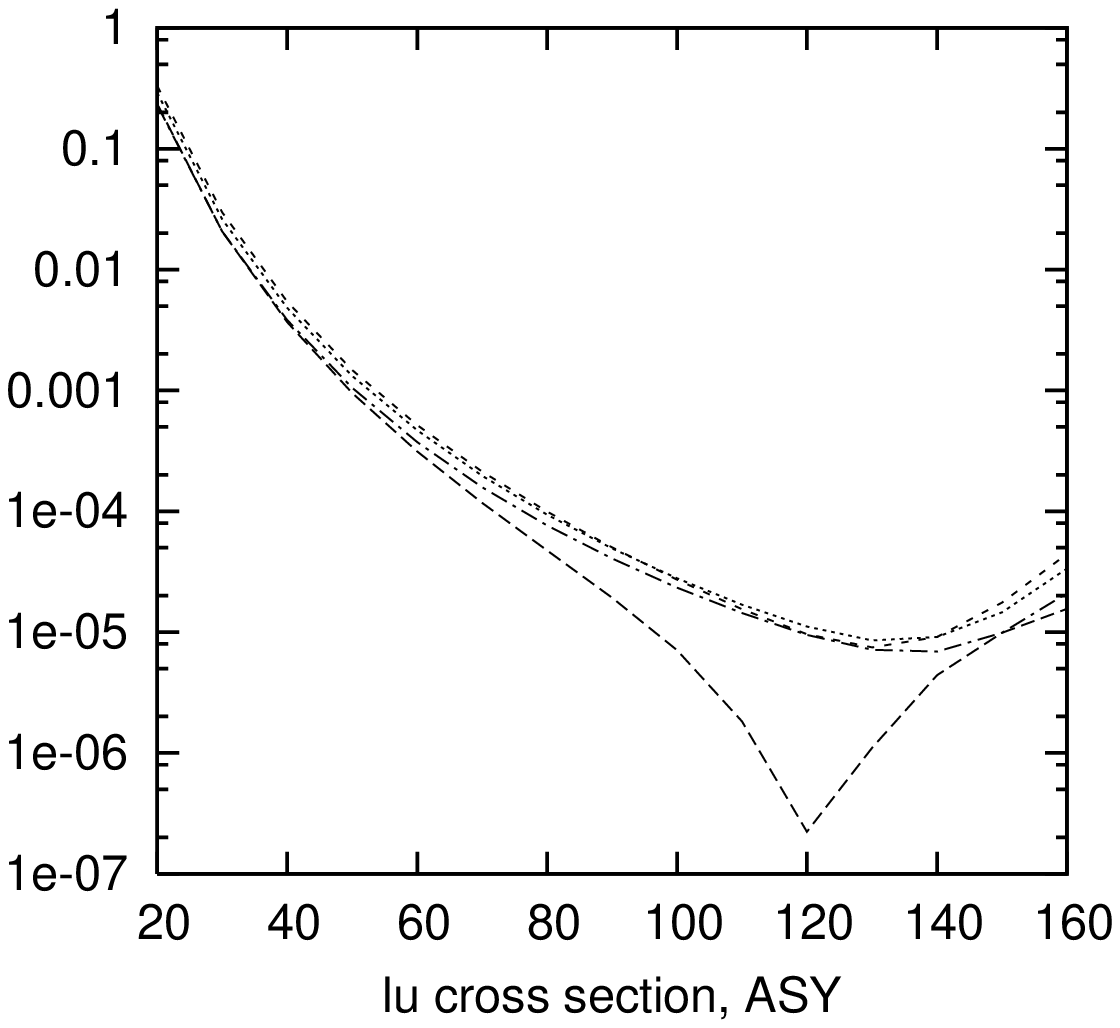}
\includegraphics[angle=270,width=3.3in]{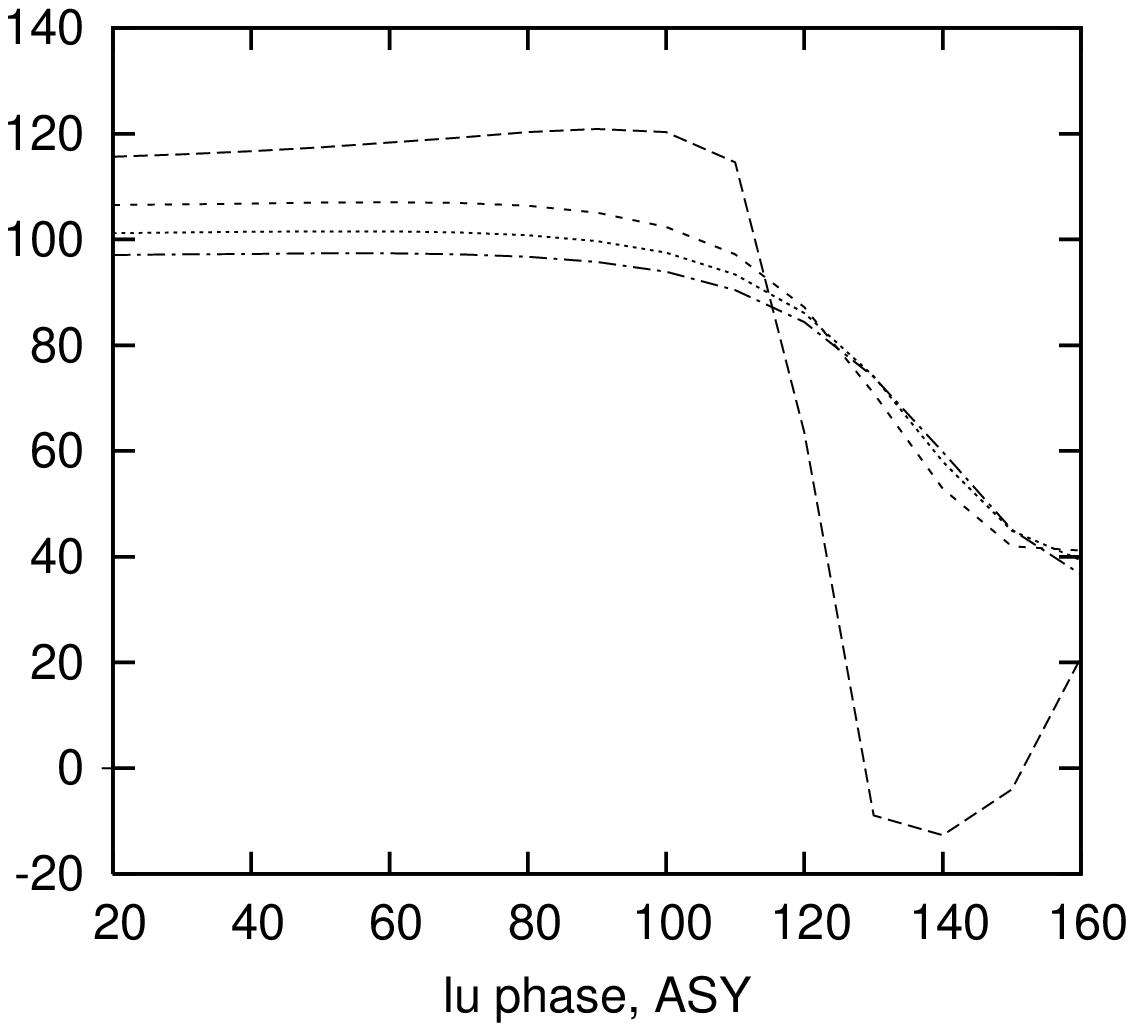}
\includegraphics[angle=270,width=3.3in]{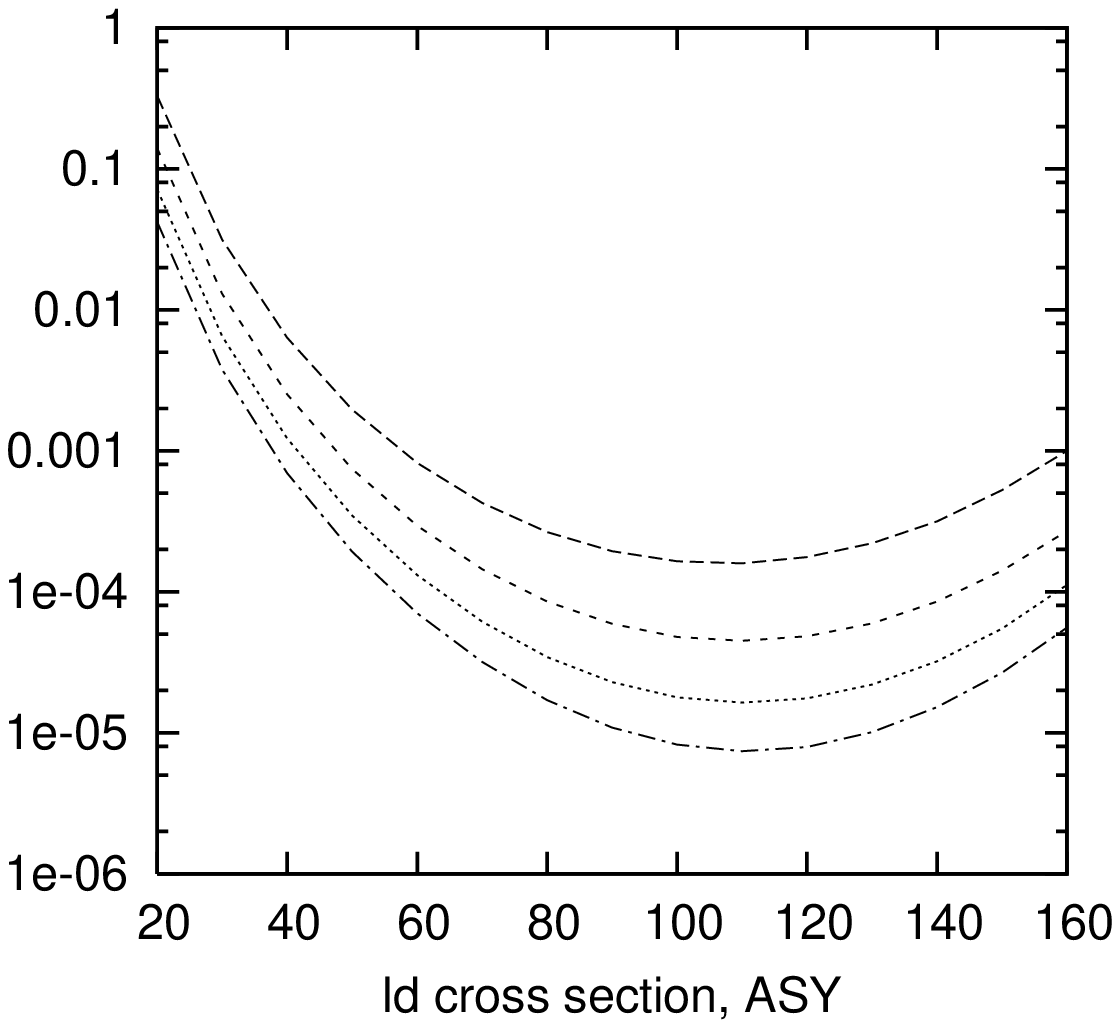}
\includegraphics[angle=270,width=3.3in]{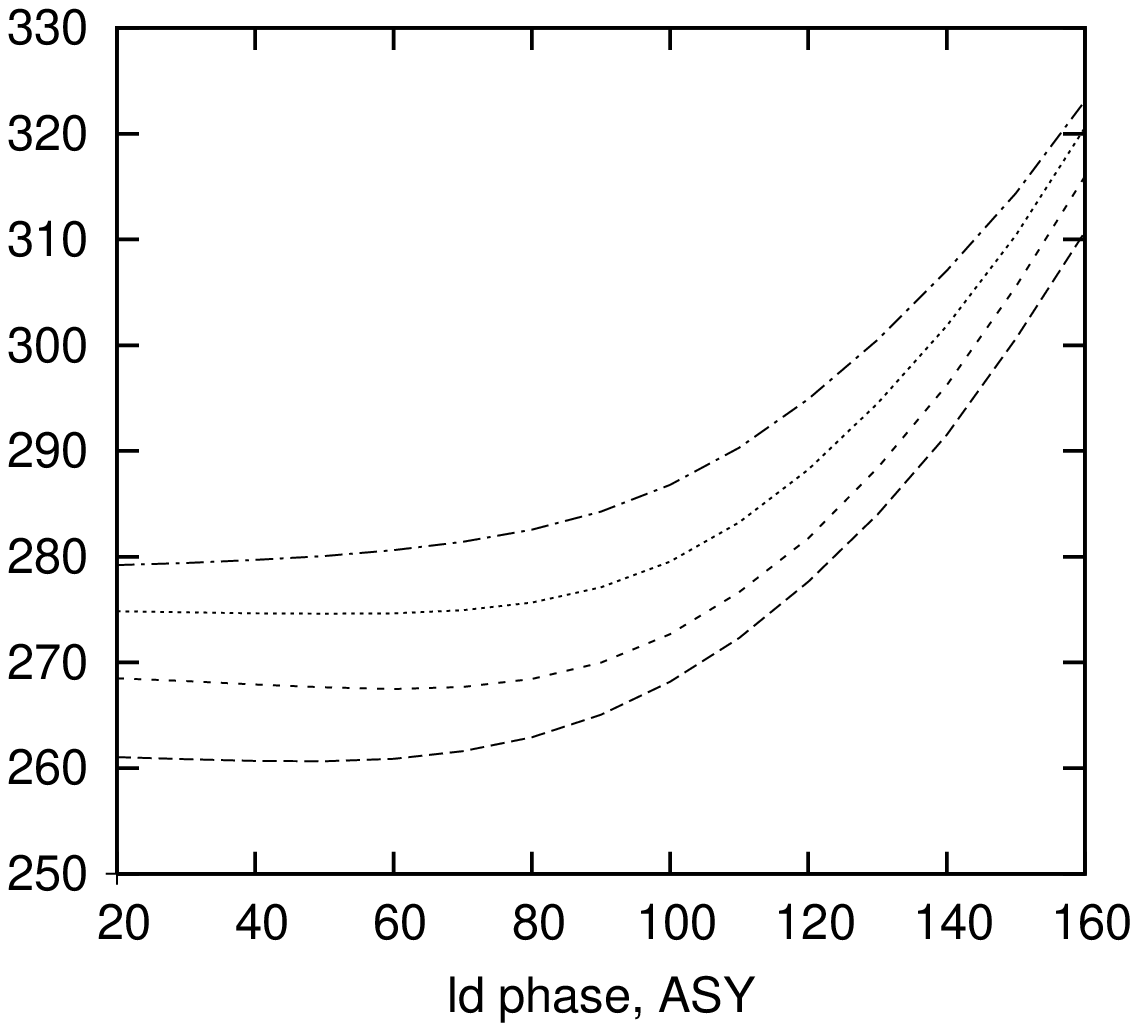}
\caption{Cross section and phase for the ASY distribution amplitude (2 of 2). The vertical axis is 
$s^6d\sigma/dt$ ($10^4$ nb GeV$^{10}$) for cross section plots  and angle in degrees for phase plots. The horizontal 
axis is center of mass scattering angle for all plots. Different values of R are shown as follows: R=1.00 (solid),
R=1.25 (longer dashes), R=1.50 (shorter dashes), R=1.75 (dots), R=2.00 (dashes/dots). 
\label{asy2}}
\end{figure}

\noindent
Plots of cross section and phase against the center-of-mass scattering angle $\theta$ are shown in 
Fig. \ref{coz1} to Fig. \ref{asy2} for different polarizations, for three different distribution 
amplitudes: Chernyak, Oglobin, and Zhitnitsky (COZ), King and Sachrajda (KS), and the asymptotic (ASY)
distribution amplitude. As for the real case, we computed the amplitudes in steps of $10^o$ for
$20^o\leq\theta\leq 160^o$. We include results for different values of $R$, 
the virtuality parameter: 
$R=1.25$ longer dashes; $R=1.50$ shorter dashes; $R=1.75$ dots; 
and $R=2.00$ dashes/dots. Also,  
the real photon ($R=1$) results are shown as a solid line. For the purpose of displaying 
our results graphically, we have again employed the values used by Brooks and Dixon for $\alpha_s$, 
DA evolution, and $f_N$. 

\clearpage

Our results differ somewhat from those of Farrar and Zhang \cite{FZ}
which we do not show in our figures for simplicity.
However, there are some good 
arguments to suggest that our results are more reliable. 
First, it has been shown by Kronfeld and 
Nizic (see Appendix D of \cite{KN}) that the integration method used by Farrar and Zhang is unreliable. 
By comparison, we use the folding method, developed by Kronfeld and Nizic, and we can argue that its use has been 
validated by the fact that we were able to reproduce the results of Brooks and Dixon for the real photon case. 
Second, the FeynComp package has generated the hard scattering amplitudes for each diagram as a function of $R$. 
By setting $R$ equal to 1, we have been able to confirm that these expressions reduce to the 
expressions given in Table III and Table VI of Ref. \cite{KN}, with the noted corrections in Ref. \cite{BD}. 
This gives considerable confidence that FeynComp does indeed accurately calculate the hard scattering 
amplitudes. Unfortunately, we were not able to obtain a copy of the hard scattering amplitudes used by Farrar 
and Zhang for the virtual case. 
Third, one important point to note relates to our implementation of the integrations. 
We use essentially the same Fortran code for both the real ($R=1$) calculation 
and the virtual ($R>1$) calculation and, for a sample of diagrams, we have examined the behavior as 
$R\rightarrow1$ to confirm that the amplitude does approach the $R=1$ result. Since the code for the real case 
appears to be correct, the possibility of errors for the virtual case is diminished. 

Our results for COZ and KS distribution amplitudes are very similar. The most noticable difference is that the KS 
cross sections are, consistently, slightly larger than those of COZ. Our COZ results have also been compared with the 
COZ results of Farrar and Zhang (FZ). (COZ was the only distribution amplitude analyzed by Farrar and Zhang.) The differences 
can be summarized as follows.\\
(1) The normalization used by FZ appears to be different to ours. Their cross sections 
are typically 10 or 100 times greater than ours. Our normalization is the same as used by Kronfeld and Nizic and by 
Brooks and Dixon.\\
(2) For certain helicities, there appears to be a $180^o$ difference in the phase as calculated by FZ compared to
the phase as calculated by us. This is seen for the cases where just one photon has helicity down, i.e. for 
up$\rightarrow$down, down$\rightarrow$up, and longitudinal/temporal$\rightarrow$down. 
We do not know the reason for this difference.\\
(3) up$\rightarrow$up. Cross section is similar for $\theta < 90^o$. However, as $\theta\rightarrow160^o$, in our results 
the cross section reaches a minimum around $120^o$ and then grows larger, whereas in the FZ case, it continues to decrease. 
Phase is also similar for 
$\theta < 90^o$, but, as $\theta\rightarrow160^o$, phase goes to zero in our case, but not in the FZ case.\\
(4) up$\rightarrow$down. Cross section is similar for $\theta < 90^o$. However, as $\theta\rightarrow160^o$, in our results 
the cross section reaches a minimum and then grows larger, whereas in the FZ case, it flattens out. 
Phase plots are similar, apart from the noted $180^o$ difference.\\
(5) down$\rightarrow$up. Cross section has similar shape in both cases, but in our results it grows larger for small 
$\theta$ than in the FZ case. Phase plots have some similarity, but some substantial differences, e.g. in our results, 
as $\theta\rightarrow160^o$, phase approaches $180^o$ for all $R>1$, whereas in the FZ case, this convergence does not 
occur.\\
(6) down$\rightarrow$down. Cross section is similar for small $\theta$, but FZ shows a minimum around 
$\theta = 90^o$, whereas our results show it flattens out as $\theta$ increases. Phase plots are quite different. \\
(7) longitudinal/temporal$\rightarrow$up. Cross section is similar, but the minimum shown in our plot dips lower than 
in the FZ case, as $R$ increases. Phase matches for small and large $\theta$, where the cross section is not close to the 
minimum. Around the cross section minimum, it appears that our amplitude goes to the left of the origin in the 
complex plane, whereas FZ goes to the right.\\
(8) longitudinal/temporal$\rightarrow$down. Cross section and phase match quite well, apart from 
the noted $180^o$ difference in the phase.

We examined another aspect of our results: the relative magnitudes of the contributions coming from 
different diagrams. For real Compton scattering ($R=1$), using the COZ distribution 
amplitude, it was found that, for all forward scattering angles, the amplitude is 
dominated by diagrams in which the two photons attached to the same quark, as already noted in \cite{2photonex}.
As $R$ increases from $1$ to $2$, the range of angles for which dominance occurs decreases, so that, for 
DVCS ($R=2$), 'same-quark' diagrams only dominate for a scattering angle up to $20^o$. At larger angles, for 
$R=2$, the dominance is not observed.
 
\begin{figure}
\centerline{\epsfig{figure=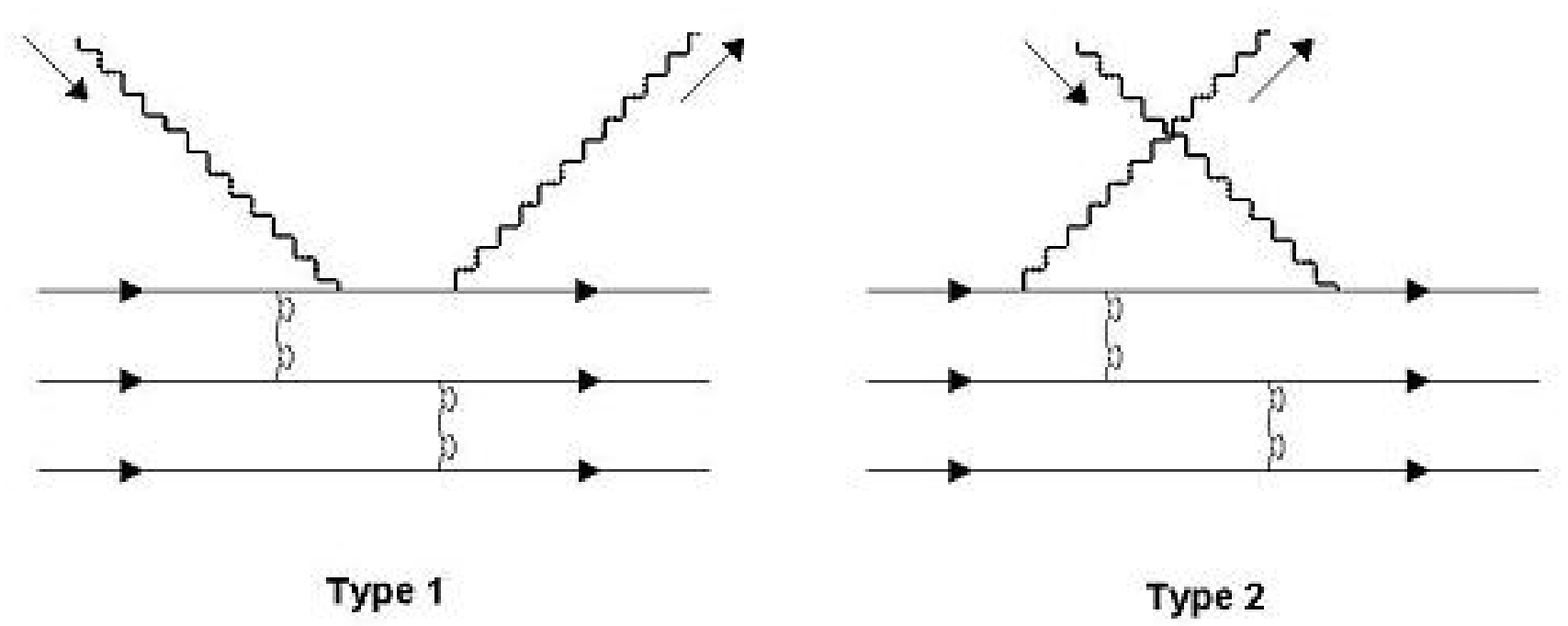, width=6in}}
\caption{Diagrams where both photons are attached to the same quark 
\label{DVCS}}
\end{figure}

The 'same-quark' dominance is of interest because of its relationship to the handbag diagram 
approach to DVCS, using generalized parton distributions \cite{V}. This approach is 
applicable in the case of small momentum transfer and large $Q^2$, which implies small 
scattering angle. The condition that the same quark should be struck by both 
initial and final photons is necessary for the handbag diagrams. However, 
this is not a sufficient condition for the handbag diagrams. In general,  
the diagrams in which the same quark is struck by the two photons include 
some where there is no gluon exchange between the two photons ({\it type 1} in Fig. 
\ref{DVCS}) and some where there is a gluon exchange between the two photons ({\it type 
2} in Fig. \ref{DVCS}).  
In the handbag diagram, the two photons are attached to the same quark, but there 
is no gluon exchange in between. This corresponds to the {\it type 1} diagram in Fig. \ref{DVCS}. How do we account for the 
presence of {\it type 2} diagrams in our result? We believe that it can be explained by noting that the GPD calculation should 
be performed in the Light Front gauge, whereas our calculation has been performed in the Feynman gauge. The total amplitude should be 
gauge invariant, but the contributions from individual diagrams will vary according to the gauge used. If we recalculate in 
the Light Front gauge, it is expected that contributions from {\it type 2} diagrams will become much smaller, while 
contributions from {\it type 1} diagrams will grow correspondingly to leave the total amplitude unaltered \cite{SB}. 
Therefore we speculate that, in the Light Front gauge, for a scattering angle of $20^o$, the amplitude is 
dominated by {\it type 1} diagrams only.

\subsection{Normalization of Results and the Proton Form Factor}
\noindent
Due to uncertainty in the normalization of the pQCD cross section, following the example of Reference \cite{BD}, 
it is recommended to normalize using the Dirac form factor for the proton, $F^p_1$, as follows:
\be\label{Ratio}
\frac{s^6\frac{d\sigma}{dt}}{[Q^4F^p_1(Q^2)]^2},
\ee
where $Q$ represents spacelike momentum transfer. This dimensionless ratio is independent of the normalization. 

To normalize the calculated Compton data, we therefore need to use a calculated form 
factor. Brooks and Dixon have calculated this form factor and found a discrepancy with 
previous results \cite{FF,JSL}. 
They obtained a result exactly half of the previous results. To try to resolve this 
discrepancy, we recalculated the form factor ourselves. Using the
COZ and KS distribution amplitude, we found that we got the same results as 
Brooks and Dixon. Details of our calculation can be found in Appendix C. 

We reviewed the earlier paper of Ji, Sill and Lombard (JSL) \cite{JSL} and identified the cause for this discrepancy. 
It turns out that they had used a different normalization for the distribution amplitude. 
JSL included a factor of $8\sqrt{3}$ in the denominator 
of the distribution amplitude, rather than the factor of $8\sqrt{6}$ we have used, as in Eq. (\ref{flip}). 
If the JSL calculation is modified to use a factor of $8\sqrt{6}$ instead of $8\sqrt{3}$, then the 
result concurs with ours. Further analysis (by ourselves and by Hsiang-nan Li \cite{LImail}) of other calculations of the 
form factor \cite{FF, LI, BKBS, KLSJ} indicates that they
have all used a normalization consistent with JSL. On the other hand, Compton scattering calculations \cite{KN,BD,VG} appear to have 
all used the same normalization as we have used. We do not know which of these two normalization choices is correct. However, if one 
forms the ratio (\ref{Ratio}), as suggested by Brooks and Dixon, this dependency disappears.   
\section{Comparison with Experiment}
\label{sect.IV}
\noindent
\subsection{Experimental Results for Real Compton Scattering}
\begin{figure}
\centering
\includegraphics[angle=270,width=6in]{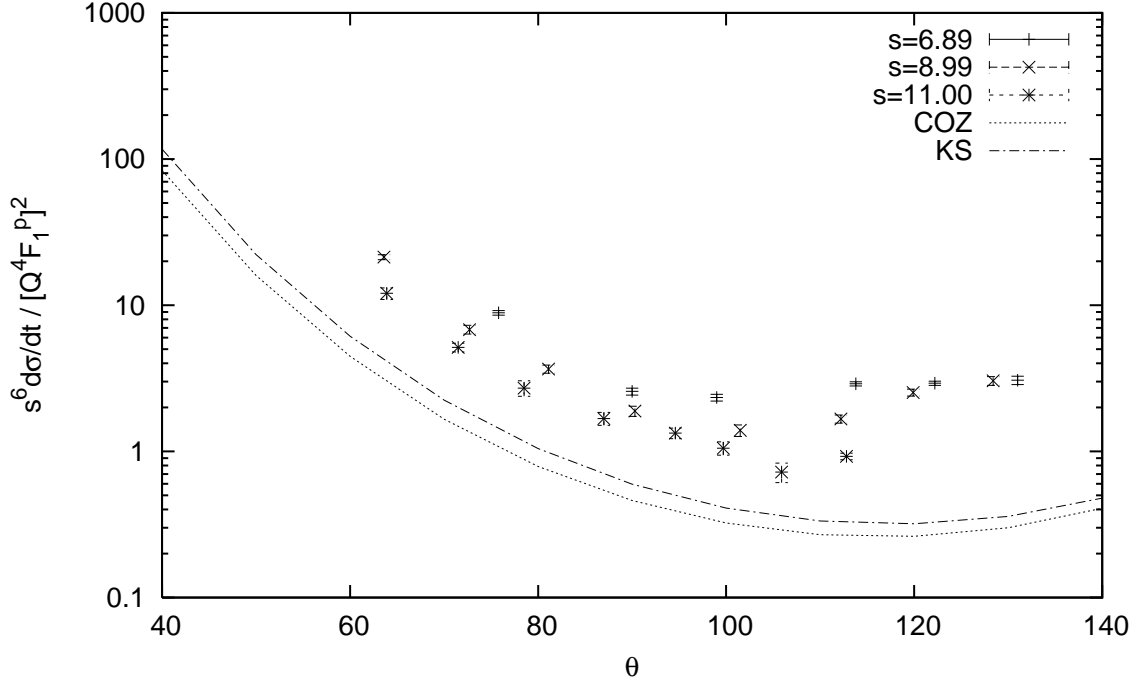}
\caption{Scaled unpolarized real Compton cross section: comparing our results with JLAB experiment. Note that the results have been
normalized by the scaled form factor, $F_1^p$. See the text for full explanation.}  
\label{realexp}
\end{figure}

Fig. \ref{realexp} shows the unpolarized, real Compton cross section for the proton, comparing our calculations with recent, 
preliminary experimental results from JLAB \cite{BW}. The figure plots our results for the COZ and KS distribution amplitudes and the 
experimental results for $s=6.89$ GeV$^2$, $s=8.99$ GeV$^2$, and $s=11.00$ GeV$^2$. All results are scaled, by multiplying by $s^6$ 
and then normalized using the Dirac form factor, as explained in Section \ref{sect.III}. For the COZ and KS distribution 
amplitudes, we have used the form factor we calculated ourselves following the method described in Appendix C. For the 
experimental results, we have used 
\be
Q^4F^p_1(Q^2)=1.0\text{ GeV}^4
\ee
for $Q^2$ in the range $7-15$ GeV$^2$, as suggested by Brooks and Dixon.

Some interesting points emerge from the figure. First, it can be seen quite clearly that the experimental data does 
indicate that the scaled cross section is decreasing as $s$ increases. Thus the data is consistent with the 
expectation that it will asymptotically approach the pQCD curve. Secondly, the difference between the $s=11.00$ GeV$^2$ data 
and the pQCD curves represents a factor of 2-3, for KS, and 3-4, for COZ. This is a much smaller factor than Brooks and 
Dixon found, when comparing with the earlier data of Shupe {\it et al} \cite{Shupe}. We are therefore optimistic 
that the $12$ GeV upgrade of the CEBAF facility can produce results that are even closer to our pQCD curves, when $s$ 
is increased still further. A third point that can be deduced from the figure is that the scaling factor $s^{-n}$ for the 
experimental data corresponds to a value of $n>6$. Clearly, if this were not the case, we would not see the scaled 
cross section decreasing as $s$ increases. Since the $s^{-6}$ scaling law is predicted by counting the quarks and photons 
involved, we can also speculate that there are additional virtual particles involved in the interaction that cause 
$n$ to increase. Finally, we note that, at larger angles ($>110^o$), the experimental data deviates more strongly from 
the pQCD curve. In this region, the Mandelstam u becomes quite small (about $-1$ GeV$^2$), so the applicability of the pQCD 
curve is questionable.

As was noted in the Introduction, Section \ref{sect.I}, there are serious doubts raised about the validity of the 
Brodsky-Lepage approach we have used. Therefore we ask the question, what can the experimental data tell us about the 
validity of this approach? Since our calculations represent an asymptote, we need to look for asymptotic behavior also 
in the experimental data. This means experimentalists should seek a value of $s$ where further increases in $s$ do not 
substantially change the scaled cross section plot. Several experimental outcomes are possible and each has some 
consequences for our model.\\
(a) Suppose that, as $s$ increases, the scaled cross section continues to decrease and goes lower than the pQCD 
prediction. In this case, it is likely that the COZ and KS models are not valid. Probably a more symmetric distribution 
amplitude would be needed.\\
(b) Suppose that, as $s$ increases, the scaled cross section stops decreasing, but is still higher than the pQCD 
prediction. In this case, if we suppose that the asymptote has been reached, then it suggests that there is still 
some soft contribution that needs to be considered, even asymptotically.\\
(c) Suppose that, as $s$ increases, the scaled cross section stops decreasing and approximates to our pQCD prediction. 
In this case, the experimental data will give some support for our approach and the asymmetric form of the distribution 
amplitudes. However, even if the ratio (\ref{Ratio}) is agreeing well with experiment, it is still necessary to 
successfully predict the normalization of the Compton scattering cross section and the form factor separately 
before success of the Brodsky-Lepage approach can be claimed.\\
We are hopeful that the $12$ GeV upgrade of the CEBAF facility can shed light on which of these three outcomes actually 
occurs in the real world.
 
Another recently published result from JLAB \cite{JLABR} discusses the polarization transfer observable $K_{LL}$, at a scattering 
angle of $120^o$, $s=6.9$ GeV$^2$ and $t=-4.0$ GeV$^2$, obtaining a result of $0.678\pm 0.083\pm 0.04$.  We note that this experiment 
has a value of $u=-1.1$ GeV$^2$, which is barely more than the square of the proton mass. It thus appears that the kinematical 
requirements of the Brodsky-Lepage approach are not met, so it should not be a surprise that the value predicted by our pQCD 
calculations is quite different from the experimental value. We use the formula
\be
\label{KLL}
K_{LL}=\frac{\frac{d\sigma^+_+}{dt}-\frac{d\sigma^-_+}{dt}}{\frac{d\sigma^+_+}{dt}+\frac{d\sigma^-_+}{dt}},
\ee
where $d\sigma^\lambda_h$/$dt$ is the cross section for proton helicity $h$ and incoming 
photon helicity $\lambda$. 
Using the COZ and KS distribution amplitudes gives the value of $K_{LL}$ as -0.134 and 
-0.298, respectively, in agreement with the similar calculation of Brooks and Dixon, but 
clearly very different from the experimental value. However, for kinematics where hard physics is dominant, we 
expect that experimental results would agree with our calculation of $K_{LL}$.

We note that the handbag approach has been applied to wide-angle Compton scattering 
\cite{Radyushkin, DFJK, HKM, Kroll1, Kroll2} and has provided, in the case of RCS, 
predictions which are in good agreement with 
the JLAB data. This approach argues that, at currently accessible kinematics, in the 
domain where $s$, $|t|$, and $|u|$ are all large on a hadronic scale of $~1$ GeV$^2$, 
the scattering amplitude is dominated by soft overlap contributions. It is further argued 
that the hard scattering contributions are very small in comparison, resulting 
from a symmetric distribution amplitude similar to the asymptotic form rather than the 
asymmetric forms such as COZ or KS. In the handbag 
approach, the Compton helicity amplitudes can be written \cite{HKM}
\begin{eqnarray}
\label{HBAGRCS}
{\cal M}^{\lambda\lambda^\prime}_{++^\prime}(s,t)&=&
2\pi \alpha_{em}[{\cal H}^{\lambda\lambda^\prime}_{++^\prime}(s,t)(R_V(t)+R_A(t))
+ {\cal H}^{\lambda\lambda^\prime}_{--^\prime}(s,t)(R_V(t)-R_A(t))], \text{ and}\\
{\cal M}^{\lambda\lambda^\prime}_{+-^\prime}(s,t)&=&
-\pi \alpha_{em}\frac{\sqrt{-t}}{m}[{\cal H}^{\lambda\lambda^\prime}_{++^\prime}(s,t)
+ {\cal H}^{\lambda\lambda^\prime}_{--^\prime}(s,t)]R_T(t),
\end{eqnarray}
where ${\cal H}^{\lambda\lambda^\prime}_{hh^\prime}$ denotes the helicity amplitudes for the 
subprocess $\gamma q \rightarrow \gamma q$ (i.e. the handle of the handbag) and 
$R_V$, $R_A$, and $R_T$ are soft form factors which can be defined in terms of moments of GPDs. 
For example, $R_V$ is defined via the equations
\begin{eqnarray}
R_V^a(t)&=&\int^1_{-1}\frac{d\bar{x}}{\bar{x}}H^a(\bar{x},0;t), \text{ and}\\
R_V(t)&=& \sum_a e^2_aR_V^a(t).
\end{eqnarray}
Here, if the $j$th quark is struck, and $p$ ($p^\prime$) and $k_j$ ($k_j^\prime$) denote the 
incoming (outgoing) nucleon and struck quark momenta, then 
$\bar{x}=(k_j+k_j^\prime)^+/(p+p^\prime)^+$, $a$ is the flavor of the quark, 
$e_a$ is the charge of the quark and the sum is over all quark flavors.
 
Since the exact form of the GPDs is not known from first principles, a model is 
used. See \cite{Kroll1,Kroll2}, for example, for such a model. Based on this model, 
the handbag approach is able to successfully predict \cite{Kroll2} the JLAB data in Fig. \ref{realexp}, 
except for larger angles where $u$ becomes small, as noted previously. The handbag approach 
also gives a prediction of $K_{LL}$ based on the equation \cite{JLABR}
\begin{equation}
K_{LL}\simeq \frac{R_A}{R_V}K_{LL}^{KN}\left[ 1-\frac{t^2}{2(s^2+u^2)}
      \left( 1-\frac{R_A^2}{R_V^2} \right) \right]^{-1}, 
\end{equation}
where $K_{LL}^{KN}$ is the Klein-Nishina asymmetry for a structureless proton. 
Using a similar GPD model, this gives a result which is in good agreement with the JLAB measurement. 
While the success of the handbag approach and the doubts about the validity of the asymmetric distribution 
amplitudes imply that experiments may need to go to much higher energy before the onset of asymptotic 
behavior occurs, further comparisons between available data and theoretical calculations (GPD model 
and pQCD) are necessary to make more firm conclusions. 

\subsection{Experimental Results for Virtual Compton Scattering}
For virtual Compton scattering (VCS) there is also data available from two recent experiments conducted at JLAB. The first experiment 
\cite{JLABV1} gives measurements of beam-spin asymmetry for the process $ep\rightarrow ep\gamma$. The beam-spin asymmetry is 
given by 
\be
\label{BSA1}
A=\left(\frac{d^4\sigma^+}{dQ^2dx_{B}dtd\phi}-\frac{d^4\sigma^-}{dQ^2dx_{B}dtd\phi}\right)/
        \left(\frac{d^4\sigma^+}{dQ^2dx_{B}dtd\phi}+\frac{d^4\sigma^-}{dQ^2dx_{B}dtd\phi}\right),
\ee
where $\sigma^+$ and $\sigma^-$ refer to the cross section for incoming electron beams with positive and negative helicity, respectively; 
$Q^2$, as before, is $-q^2$; $x_{B}$ is the Bjorken scaling variable given by $Q^2$/$2p.q$ where $p$ and $q$ are the 
4-momenta of the incoming proton and photon, respectively; $t$, as before, is $(p-p^\prime)^2$; and $\phi$ is the angle
between the proton and electron planes. The JLAB paper compares the experimental result with theoretical calculations based on Generalized 
Parton Distributions (GPDs) with $Q^2=1.25$ (GeV/c)$^2$, $x_B=0.19$, and $t=-0.19$ (GeV/c)$^2$. We note, unfortunately, that the 
kinematics of this experiment are outside the range of applicability for the Brodsky-Lepage approach. Therefore, we do not consider it 
further.  

The second experiment \cite{JLABV2} measures the cross section for VCS at a large backward angle, for various energies up to about
$s=3.6$ GeV$^2$, with a photon virtuality of $Q^2=1$ GeV$^2$. Again, the kinematics of the experiment are not suitable for applying the 
Brodsky-Lepage approach. However, there is one point discussed in the paper, where our calculations can give some insight. The 
paper finds at around $s=3.2$ GeV$^2$, that the virtual cross section and the real cross section become equal, thus providing evidence 
of $Q^2$-independence. The $Q^2$-independence, if true, would presumably also exist at higher energies where pQCD becomes 
applicable, so we should be able to see evidence of it in our own results. Referring to the plots of the COZ and KS cross sections 
(Fig. \ref{coz1} to Fig. \ref{ks2}), one can see for VCS ($R>1$) that, at large backward angles, the unpolarized cross section is 
dominated by contributions where the incoming photon polarization is aligned with that of the proton ('uu' and 'ud' in the figures). 
However, the cross section is much larger than for RCS ($R=1$). Therefore, our results for COZ and KS are not consistent with the 
conjecture of $Q^2$-independence. The same exercise can be performed using the ASY distribution amplitude (Fig. \ref{asy1} and 
Fig. \ref{asy2}). Once again, our results for ASY are not consistent with $Q^2$-independence. However, for ASY, if one considers 
only VCS, the 'ud' cross section dominates at large backward angles, and does show $Q^2$-independence. 

Based on these results, we believe that it is very important to repeat this experiment at larger values of $s$ and at different 
$Q^2$, as suggested in the paper. If future experiments do verify the existence of $Q^2$-independence, it would, at a minimum, 
place some strong restrictions on the possible form that the distribution amplitude could take. Furthermore, if $Q^2$-independence 
is observed for virtual photons but not for real photons, it indicates that the COZ and KS distribution amplitudes are not valid, 
but gives support to the symmetric ASY distribution amplitude. 

We also note that the handbag approach may not be suited to the kinematics of VCS at large backward angle. An 
alternative theoretical framework has recently been proposed \cite{PS} which may be able to provide predictions in 
this kinematic regime.  

\section{Conclusions}
\label{sect.V}
\noindent 
We have presented a leading order pQCD calculation of cross section and phase for real Compton scattering.
We are quite confident in the accuracy of our results, but the experimental data indicates 
that, at available energies, our pQCD prediction only represents a fraction of the total cross section. 
However, the experimental data also shows that the scaled cross section is decreasing as $s$ increases, 
which is consistent with the view that it will approach an asymptote as $s$ increases further.  
The authors are optimistic that the planned upgrade of JLAB will allow further experiments that can 
get close to this asymptotic region. If this proves to be the case, it will provide support for 
the Brodsky-Lepage approach and will allow us to determine the validity of various proposed distribution 
amplitudes.

One improvement that can be made to the real Compton calculation is to add the effects of 
higher twist. It is noted that, in addition to the measurement of $K_{LL}$, JLAB \cite{JLABR} 
has also measured the value of $K_{LS}$ as $0.114\pm 0.078\pm .04$, 
which implies a non-negligible proton helicity flip process. In the leading twist calculation, 
this helicity flip cannot occur. Therefore the authors believe that it is important to extend 
the pQCD calculation to include the higher twist effects in order to 
provide a mechanism whereby the helicity flip may occur. 

We have also presented leading order pQCD calculation of cross section and phase for virtual Compton 
scattering. Since the calculation uses the same method and Fortran code as the real 
calculation, we have some confidence also in these results. Unfortunately, the available experimental data 
for VCS is outside the kinematic range where pQCD is applicable, so we have not been able to make a 
meaningful comparison between our results and experimental data. However, we are able to give some insight 
into the '$Q^2$-independence' phenomenon, proposed in Ref. \cite{JLABV2}. 

One other area where our virtual results are of interest is DVCS. Our calculations show that, for angles 
where $(4$-momentum transfer$)^2$ is small compared to the $($total $4$-momentum$)^2$, the amplitude is 
dominated by diagrams in which the incoming and outgoing photons attach to the same quark. This same scenario can be 
calculated in the Light Front gauge using the GPD method applied to the 'handbag' diagram. We expect that, if our 
calculations are redone in the Light Front gauge, the amplitude will be dominated by handbag-type diagrams. 
If this expectation turns out to be correct, it indicates some commonality between the pQCD and GPD approaches. 
Furthermore, by examining the high energy limit, we expect to see a correlation between the GPD and pQCD approaches. 
This expected correlation can be used to test the validity of various GPD models proposed in the literature 
by looking at how closely they reproduce the pQCD predictions. This can also shed some light on the question 
concerning the energy level at which the pQCD approximation will become valid.
  
\acknowledgments
\noindent
We would like to thank the JLAB spokespersons for the E99-114 experiment (C. Hyde-Wright, A. Nathan, and B. Wojtsekhowski) for making available the results of their real Compton scattering experiment. We would like to thank Lance Dixon for making available details of his calculations for real Compton scattering. We would also like to thank Stan Brodsky for valuable discussions and encouragement at the Lightcone 2004 conference. This work has been supported in part by DOE grant DE-FG02-96ER40947, NSF grant INT-99063384, and R. Thomson's SURA fellowship.

\appendix

\section{Details of Calculation for Diagram A51}
\label{sect.A}
\begin{figure}
\centerline{\epsfig{figure=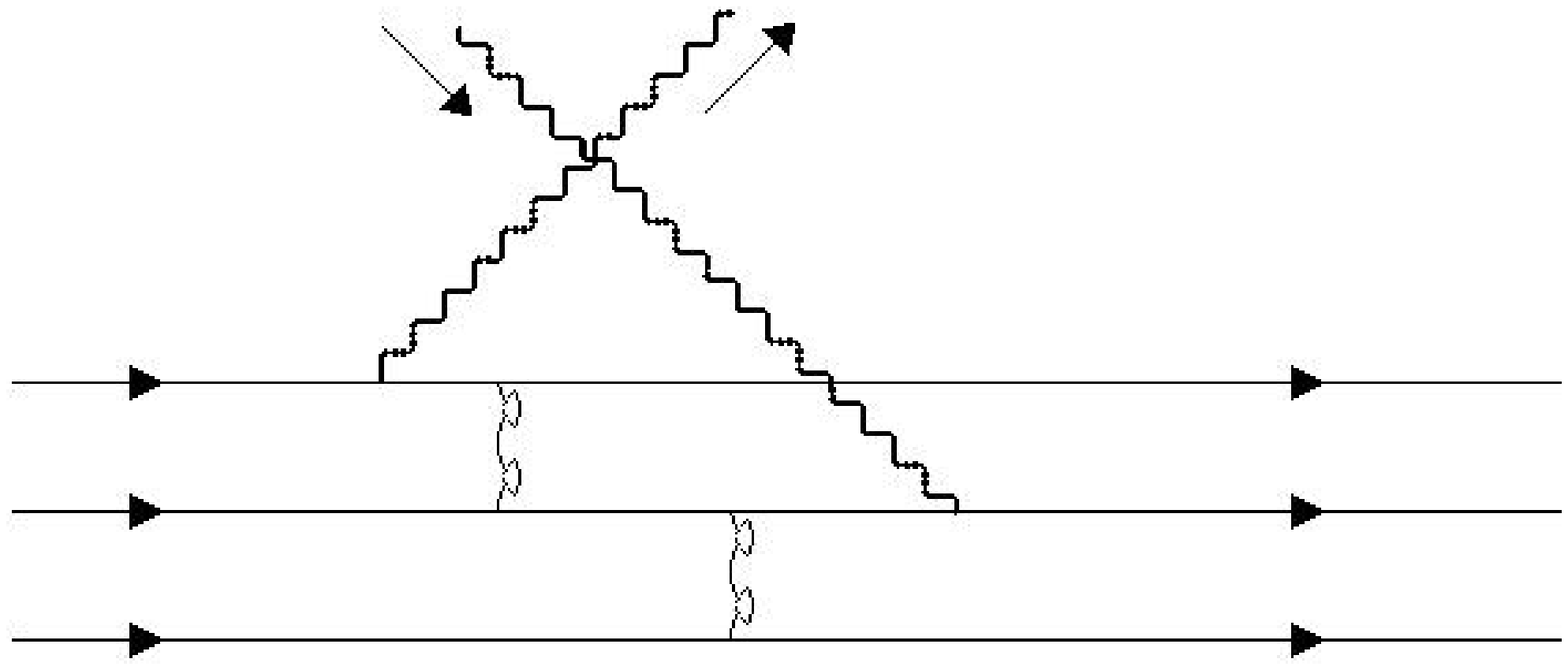,width=5in}}
\caption{Feynman Diagram A51
\label{compute_a51}}
\end{figure}

\noindent
The diagram $A51$ is shown in 
Fig.\ref{compute_a51}.
From the output of FeynComp, A51 is found to have a pole structure of 
$<\bar{y}_3,1><y_1,x_1><y_1,\bar{x}_3>$ belonging to the group $3e$ in
Table \ref{pole}.
The integral $I$ that has to be considered is
\begin{equation}
I(k_1,\bar{k}_3;l_1,\bar{l}_3)
=\int^1_0 dx_1dx_3\int^1_0 dy_1dy_3 
{ x^{k_1}_1 \bar{x}^{\bar{k}_1}_1
x^{k_3}_3\bar{x}^{\bar{k}_3}_3 y^{l_1}_1\bar{y}^{\bar{l}_1}_1
y^{l_3}_3\bar{y}^{\bar{l}_3}_3
\over <1-\bar{y}_1y_3,1><y_1,\bar{x}_3x_1><y_1,\bar{x}_3>}.
\end{equation}
Since $<\bar{y}_3,1>=\bar{y}_3(1-Rs^2)-Rc^2$ do not vanish for
all kinematics we consider, it is not a real pole and $I$
has in fact only two real poles.
Moreover $\delta(<y_1,\bar{x}_3x_1>)\delta(<y_1,\bar{x}_3>)$ has
contribution only from the region $x_1=1$  which has a measure zero.
Therefore the real part of $I$ has contribution coming from
the two principal value integral:
\begin{eqnarray}
Re[I(k_1,\bar{k}_3;l_1,\bar{l}_3)] 
&=&\int^1_0 dx_1dx_3\int^1_0 dy_1dy_3 
{x^{k_1}_1 \bar{x}^{\bar{k}_1}_1
x^{k_3}_3\bar{x}^{\bar{k}_3}_3 y^{l_1}_1\bar{y}^{\bar{l}_1}_1
y^{l_3}_3\bar{y}^{\bar{l}_3}_3
\over <1-\bar{y}_1y_3,1>}  \\ \nonumber
& & P{1\over <y_1,\bar{x}_3x_1>} P{1\over <y_1,\bar{x}_3>}.
\end{eqnarray}
We denote $Re[I]$ by $J_1$. It can be reduced to
\begin{equation}
\label{J1}
J_1={1\over R^2}\int dx_3dy_1dy_3 
{ x^{k_3}_3\bar{x}^{\bar{m_3}-1}_3
y^{l_1}_1\bar{y}^{\bar{l}_1}_1 y^{l_3}_3\bar{y}^{\bar{l}_3}_3
\Omega^{\bar{m}_1}_{m_1}\left({y_1\over R\bar{x}_3(1-\bar{y}_1s^2)}\right)
\over
<1-\bar{y}_1y_3,1>(1-\bar{y}_1s^2)^2} 
P{1\over \bar{x}_3-{y_1\over R(1-\bar{y}_1s^2)}}.
\end{equation}
Here we have used the function $\Omega^n_m(a)$ defined by
\begin{equation}
\Omega^n_m(a)=\int^1_0 dx{x^m\bar{x}^n\over x-a}.
\end{equation}
It can be implemented via the recursive relations:
\begin{eqnarray}
\Omega^0_0(a) &=& ln|(1-a)/a|, \\
\Omega^0_m(a) &=& a\Omega^0_{m-1}(a)+1/m, \\
\Omega^n_m(a) &=& (1-a)\Omega^{n-1}_m(a)-B(m+1,n),
\end{eqnarray}
for $|a|<1$  and via the Gauss hypergeometric series for $|a|>1$.
But in fact, as we only need to know $\Omega^n_m$ for a limited
finite values for $m$ and $n$,
an easier and faster way is to integrate $\Omega^n_m$ 
explicitly for each $m$ and $n$ values.
The final principle value integration over $x_3$ can be handled
using the method described in Appendix B.

The imaginary part of $I$ consists of two contributions.
The first one is 
\begin{equation}
H_1=-i\pi\int^1_0 dx_1dx_3\int^1_0 dy_1dy_3 
x^{k_1}_1\bar{x}^{\bar{k}_1}_1 x^{k_3}_3\bar{x}^{\bar{k}_3}_3
y^{l_1}_1\bar{y}^{\bar{l}_1}_1y^{l_3}_3\bar{y}^{\bar{l}_3}_3
{\delta(<y_1,x_1>)\over <1-\bar{y}_1y_3,1><y_1,1-\bar{x}_1x_3>}.
\end{equation}
It can be reduced to a two-dimensional integral:
\begin{equation}
\label{H1}
H_1={i\pi B(k_3+1,\bar{k_3})\over R^{k_1+\bar{k}_1+1}}
\int^1_0 dy_1dy_3 
{y_1^{k_1+l_1}\bar{y}_1^{l_1}\tilde{y}^{\bar{k}_1-1}y^{l_3}_3
 \bar{y}^{\bar{l}_3}_3
 \over
 (1-\bar{y}_1s^2)^{k_1+\bar{k}_1+1}<1-\bar{y}_1y_3,1>
},
\end{equation}
where $\tilde{y}=R(1-\bar{y}_1s^2)-y_1$ and the function
$B(m,n)$ is the beta function defined by
\begin{equation}
B(m,n)=\int^1_0 dx x^{m-1}\bar{x}^{n-1}=
{\Gamma (m)\Gamma (n)\over\Gamma(m+n)}.
\end{equation}

The contribution from the other $\delta$-function:
\begin{equation}
H_2=-i\pi\int^1_0 dx_1dx_3\int^1_0 dy_1dy_3
x^{k_1}_1\bar{x}^{\bar{k}_1}_1 x^{k_3}_3\bar{x}^{\bar{k}_3}_3
y^{l_1}_1\bar{y}^{\bar{l}_1}_1y^{l_3}_3\bar{y}^{\bar{l}_3}_3
{\delta(<y_1,\bar{x}_3>)\over <1-\bar{y}_1y_3,1><y_1,\bar{x}_3x_1>}
\end{equation}
can be reduced to  
\begin{equation}
\label{H2}
H_2={-i\pi B(k_1+1,\bar{k_1})\over R^{k_3+\bar{k}_3+1}}
\int^1_0 dy_1dy_3 
{y_1^{\bar{k}_3+l_1-1}\bar{y}_1^{\bar{l}_1}\tilde{y}^{k_3}
y^{l_3}_3 \bar{y}^{\bar{l}_3}_3
 \over
 (1-\bar{y}_1s^2)^{k_3+\bar{k}_3+1}<1-\bar{y}_1y_3,1>
}.
\end{equation}
Eqs.(\ref{J1}),(\ref{H1}) and (\ref{H2}) 
act as ``template'' for all
the diagrams in group $3e$. The contributions for each
diagram in the same group can be written in terms of these
``template'' expressions.
For example, the contributions from $A51$ to the $\lambda,
\lambda^\prime=\uparrow,\uparrow$ amplitude is given by
\begin{equation}
I^{A51}(\uparrow\uparrow)=\int [dx][dy] 
x^{m_1+1}_1x_2x^{m_3+1}y^{n_1+1}_1y_2y^{n_3+1}_3 \tilde{T}^{A51}
(\uparrow\uparrow),
\end{equation}
where
\begin{equation}
\tilde{T}^{A51}(\uparrow\uparrow)=
{-R^{1/2}c(1-y_2s^2)\over x_3y_3<\bar{y}_2,1><y_1,x_1><y_1,\bar{x}_3>}.
\end{equation}
By relabelling the variable $y_2\leftrightarrow y_3$ and then
eliminate $x_2$ and $y_2$ using Eq.(\ref{F}), the contribution of
$A51$ can be written as
\begin{eqnarray}
Re I^{A51}(\uparrow\uparrow)&=&-R^{1/2}c\{
J_1(m_1+1,1,m_3,m_1+3,n_1+1,n_3+2,1, \\
& & n_3)-s^2 J_1(m_1+1,1,m_3,m_1+3,n_1+1,n_3+3,2,n_3)\}. \nonumber
\end{eqnarray}
Similarly, the imaginary part can be written in terms of $H_1$ and
$H_2$ as 
\begin{eqnarray}
Im I^{A51}(\uparrow\uparrow) &=&-\pi R^{1/2}c\{
H_1(m_1+1,m_3+2,m_3,1,n_1+1,n_3+2,1, \\
& & n_3) -s^2 H_1(m_1+1,m_3+2,m_3,1,n_1+1,n_3+3,2,n_3) \nonumber \\
& &+H_2(m_1+1,1,m_3,m_1+3,n_1+1,n_3+2,1,n_3)-s^2 \nonumber\\
& &H_2(m_1+1,1,m_3,m_1+3,n_1+1,n_3+3,2,n_3)\}. \nonumber
\end{eqnarray}

\section{Monte Carlo Integration of integrals defined by principal
part prescription}
\label{sect.B}
\noindent

As mentioned in Section II, all principal part integrals encountered
in real Compton scattering take the form of $P{1\over
\xi(1-\eta s^2)-\eta c^2+i\epsilon}$ and it is enough
to consider single pole integrals of the form
\begin{equation}
J=\int^1_0 dx P{f(x)\over x-a}, \;\;\; 0<a<1, \;\;\; f(a)\not\!=0.
\end{equation}
However, in our work for virtual case, one cannot always reduce
the integral to the above form. Instead, we have to consider
the following multi-poles integral:
\begin{equation}
J_n=\int^1_0 P {f(x)\over (x-a_1)(x-a_2)...(x-a_n)},\;\; a_i<a_j\;
{\rm for } i<j,
\end{equation}
for $n$ up to four.
The idea of the folding method used in Ref.\cite{KN} is to make
a change of variable  for the integral on one side of the pole
by scaling and reflection  so that the negative large peak
on one side of the pole overlap with the positive large peak
on the other side of the pole.
This assures the local cancellation of the two peaks on both
sides of the pole.
The same idea can be generalized to multi-poles integral
\cite{Nizic}.
The generalized formula can be found in Ref.\cite{Nizic}.
Here, we list explicitly the expression for the transformed
integral $\tilde{J}$ for $n$ up to four: \\
{\it Case 1. 1-pole: }
\begin{equation}
J=\int^1_0 dx {f(x)\over (x-a_1)},
\end{equation}
\begin{eqnarray}
\tilde{J}&=&\int^1_0 dy a_1 {f(x_1)\over (x_1-a_1)} 
+ {\left(a^2_1+(1-2a_1)x_2\right)^2\over a_1(1-a_1)^2}
{f(x_2)\over (x_2-a_1)},
\end{eqnarray}
where $x_1=a_1y$ and $x_2={a_1(1-a_1y)\over a_1+(1-2a_1)y}$. \\
{\it Case 2. 2-poles:}
\begin{equation}
J=\int^1_0 dx {f(x)\over (x-a_1)(x-a_2)},
\end{equation}
\begin{eqnarray}
\tilde{J} &=&
\begin{array}[t]{l}
\int^1_0 dy {a_1 f(x_1)\over (x_1-a_1)(x_1-a_2)}
 + {\left(a^2_1+(a_2-2a_1)x_2\right)^2\over a_1(a_2-a_1)^2}
{f(x_2)\over (x_2-a_1)(x_2-a_2)}  \\ 
 + { \left\{ (a_2-a_1)^2(1-a_2)-(x_3-a_2)\left[
(a_2-a_1)^2-a_1(1-a_2)\right]\right\}^2 \over
a_1(1-a_2)^2(a_2-a_1)^2(x_3-a_1)(x_3-a_2) } f(x_3),
\end{array}
\end{eqnarray}
where 
\begin{eqnarray}
x_1=a_1 y; \;\;\;\ x_2={a_1(a_2-a_1y)\over a_1+(a_2-2a_1)y}; \\
x_3={ a_1a_2(1-a_2)+y\left[(a_2-a_1)^2-a_1a_2(1-a_2)\right] \over
a_1(1-a_2)+y\left[(a_2-a_1)^2-a_1(1-a_2)\right] }.
\end{eqnarray}
{\it Case 3. 3-poles:}
\begin{equation}
J=\int^1_0 dx {f(x)\over (x-a_1)(x-a_2)(x-a_3)},
\end{equation}
\begin{eqnarray}
\tilde{J} &=&
\begin{array}[t]{l}
\int^1_0 dy {a_1 f(x_1)\over (x_1-a_1)(x_1-a_2)(x_1-a_3)}
 + {\left(a^2_1+(a_2-2a_1)x_2\right)^2\over a_1(a_2-a_1)^2}
{f(x_2)\over (x_2-a_1)(x_2-a_2)(x_2-a_3)}  \\ 
+ { \left\{ (a_2-a_1)^2(a_3-a_2)-(x_3-a_2)\left[
(a_2-a_1)^2-a_1(a_3-a_2)\right]\right\}^2 \over
a_1(a_3-a_2)^2(a_2-a_1)^2(x_3-a_1)(x_3-a_2)(x_3-a_3)} f(x_3), \\ 
+ {\left\{ a_1(a_3-a_2)^2(1-a_3)-(x_4-a_3)
\left[ a_1(a_3-a_2)^2-(1-a_3)(a_2-a_1)^2\right]\right\}^2 \over
a_1 (a_2-a_1)^2(a_3-a_2)^2(1-a_3)^2(x_4-a_1)(x_4-a_2)(x_4-a_3)}
f(x_4),
\end{array}
\end{eqnarray}
where
\begin{eqnarray}
x_1&=&a_1 y; \;\;\;\ x_2={a_1(a_2-a_1y)\over a_1+(a_2-2a_1)y};\\
x_3 &=&{ a_1a_2(a_3-a_2)+y\left[a_3(a_2-a_1)^2-a_1a_2(a_3-a_2)\right] \over
a_1(a_3-a_2)+y\left[(a_2-a_1)^2-a_1(a_3-a_2)\right] }; \\
x_4 &=&{ a_1(a_3-a_2)^2+y\left[a_3(1-a_3)(a_2-a_1)^2-a_1(a_3-a_2)^2\right]
\over a_1(a_3-a_2)^2+y\left[(1-a_3)(a_2-a_1)^2-a_1(a_3-a_2)^2\right] }.
\end{eqnarray}
{\it Case 4. 4-poles:}
\begin{equation}
J=\int^1_0 dx {f(x)\over (x-a_1)(x-a_2)(x-a_3)(x-a_4)},
\end{equation}
\begin{eqnarray}
\tilde{J} &=&
\begin{array}[t]{l}
\int^1_0 dy {a_1 f(x_1)\over (x_1-a_1)(x_1-a_2)(x_1-a_3)}
+{\left(a^2_1+(a_2-2a_1)x_2\right)^2\over a_1(a_2-a_1)^2}
{f(x_2)\over (x_2-a_1)(x_2-a_2)(x_2-a_3)}  \\ 
+ { \left\{ (a_2-a_1)^2(a_3-a_2)-(x_3-a_2)\left[
(a_2-a_1)^2-a_1(a_3-a_2)\right]\right\}^2 \over
a_1(1-a_2)^2(a_2-a_1)^2(x_3-a_1)(x_3-a_2)(x_3-a_3)} f(x_3) \\
 + {\left\{ a_1(a_3-a_2)^2(a_4-a_3)-(x_4-a_3)
\left[ a_1(a_3-a_2)^2-(a_4-a_3)(a_2-a_1)^2\right]\right\}^2 \over
a_1 (a_2-a_1)^2(a_3-a_2)^2(a_4-a_3)^2(x_4-a_1)(x_4-a_2)(x_4-a_3)}
f(x_4) \\ 
 + {\left\{ (a_2-a_1)^2(a_4-a_3)^2(1-a_4)-
(x-a_4)\left[(a_2-a_1)^2(a_4-a_3)^2-a_1(a_3-a_2)^2(1-a_4)\right]\right\}^2
\over a_1(a_3-a_2)^2(a_2-a_1)^2(a_4-a_3)^2(1-a_4)^2 } f(x_5),
\end{array}
\end{eqnarray}
where
\begin{eqnarray}
x_1&=&a_1 y; \;\;\;\ x_2={a_1(a_2-a_1y)\over a_1+(a_2-2a_1)y};\\
x_3 &=&{ a_1a_2(a_3-a_2)+y\left[(a_2-a_1)^2-a_1a_2(a_3-a_2)\right] \over
a_1(a_3-a_2)+y\left[(a_2-a_1)^2-a_1(a_3-a_2)\right] }; \\
x_4 &=&{ a_1(a_3-a_2)^2+y\left[a_3(1-a_3)(a_2-a_1)^2-a_1(a_3-a_2)^2\right]
\over a_1(a_3-a_2)^2+y\left[(a_4-a_3)(a_2-a_1)^2-a_1(a_3-a_2)^2\right] }; \\
x_5 &=&
\begin{array}[t]{l}
{ a_1a_4(a_3-a_2)^2(1-a_4)+ y\left[(a_2-a_1)^2(a_4-a_3)^2
-a_1a_4(1-a_4)(a_3-a_2)^2\right] \over
a_1(a_3-a_2)^2(1-a_4)+y\left[(a_2-a_1)^2(a_4-a_3)^2-a_1(1-a_4)(a_3-a_2)^2
\right] }.

\end{array}
\end{eqnarray}

\section{Dirac Proton Form Factor, details of the calculation}
\label{sect.C}
\noindent

The method used is essentially the same as used for the Compton scattering as described 
in Section II, Appendix A, and Appendix B. Diagrams of the 6 types shown in Fig. \ref{vnc_gluon} 
are considered. However, for the form factor, the outgoing photon is omitted from the diagram. As for
Compton scattering, diagrams of types B, D and F can be accounted for by multiplying the result 
for diagrams A, C and E by 2. Since the single photon can be attached in 7 different places in each 
of diagrams A, C and E, there are thus $7\times3=21$ diagrams that need to be considered. FeynComp has 
been used to calculate the hard scattering amplitudes for each of these diagrams, taking the photon 
polarized in the light-cone + direction. Out of the 21 diagrams, it turns out that only 7 have 
non-zero amplitude: A1, A4, A7, C2, C7, E1, E5.

Comparing the expression for the Compton amplitude given in Eq.(\ref{Mexpanded}), we can write our amplitude 
as
\begin{eqnarray}
2\times{4\over 9}(4\pi\alpha_s)^2 
\left[{120f_N\over 8\sqrt{6}}\right]^2 \sum_{d\in A\cup C\cup E}
\sum_{m_1,m_3,n_1,n3}{\cal C}^{(d)}(m_1,m_3|n_1,n_3)
I^{(d)}(m_1,m_3;n_1,n_3),
\end{eqnarray}
where variable names have the same meaning as for Compton scattering. Note that the factor of $4\pi\alpha_{em}$ 
has been omitted. This amplitude can also be written in terms of the proton form factors:
\begin{eqnarray}
\bar{u}(p^{\prime})\left[\gamma^{\mu}F_1(q^2)+\frac{i\sigma^{\mu\nu}}{2m}q_{\nu}F_2(q^2)\right]u(p)\epsilon_{\mu}.
\end{eqnarray}
In the leading twist approximation the $F_2$ term is omitted. Again taking the photon polarized in the light-cone 
+ direction, we arrive at the equation 
\begin{eqnarray}
\bar{u}(p^{\prime})\gamma^{+}F_1(q^2)u(p)=
2\times{4\over 9}(4\pi\alpha_s)^2 
\left[{120f_N\over 8\sqrt{6}}\right]^2 \sum_{d\in A\cup C\cup E}
\sum_{m_1,m_3,n_1,n3}{\cal C}^{(d)}(m_1,m_3|n_1,n_3)
I^{(d)}(m_1,m_3;n_1,n_3).
\end{eqnarray}
Since $\bar{u}(p^{\prime})\gamma^{+}u(p)=2$, we arrive at the following formula for $F_1$: 
\begin{eqnarray}
F_1(q^2)=
{4\over 9}(4\pi\alpha_s)^2 
\left[{120f_N\over 8\sqrt{6}}\right]^2 \sum_{d\in A\cup C\cup E}
\sum_{m_1,m_3,n_1,n3}{\cal C}^{(d)}(m_1,m_3|n_1,n_3)
I^{(d)}(m_1,m_3;n_1,n_3).
\end{eqnarray}
For the COZ distribution amplitude, explicit calculation, following the same procedure used for the Compton
scattering, yields the result
\begin{eqnarray}
F_1(q^2)=
{4\over 9}(4\pi\alpha_s)^2 
\left[{120f_N\over 8\sqrt{6}}\right]^2 \times \frac{80.47}{q^4}.
\end{eqnarray}
This result is in complete agreement with that of Brooks and Dixon \cite{BD}.

\section{Results for low order Appell Polynomials}
\label{sect.D}
\noindent

This section includes results for each of the first four Appell polynomials, $A_0(x_1,x_2,x_3)$, 
$A_1(x_1,x_2,x_3)$, $A_2(x_1,x_2,x_3)$, and $A_0(x_1,x_2,x_3)$. Amplitudes are tabulated for 
$A_i(x)A_i(y)$ and $A_i(x)A_j(y)+A_j(x)A_i(y)$. Real and imaginary parts are given for each 
helicity combination, for angles from $20^o$ to $160^o$, for $R=1.00, 1,25, 1.50, 1.75, 2.00$.

These tables can be used to calculate cross sections and phases for any distribution amplitude
written as
\begin{eqnarray}
\alpha_0A_0+\alpha_1A_1+\alpha_2A_2+\alpha_3A_3.
\end{eqnarray} 
Real and imaginary components of the amplitude can be calculated using the formula
\begin{eqnarray}
\sum\alpha_i^2T_{A_i^2}+\sum_{i<j}\alpha_i\alpha_jT_{A_iA_j},
\end{eqnarray}
where $T_{A_i^2}$ is the value given in the table for $A_i(x)A_i(y)$ and $T_{A_iA_j}$ 
is the value given in the table for $A_i(x)A_j(y)+A_j(x)A_i(y)$.

\clearpage
\begin{table}
\scriptsize
\caption{$A_0(x)A_0(y)$}
\label{Apptab1}
\begin{tabular}{lllllllllll}
\hline
angle &R=1.00 &R=1.25 &R=1.50 &R=1.75 &R=2.00 &R=1.00 &R=1.25 &R=1.50 &R=1.75 &R=2.00\\
\hline
$\uparrow$ to $\uparrow$&Real&&&&&Imag\\
\hline
20&-8.88e-01&7.41e+00&7.28e+00&5.93e+00&4.64e+00&5.11e+00&4.77e+00&1.29e+00&-4.96e-01&-1.30e+00\\
40&-6.48e-02&4.74e-01&4.79e-01&4.09e-01&3.41e-01&2.81e-01&2.91e-01&8.26e-02&-2.58e-02&-7.55e-02\\
60&-1.47e-02&1.10e-01&1.19e-01&1.14e-01&1.10e-01&4.09e-02&5.42e-02&1.66e-02&-3.54e-03&-1.30e-02\\
80&-4.58e-03&4.89e-02&5.70e-02&6.14e-02&6.62e-02&5.22e-03&1.52e-02&5.09e-03&-5.47e-04&-3.26e-03\\
100&-9.46e-04&3.08e-02&3.74e-02&4.28e-02&4.86e-02&-2.95e-03&5.13e-03&1.94e-03&3.56e-05&-8.81e-04\\
120&8.29e-04&2.33e-02&2.80e-02&3.24e-02&3.72e-02&-5.41e-03&2.55e-03&1.19e-03&3.90e-04&-4.72e-05\\
\end{tabular}
\normalsize
\end{table}

\clearpage
\begin{table}
\scriptsize
\caption{$A_1(x)A_1(y)$}
\label{Apptab2}
\begin{tabular}{lllllllllll}
\hline
angle &R=1.00 &R=1.25 &R=1.50 &R=1.75 &R=2.00 &R=1.00 &R=1.25 &R=1.50 &R=1.75 &R=2.00\\
\hline
$\uparrow$ to $\uparrow$&Real&&&&&Imag\\
\hline
20&-2.01e-01&-1.47e-02&-3.31e-03&2.32e-03&1.29e-02&-7.16e-02&-3.82e-02&-5.34e-02&-4.80e-02&-3.83e-02\\
40&-1.41e-02&2.67e-02&3.68e-02&4.72e-02&5.85e-02&-5.83e-03&-2.21e-03&-3.16e-03&-2.89e-03&-2.34e-03\\
60&-3.39e-03&2.70e-02&3.61e-02&4.57e-02&5.61e-02&-1.78e-03&-3.94e-04&-5.61e-04&-5.23e-04&-4.32e-04\\
80&-1.37e-03&2.43e-02&3.23e-02&4.09e-02&5.01e-02&-1.03e-03&-1.29e-04&-1.52e-04&-1.38e-04&-1.14e-04\\
100&-6.78e-04&2.04e-02&2.72e-02&3.44e-02&4.19e-02&-8.94e-04&-9.74e-05&-6.83e-05&-4.86e-05&-3.42e-05\\
120&-1.79e-04&1.60e-02&2.11e-02&2.67e-02&3.27e-02&-1.00e-03&-1.43e-04&-9.21e-05&-5.32e-05&-2.77e-05\\
\end{tabular}
\normalsize
\end{table}

\clearpage
\begin{table}
\scriptsize
\caption{$A_2(x)A_2(y)$}
\label{Apptab3}
\begin{tabular}{lllllllllll}
\hline
angle &R=1.00 &R=1.25 &R=1.50 &R=1.75 &R=2.00 &R=1.00 &R=1.25 &R=1.50 &R=1.75 &R=2.00\\
\hline
$\uparrow$ to $\uparrow$&Real&&&&&Imag\\
\hline
20&-1.50e+00&1.67e+00&1.24e+00&6.67e-01&2.89e-01&1.60e+00&7.71e-01&-4.64e-01&-7.73e-01&-7.60e-01\\
40&-9.85e-02&1.25e-01&1.10e-01&8.60e-02&7.40e-02&8.52e-02&4.82e-02&-2.55e-02&-4.49e-02&-4.47e-02\\
60&-2.03e-02&4.41e-02&4.92e-02&5.24e-02&5.86e-02&1.14e-02&9.07e-03&-4.13e-03&-7.85e-03&-7.97e-03\\
80&-6.35e-03&2.92e-02&3.60e-02&4.27e-02&5.05e-02&9.92e-04&2.34e-03&-1.09e-03&-2.12e-03&-2.19e-03\\
100&-2.17e-03&2.29e-02&2.90e-02&3.53e-02&4.21e-02&-1.07e-03&4.72e-04&-4.88e-04&-7.83e-04&-7.84e-04\\
120&-4.87e-04&1.86e-02&2.31e-02&2.78e-02&3.32e-02&-1.42e-03&-1.52e-04&-3.67e-04&-4.19e-04&-4.04e-04\\
\end{tabular}
\normalsize
\end{table}

\clearpage
\begin{table}
\scriptsize
\caption{$A_3(x)A_3(y)$}
\label{Apptab4}
\begin{tabular}{lllllllllll}
\hline
angle &R=1.00 &R=1.25 &R=1.50 &R=1.75 &R=2.00 &R=1.00 &R=1.25 &R=1.50 &R=1.75 &R=2.00\\
\hline
$\uparrow$ to $\uparrow$&Real&&&&&Imag\\
\hline
20&-2.52e+00&-8.13e-01&-1.42e+00&-1.18e+00&-7.43e-01&7.48e-01&-5.30e-01&-7.85e-02&5.15e-01&7.63e-01\\
40&-1.71e-01&-3.22e-02&-5.54e-02&-2.99e-02&7.43e-03&3.72e-02&-3.33e-02&-5.17e-03&3.15e-02&4.70e-02\\
60&-3.86e-02&1.20e-02&1.68e-02&2.96e-02&4.46e-02&3.87e-03&-6.66e-03&-1.08e-03&5.95e-03&8.94e-03\\
80&-1.45e-02&1.77e-02&2.58e-02&3.54e-02&4.56e-02&-5.18e-04&-2.07e-03&-3.00e-04&1.76e-03&2.61e-03\\
100&-7.47e-03&1.65e-02&2.41e-02&3.16e-02&3.92e-02&-1.21e-03&-6.16e-04&9.77e-05&7.38e-04&9.58e-04\\
120&-5.04e-03&1.39e-02&1.84e-02&2.36e-02&2.95e-02&-1.24e-03&1.58e-04&5.89e-04&6.57e-04&5.58e-04\\
\end{tabular}
\normalsize
\end{table}

\clearpage
\begin{table}
\scriptsize
\caption{$A_0(x)A_1(y)+A_1(x)A_0(y)$}
\label{Apptab5}
\begin{tabular}{lllllllllll}
\hline
angle &R=1.00 &R=1.25 &R=1.50 &R=1.75 &R=2.00 &R=1.00 &R=1.25 &R=1.50 &R=1.75 &R=2.00\\
\hline
$\uparrow$ to $\uparrow$&Real&&&&&Imag\\
\hline
20&-1.29e-01&-4.93e-01&-2.09e-01&-4.37e-02&2.91e-02&-3.02e-01&1.71e-01&3.58e-01&3.28e-01&2.52e-01\\
40&-8.94e-03&-6.11e-02&-5.29e-02&-5.18e-02&-5.69e-02&-2.05e-02&9.18e-03&2.18e-02&2.04e-02&1.59e-02\\
60&-2.60e-03&-3.50e-02&-4.04e-02&-4.74e-02&-5.65e-02&-4.84e-03&1.22e-03&4.06e-03&3.99e-03&3.20e-03\\
80&-1.48e-03&-2.77e-02&-3.44e-02&-4.21e-02&-5.07e-02&-2.24e-03&1.05e-05&1.12e-03&1.22e-03&1.03e-03\\
100&-1.26e-03&-2.26e-02&-2.87e-02&-3.54e-02&-4.26e-02&-1.84e-03&-3.69e-04&2.61e-04&4.26e-04&4.10e-04\\
120&-1.16e-03&-1.78e-02&-2.28e-02&-2.80e-02&-3.37e-02&-2.46e-03&-7.06e-04&-1.89e-04&2.15e-05&1.09e-04\\
\end{tabular}
\normalsize
\end{table}

\clearpage
\begin{table}
\scriptsize
\caption{$A_0(x)A_2(y)+A_2(x)A_0(y)$}
\label{Apptab6}
\begin{tabular}{lllllllllll}
\hline
angle &R=1.00 &R=1.25 &R=1.50 &R=1.75 &R=2.00 &R=1.00 &R=1.25 &R=1.50 &R=1.75 &R=2.00\\
\hline
$\uparrow$ to $\uparrow$&Real&&&&&Imag\\
\hline
20&-1.56e+00&1.91e+00&1.23e+00&4.28e-01&-5.09e-02&4.01e+00&1.22e+00&-4.52e-01&-8.04e-01&-7.24e-01\\
40&-9.81e-02&8.98e-02&4.31e-02&-1.41e-02&-5.41e-02&2.59e-01&8.26e-02&-2.08e-02&-4.41e-02&-4.08e-02\\
60&-2.00e-02&-4.28e-03&-1.84e-02&-3.66e-02&-5.27e-02&5.50e-02&1.86e-02&-1.39e-03&-6.46e-03&-6.38e-03\\
80&-7.01e-03&-1.69e-02&-2.56e-02&-3.67e-02&-4.76e-02&2.01e-02&6.99e-03&9.50e-04&-8.60e-04&-1.10e-03\\
100&-3.90e-03&-1.72e-02&-2.38e-02&-3.19e-02&-4.01e-02&1.08e-02&3.47e-03&1.22e-03&4.07e-04&1.58e-04\\
120&-3.78e-03&-1.45e-02&-1.94e-02&-2.54e-02&-3.17e-02&8.14e-03&1.85e-03&1.02e-03&6.79e-04&5.38e-04\\
\end{tabular}
\normalsize
\end{table}

\clearpage
\begin{table}
\scriptsize
\caption{$A_0(x)A_3(y)+A_3(x)A_0(y)$}
\label{Apptab7}
\begin{tabular}{lllllllllll}
\hline
angle &R=1.00 &R=1.25 &R=1.50 &R=1.75 &R=2.00 &R=1.00 &R=1.25 &R=1.50 &R=1.75 &R=2.00\\
\hline
$\uparrow$ to $\uparrow$&Real&&&&&Imag\\
\hline
20&-7.62e-01&1.88e+00&-1.11e+00&-1.14e+00&-4.30e-01&3.76e+00&-1.88e+00&-7.05e-01&9.57e-01&1.66e+00\\
40&-6.50e-02&7.24e-02&-1.20e-01&-1.34e-01&-9.90e-02&2.26e-01&-1.14e-01&-4.68e-02&5.48e-02&9.92e-02\\
60&-2.08e-02&-1.66e-02&-6.03e-02&-7.02e-02&-7.07e-02&4.06e-02&-2.09e-02&-9.59e-03&9.39e-03&1.80e-02\\
80&-1.05e-02&-2.78e-02&-4.55e-02&-5.34e-02&-5.88e-02&1.01e-02&-5.57e-03&-2.74e-03&2.56e-03&5.01e-03\\
100&-6.97e-03&-2.77e-02&-3.75e-02&-4.34e-02&-4.88e-02&1.66e-03&-1.58e-03&-5.51e-04&1.12e-03&1.83e-03\\
120&-5.28e-03&-2.26e-02&-2.90e-02&-3.32e-02&-3.76e-02&-1.66e-03&-3.14e-04&4.63e-04&9.36e-04&9.64e-04\\
\end{tabular}
\normalsize
\end{table}

\clearpage
\begin{table}
\scriptsize
\caption{$A_1(x)A_2(y)+A_2(x)A_1(y)$}
\label{Apptab8}
\begin{tabular}{lllllllllll}
\hline
angle &R=1.00 &R=1.25 &R=1.50 &R=1.75 &R=2.00 &R=1.00 &R=1.25 &R=1.50 &R=1.75 &R=2.00\\
\hline
$\uparrow$ to $\uparrow$&Real&&&&&Imag\\
\hline
20&-6.27e-02&3.65e-02&7.22e-02&5.83e-02&4.02e-02&2.72e-02&1.77e-01&1.18e-01&6.83e-02&3.61e-02\\
40&-1.27e-03&-2.55e-02&-3.28e-02&-4.33e-02&-5.51e-02&1.46e-03&1.06e-02&7.20e-03&4.20e-03&2.26e-03\\
60&5.47e-04&-2.69e-02&-3.53e-02&-4.48e-02&-5.53e-02&3.57e-04&1.99e-03&1.37e-03&8.07e-04&4.40e-04\\
80&5.80e-04&-2.43e-02&-3.20e-02&-4.05e-02&-4.97e-02&2.51e-04&6.28e-04&4.25e-04&2.53e-04&1.35e-04\\
100&5.47e-04&-2.05e-02&-2.70e-02&-3.41e-02&-4.16e-02&2.58e-04&3.04e-04&2.03e-04&1.19e-04&6.52e-05\\
120&6.38e-04&-1.63e-02&-2.10e-02&-2.63e-02&-3.22e-02&3.09e-04&2.40e-04&1.72e-04&1.11e-04&6.66e-05\\
\end{tabular}
\normalsize
\end{table}

\clearpage
\begin{table}
\scriptsize
\caption{$A_1(x)A_3(y)+A_3(x)A_1(y)$}
\label{Apptab9}
\begin{tabular}{lllllllllll}
\hline
angle &R=1.00 &R=1.25 &R=1.50 &R=1.75 &R=2.00 &R=1.00 &R=1.25 &R=1.50 &R=1.75 &R=2.00\\
\hline
$\uparrow$ to $\uparrow$&Real&&&&&Imag\\
\hline
20&-2.62e-01&-2.96e-01&-1.64e-01&-9.82e-02&-7.90e-02&-3.54e-02&9.31e-02&1.14e-01&8.19e-02&4.96e-02\\
40&-1.91e-02&-4.92e-02&-4.92e-02&-5.46e-02&-6.36e-02&-3.55e-03&5.22e-03&7.00e-03&5.19e-03&3.24e-03\\
60&-4.83e-03&-3.25e-02&-3.92e-02&-4.78e-02&-5.77e-02&-1.34e-03&7.18e-04&1.26e-03&1.01e-03&6.68e-04\\
80&-2.13e-03&-2.63e-02&-3.36e-02&-4.20e-02&-5.10e-02&-9.02e-04&-3.42e-05&2.53e-04&2.54e-04&1.93e-04\\
100&-1.32e-03&-2.08e-02&-2.77e-02&-3.51e-02&-4.28e-02&-8.78e-04&-2.97e-04&-1.00e-04&-1.32e-05&1.50e-05\\
120&-9.69e-04&-1.31e-02&-1.99e-02&-2.70e-02&-3.38e-02&-1.13e-03&-5.24e-04&-3.76e-04&-2.42e-04&-1.45e-04\\
\end{tabular}
\normalsize
\end{table}

\clearpage
\begin{table}
\scriptsize
\caption{$A_2(x)A_3(y)+A_3(x)A_2(y)$}
\label{Apptab10}
\begin{tabular}{lllllllllll}
\hline
angle &R=1.00 &R=1.25 &R=1.50 &R=1.75 &R=2.00 &R=1.00 &R=1.25 &R=1.50 &R=1.75 &R=2.00\\
\hline
$\uparrow$ to $\uparrow$&Real&&&&&Imag\\
\hline
20&1.18e+00&2.67e-01&6.20e-02&3.80e-02&7.05e-02&1.27e-01&-6.33e-02&2.03e-01&3.36e-01&3.79e-01\\
40&8.89e-02&-9.94e-03&-3.41e-02&-4.37e-02&-5.10e-02&1.05e-02&-8.08e-03&1.00e-02&1.97e-02&2.32e-02\\
60&2.29e-02&-2.42e-02&-3.63e-02&-4.50e-02&-5.40e-02&3.53e-03&-2.85e-03&1.06e-03&3.46e-03&4.43e-03\\
80&9.91e-03&-2.52e-02&-3.40e-02&-4.15e-02&-4.95e-02&2.22e-03&-1.37e-03&-1.06e-04&8.68e-04&1.37e-03\\
100&5.77e-03&-2.49e-02&-3.15e-02&-3.70e-02&-4.29e-02&1.90e-03&-7.84e-04&-3.40e-04&2.24e-04&5.69e-04\\
120&4.31e-03&-2.92e-02&-3.33e-02&-3.54e-02&-3.84e-02&2.11e-03&-5.30e-04&-5.18e-04&-7.42e-05&2.80e-04\\
\end{tabular}
\normalsize
\end{table}

\end{document}